\documentstyle[12pt,aasms4]{article}

\slugcomment{Accepted to the Astrophysical Journal, Part I}

\lefthead{Freedman {\it et~al.}}
\righthead{HST Key Project Summary}

\begin{document}

\title{Final Results from the   Hubble Space Telescope Key Project  to
Measure the Hubble Constant
\footnote{Based on observations 
with the 
NASA/ESA \it Hubble Space Telescope\rm, obtained at the Space Telescope 
Science Institute, which is operated by AURA, Inc., under NASA Contract No. 
NAS 5-26555.}}

\author{Wendy L. Freedman\altaffilmark{2}, 
Barry F. Madore\altaffilmark{2,3},
Brad K. Gibson\altaffilmark{4}, 
Laura Ferrarese\altaffilmark{5}, 
Daniel D. Kelson\altaffilmark{6},
Shoko Sakai\altaffilmark{7},
Jeremy R. Mould\altaffilmark{8},
Robert C. Kennicutt, Jr.\altaffilmark{9},
Holland C. Ford\altaffilmark{10}, 
John A. Graham\altaffilmark{6}, 
John P. Huchra\altaffilmark{11}, 
Shaun M.G. Hughes\altaffilmark{12},
Garth D. Illingworth\altaffilmark{13}, 
Lucas M. Macri\altaffilmark{11} and
Peter B. Stetson\altaffilmark{14},\altaffilmark{15} }

\altaffiltext{2}{The Observatories, Carnegie Institution of Washington, Pasadena, CA, USA  91101}
\altaffiltext{3}{NASA/IPAC Extragalactic Database, California Institute of Technology, Pasadena, CA, USA  91125}
\altaffiltext{4}{Centre for Astrophysics \& Supercomputing, Swinburne University of Technology, Hawthorn, Victoria 3122, Australia}
\altaffiltext{5}{Rutgers University, New Brunswick, NJ, 08854}
\altaffiltext{6}{Department of Terrestrial Magnetism, Carnegie Institution of Washington, 5241 Broad Branch Rd. N.W., Washington, D.C., USA  20015}
\altaffiltext{7}{National Optical Astronomy Observatories, P.O. Box 26732, Tucson, AZ, USA  85726}
\altaffiltext{8}{Research School of Astronomy \& Astrophysics, Australian National University, Weston Creek Post Office, Weston, ACT, Australia  2611}
\altaffiltext{9}{Steward Observatory, University of Arizona, Tucson, AZ, USA  85721}
\altaffiltext{10}{Department of Physics \& Astronomy, Bloomberg 501, Johns Hopkins University, 3400 N. Charles St., Baltimore, MD, USA  21218}
\altaffiltext{11}{Harvard-Smithsonian Center for Astrophysics, 60 Garden St., Cambridge, MA, USA  02138}
\altaffiltext{12}{Institute of Astronomy, Madingley Road., Cambridge, UK  CB3~0HA}
\altaffiltext{13}{Lick Observatory, University of California, Santa Cruz, CA, USA 95064}
\altaffiltext{14}{Dominion Astrophysical Observatory, Herzberg Institute of Astrophysics, National Research Council, 5071 West Saanich Rd., Victoria, BC, Canada  V8X~4M6}
\altaffiltext{15} {Guest User, Canadian Astronomy Data Centre, which is operated
by the Herzberg Institute of Astrophysics, National Research Council of Canada.}


\def\eg{{\rm e.g., }}
\def\ie{{\rm i.e., }}
\def\etal{{\rm et~al.~}}
\def\hub{{\rm H$_0$~}}
\def\h0units{{\rm km\,s$^{-1}$\,Mpc$^{-1}$}}
\def\tf{{\rm Tully-Fisher }}
\def\fp{{\rm fundamental plane }}
\def\snia{{\rm Type Ia supernovae }}
\def\snii{{\rm Type II supernovae }}
\def\sbf{{\rm surface--brightness fluctuations }}
\def\ltsim{\mathrel{\hbox{\rlap{\hbox{\lower4pt\hbox{$\sim$}}}\hbox{$<$}}}}
\let\la=\simlt
\def\gtsim{\mathrel{\hbox{\rlap{\hbox{\lower4pt\hbox{$\sim$}}}\hbox{$>$}}}}
\let\ga=\gtsim


\begin{abstract}

We present here the final results of the  Hubble Space Telescope (HST)
Key Project to measure the Hubble constant.   We summarize our method,
the results and the uncertainties, tabulate our revised distances, and
give the implications of these results for cosmology.  Our results are
based on a  Cepheid calibration of  several secondary distance methods
applied over the range of about 60 to 400 Mpc.  The analysis presented
here benefits  from a number of   recent improvements and refinements,
including  (1)  a larger  LMC Cepheid  sample   to define the fiducial
period--luminosity (PL)  relations, (2) a  more recent  HST Wide Field
and  Planetary  Camera   2 (WFPC2)   photometric  calibration, (3)   a
correction  for   Cepheid  metallicity,  and  (4)  a   correction  for
incompleteness  bias in the observed Cepheid   PL samples.  We adopt a
distance  modulus to the   LMC (relative  to  which  the more  distant
galaxies are  measured) of $\mu_0$(LMC) =  18.50 $\pm$ 0.10 mag, or 50
kpc.  New, revised distances are given  for the 18 spiral galaxies for
which Cepheids  have been  discovered as  part of  the Key Project, as
well as for  13 additional galaxies with  published Cepheid data.  The
new calibration results in  a Cepheid distance to  NGC 4258  in better
agreement with  the  maser distance to this   galaxy.   Based on these
revised Cepheid distances, we find values (in km/sec/Mpc) of \hub = 71
$\pm$ 2$_r$ (random)  $\pm$  6$_s$ (systematic) (type  Ia supernovae),
\hub =  71 $\pm$3$_r$ $\pm$  7$_s$ (Tully--Fisher relation), \hub = 70
$\pm$ 5$_r$ $\pm$ 6$_s$  (surface brightness fluctuations), \hub =  72
$\pm$ 9$_r$ $\pm$ 7$_s$ (type II supernovae), and 82 $\pm$ 6$_r$ $\pm$
9$_s$ (fundamental plane).  We combine these results for the different
methods with 3  different  weighting schemes, and find  good agreement
and consistency with \hub = 72 $\pm$ 8  \h0units.  Finally, we compare
these results with other, global methods for measuring \hub.

\end{abstract}


\clearpage

\section{Introduction}

In standard  Big Bang cosmology,  the universe expands  uniformly; and
locally, according to  the Hubble law, v  = H$_0$  d,  where v  is the
recession velocity  of a galaxy  at  a distance  d,  and H$_0$ is  the
Hubble constant,  the expansion rate at the  current epoch.  More than
seven decades have now  passed since Hubble (1929) initially published
the correlation between the distances  to galaxies and their recession
velocities, thereby providing    evidence  for the expansion   of  the
universe.  But pinning down an  accurate value for the Hubble constant
has  proved  extremely challenging.  There  are many  reasons for this
difficulty,  but primary  among   them  is  the basic difficulty    of
establishing accurate   distances over    cosmologically   significant
scales.

The  Hubble constant   enters   in  a  practical  way  into   numerous
cosmological and  astrophysical calculations.  H$_0^{-1}$ sets the age
of the   universe, t$_0$, and   the size  of  the observable universe,
R$_{obs}$ = ct$_0$,  given a knowledge of  the total energy density of
the universe.  The square  of the  Hubble  constant relates  the total
energy density of  the universe to its  geometry (Kolb \& Turner 1990;
Peacock 1999).  In addition, the  Hubble constant defines the critical
density of  the universe, $\rho_{crit} = {  3 H^2 \over  8  \pi G } $.
The critical density  further specifies the epoch  in  the universe at
which  the density  of  matter and radiation  were  equal, so that the
growth of structure in the universe is also dependent on the expansion
rate.  The  determination of many physical  properties of galaxies and
quasars (\eg mass, luminosity,  energy density) all  require knowledge
of the  Hubble constant, as  does  the proportion of  primordial light
elements (H, D,  $^3$He, $^4$He and Li) synthesized  in the  first few
minutes after the Big Bang.

Measuring an accurate value of H$_0$ was one of the motivating reasons
for building the NASA/ESA Hubble Space Telescope  (HST).  Thus, in the
mid-1980's, measurement  of H$_0$ with  the goal of 10\%  accuracy was
designated as one of three ``Key Projects'' of the HST, and teams from
the  astronomical community were   encouraged to propose  to undertake
these initiatives \footnote{The other  two Key Projects selected  were
Quasar Absorption Lines, and the  Medium-Deep Survey.}.  A team headed
by the  late Dr.  Marc Aaronson  began preparing our proposal in 1984;
following peer review  (subsequent   to the Challenger   explosion  in
1986), our  group  was awarded the Key  Project   on the Extragalactic
Distance  Scale in 1986.   Very sadly, Marc  met a tragic and untimely
death in  1987.  We  began our  initial  observations of   the closest
galaxies in our sample  in 1991, shortly  after the launch of HST, but
most of the project was   carried out after the refurbishment  mission
(in December 1993)  when a new  camera with optics that  corrected for
the spherical aberration of the primary mirror was installed.

The overall goal of the H$_0$ Key Project (hereafter, Key Project) was
to measure H$_0$   based  on a  Cepheid  calibration  of  a number  of
independent,   secondary  distance determination  methods.   Given the
history of   systematic errors  dominating the accuracy    of distance
measurements, the approach we adopted was to avoid relying on a single
method alone, and   instead    to average  over   the  systematics  by
calibrating and  using  a number of    different methods.  Determining
H$_0$ accurately requires the measurement of distances far enough away
that both the small and  large--scale motions of galaxies become small
compared to   the overall Hubble  expansion.   To extend  the distance
scale beyond   the range of the  Cepheids,  a number of   methods that
provide relative distances were chosen.  We  have used the HST Cepheid
distances to  provide an absolute  distance scale  for these otherwise
independent   methods,  including    the   Type  Ia    supernovae, the
Tully-Fisher relation,  the fundamental plane for elliptical galaxies,
surface-brightness fluctuations, and Type II supernovae.

The previous  29 papers in this  series have provided the distances to
individual  galaxies  based   on  the discovery   and  measurement  of
Cepheids,  discussed  the calibration  of  the data, presented interim
results  on the   Hubble constant, and  provided   the calibration  of
secondary methods,  and their individual  determinations of the Hubble
constant.   A  recent  paper by Mould  et  al.    (2000a) combines the
results   for secondary methods (Gibson   \etal  2000; Ferrarese \etal
2000a; Kelson  \etal 2000; Sakai \etal 2000)  with  a weighting scheme
based on  numerical simulations of the  uncertainties.  In this paper,
we present  the  final, combined  results of the   Key  Project.  This
analysis benefits from significant recent refinements and improvements
to  the Cepheid period-luminosity  relation, as well  as the HST WFPC2
photometric  scale, and puts all  of the data for  the Key Project and
other efforts  onto a new common  zero point.   Establishing plausible
limits  for the  Hubble  constant requires a  careful investigation of
systematic errors.   We   explicitly note  where  current   limits  in
accuracy have   been  reached.  We  intend  this paper  to  provide an
assessment of the status of the global value of H$_0$.

In  this paper, we summarize  our method  and determination of Cepheid
distances  in   \S\ref{kpdescribe}      and  \S\ref{cephdist}.      In
\S\ref{flowfd}  and \S\ref{cephh0},  we apply   a correction for   the
nearby flow field and compare the value of H$_0$ obtained locally with
that determined   at greater distances.   Secondary  methods,  and the
determination of \hub on large  scales are discussed in \S\ref{h0} and
\S\ref{combineh0}.    The  remaining sources   of  uncertainty in  the
extragalactic distance scale  and determination of H$_0$ are discussed
in \S\ref{systematics}. In \S\ref{othermethods} we compare our results
to methods      that  can be  applied    directly  at  high redshifts,
specifically  the   Sunyaev--Zel'dovich    and  gravitational lensing
techniques.  In \S\ref{cosmology},  we give the implications  of these
results for cosmology.

\section{ Description of the Key Project  }
\label{kpdescribe}

\subsection{ Goals }

The   main aims  of the  Key  Project were   (Aaronson \& Mould  1986;
Freedman \etal  1994a; Kennicutt, Freedman \& Mould  1995): (1) To use
the high resolving power of HST to discover Cepheids in, and determine
distances to,   a sample of  nearby  ($\ltsim$  20 Mpc) galaxies,  and
establish an accurate local distance scale.  (2) To determine H$_0$ by
applying  the  Cepheid    calibration to   several  secondary distance
indicators operating further     out  in the Hubble    flow.    (3) To
intercompare  the Cepheid and  other distances to provide estimates of
the external uncertainties for all   of the methods.  (4) To   conduct
tests of the  universality of the Cepheid period--luminosity relation,
in particular as a function of metal abundance.  Finally, an ancillary
aim was to measure Cepheid distances to a  small number of galaxies in
each of the two nearest clusters  (Virgo and Fornax) as an independent
check on other Hubble constant determinations.

Why  was   HST  necessary  for an  accurate   determination   of \hub?
Atmospheric seeing sets the practical limit for resolving Cepheids and
measuring  well--defined  period--luminosity relations  to only  a few
megaparsecs.  The superb and essentially  non-varying image quality of
HST extends  that limit  tenfold, and the  effective search   volume a
thousandfold.  Furthermore, HST offers  a unique capability in that it
can be scheduled  optimally  to facilitate  the discovery of   Cepheid
variables.  Observations can be  scheduled independently of  the phase
of the Moon,  the time of  day, or weather, and   there are no  seeing
variations.  Before  the launch of  HST,  most  Cepheid searches  were
confined to our own   Local Group of galaxies,   and the very  nearest
surrounding  groups  (M101, Sculptor and    M81 groups; see Madore  \&
Freedman 1991; Jacoby \etal 1992).  At that time, only 5 galaxies with
well-measured Cepheid  distances provided the absolute  calibration of
the  Tully-Fisher  relation  (Freedman   1990)  and a  single  Cepheid
distance,  that     for  M31,  provided     the  calibration   for the
surface-brightness  fluctuation method (Tonry 1991).  Moreover, before
HST {\it no} Cepheid calibrators were available for Type Ia supernovae
(although one historical, nearby  type Ia supernova, SN1885A, had been
observed in M31).

\subsection{ Choice of Target Galaxies / Observing Strategy }

In each nearby target spiral galaxy in the Key Project sample, Cepheid
searches were undertaken in regions active in  star formation, but low
in   apparent dust extinction,   based  on ground--based, photographic
images (\eg Sandage  \& Bedke 1988).  To  the largest extent possible,
we avoided high--surface--brightness   regions  in order  to  minimize
source confusion or crowding.  For each galaxy,  over a two-month time
interval,  HST images in  the visual  (V-band, 5550  \AA), and in  the
near-infrared  (I band, 8140 \AA),  were made using the corrected Wide
Field and Planetary  Camera 2 (WFPC2).  Among  the galaxies on the Key
Project observing  program, only M81 and an  outer field  in M101 were
observed with   the original Wide  Field   / Planetary camera (WF/PC),
before the first  HST  servicing mission  that restored the  telescope
capabilities. Two of the Type Ia supernova calibrators investigated by
the  Sandage,  Tammann  \etal  team and   rediscussed   here were also
observed with WF/PC:  IC 4182 and NGC 5253.   The field of view of the
WFC2 is  L--shaped with each of the  3 cameras covering 1.33 arcmin by
1.33 arcmin on the sky, and the PC 35 arcsec by 35 arcsec.

For the observations, two    wavelength bands were chosen to    enable
corrections for dust extinction, following   the precepts of  Freedman
(1988) and Madore \& Freedman (1991).  Initially, during the observing
window, 12 epochs at V (F555W), and  4 observations at I (F814W), were
obtained.  For   some of the  galaxies observed  early in the program,
some B  (F439W)  data were also obtained.   For  the  targets observed
later in the program,  observations were obtained at  both V and I  at
each of  the 12 epochs.  An  additional observation was generally made
either  one year earlier or later,  to increase the  time baseline and
reduce aliasing errors, particularly for the longer-period stars.  The
time distribution of the  observations was set  to follow a power-law,
enabling the detection and  measurement  of Cepheids  with a range  of
periods optimized  for minimum   aliasing   between 10  and 50    days
(Freedman \etal 1994b).

Since each individual secondary method is likely to be affected by its
own (independent) systematic uncertainties,  to reach a final  overall
uncertainty of $\pm$10\%,  the numbers of  calibrating  galaxies for a
given method  were chosen  initially  so that  the final (statistical)
uncertainty on the zero point for that method would be only $\sim$5\%.
(In practice, however,  some methods end up  having higher weight than
other methods, owing  to their smaller intrinsic  scatter, as well  as
how far  out into  the Hubble flow   they can be   applied --  see  \S
\ref{combineh0}).   In   Table  \ref{tbl:calibrators}, each  method is
listed  with its mean dispersion, the   numbers of Cepheid calibrators
pre-- and post--HST,  and the standard  error  of the mean.  (We  note
that   the  fundamental  plane   for   elliptical galaxies cannot   be
calibrated  directly by Cepheids; this method  was not included in our
original proposal, and it has the largest uncertainties.  As described
in \S\ref{fp},  it is calibrated by  the Cepheid distances to 3 nearby
groups and clusters.)  The calibration of Type  Ia supernovae was part
of the original Key Project proposal, but  time for this aspect of the
program was awarded to a team led by Allan Sandage.


For  the Key Project, Cepheid  distances were obtained for 17 galaxies
chosen    to  provide a   calibration  for  secondary   methods, and a
determination of H$_0$.  These galaxies lie at distances between 3 and
25 Mpc.  They are located  in the general field,  in small groups (for
example, the M81 and the Leo I groups at 3  and 10 Mpc, respectively),
and in  major clusters (Virgo and Fornax).   An additional target, the
nearby spiral galaxy, M101, was chosen to enable a test of the effects
of  metallicity on the  Cepheid  period-luminosity relation.  HST  has
also  been used to  measure Cepheid distances  to 6 galaxies, targeted
specifically  to be useful for the  calibration  of Type Ia supernovae
(\eg Sandage \etal 1996).  Finally, an HST distance to a single galaxy
in the Leo I group, NGC 3368, was measured by Tanvir and collaborators
(Tanvir \etal  1995,  1999).  Subsequently and fortuitously,  NGC 3368
was host  to  a Type Ia   supernova, useful for calibrating  \hub (Jha
\etal  1999; Suntzeff  \etal  1999).  \footnote{In addition, recently,
SN1999by occurred in NGC 2841, a galaxy for which Cepheid observations
have been taken in Cycle 9 (GO-8322).}

We list the galaxies which we have used  in the calibration of \hub in
Table 2, along with the methods that they calibrate.  To summarize the
total Cepheid calibration sample, as part of  the Key Project, we have
surveyed and analyzed data for 18 galaxies, in addition to reanalyzing
HST archival data for 8 galaxies observed by other groups.  When these
distances  are combined  with those for  5  very nearby galaxies (M31,
M33, IC  1613, NGC  300, and  NGC  2403), it results   in  a total  31
galaxies, subsets of which  calibrate individual secondary methods, as
shown in Table 2.


\subsection{ Key Project Archival Database }

As part of  our original time allocation  request for the Key Project,
we proposed  to provide all of our  data in an  archive  that would be
accessible  for the general astronomical  community. We envisaged that
the Cepheid distances  obtained  as part of    the Key Project   would
provide a  database   useful for  the   calibration of  many secondary
methods,  including those that might be  developed in the future.  For
each  galaxy  observed as part    of  the Key   Project, the   Cepheid
positions,     magnitudes,  and     periods   are   available       at
http://www.ipac.caltech.edu/H0kp/H0KeyProj.html.       In    addition,
photometry  for non-variable  stars that  can be  used for  photometry
comparisons, as well as   medianed (non-photometric) images  for these
galaxies are also  available. These images are  also archived  in NED,
and can   be    accessed    on   a galaxy-by-galaxy      basis    from
http://nedwww.ipac.caltech.edu.

\subsection{ Photometry }
\label{photometry}

As a  means of guarding  against systematic errors specifically in the
data reduction phase, each galaxy within the  Key Project was analyzed
by  two  independent  groups within  the  team, each   using different
software packages: DoPHOT (Schechter {\it  et al.}  1993; Saha {\it et
al.}  1994), and  ALLFRAME  (Stetson 1994,1996).  The latter  software
was developed specifically for the optimal  analysis of data sets like
those of the Key Project, consisting of  large numbers of observations
of a single   target field.  Only  at the  end of   the data reduction
process (including  the Cepheid selection and distance determinations)
were  the  two groups' results  intercompared.  This ``double--blind''
procedure  proved extremely valuable.  First,  it  allowed us to catch
simple (operator)  errors.  And, it also  enabled us to provide a more
realistic  estimate of  the  external data  reduction  errors for each
galaxy distance.  The limit to the accuracy of the photometry that can
be obtained in these galaxy  fields is set  by the sky (\ie unresolved
galaxy) background in  the frames, and  ultimately, the  difficulty in
determining aperture corrections.  Each of the two packages deals with
sky determination and aperture corrections  in different ways, thereby
providing  a means of  evaluating  this systematic uncertainty in  the
Cepheid photometry.   As  discussed   in \S\ref{artificial},  we  also
undertook  a  series of artificial  star  tests to better quantify the
effects of  crowding, and to  understand the limits  in each  of these
software packages (Ferrarese {\it et al.}, 2000c).

\subsection{ Calibration }
\label{cal}

The determination of accurate distances carries with  it a requirement
for  an accurate, absolute  photometric  calibration.  Ultimately, the
uncertainty in the Hubble constant from  this effort rests directly on
the  accuracy   of the  Cepheid  magnitudes themselves,    and  hence,
systematically on  the CCD zero--point  calibration.  In view   of the
importance  of  this issue for the  Key Project,  we undertook our own
program to provide  an independent  calibration of both the WF/PC  and
WFPC2 zero points, complementary to the efforts of the teams who built
these instruments, and the Space  Telescope Science Institute.   These
calibrations have been described  in Freedman \etal (1994b) and Kelson
\etal (1995)  for WF/PC  and  Hill \etal (1998),  Stetson (1998),  and
Mould \etal (2000a) for WFPC2.

As part  of an HST  program to study Galactic globular   clusters, but
also extremely  valuable  for the  photometric  calibration of  WFPC2,
hundreds  of   images of $\omega$ Cen,   NGC  2419, and M92  have been
obtained both  on the ground and with  HST over the last several years
(Stetson 1998; Mould \etal 2000a).  Despite this extensive effort, the
calibration  of  WFPC2 remains a    significant  source of  systematic
uncertainty in the determination of H$_0$.  This lingering uncertainty
results from the  difficulty  in  characterizing the  charge  transfer
efficiency (CTE) properties  of  the WFPC2,  which  turn out  to be  a
complicated function  of position on  the chip, the brightness  of the
object, the   brightness  of   the  sky, and  the wavelength   of  the
observations  (presumably because of  the differing background levels;
Stetson  1998; Whitmore,  Heyer \& Casertano  1999; Saha  \etal  2000;
Dolphin 2000).

Recent WFPC2 calibrations (Stetson 1998; Dolphin 2000) differ from our
earlier calibration    based  on Hill \etal     (1998).   Based on the
reference star  photometry published in  papers  IV to  XXI in the Key
Project       series,  Mould  \etal      (2000a)    found   that   the
reddening--corrected distance moduli on the Stetson (1998) system were
0.07 $\pm$ 0.02 mag closer, in the mean, than those published based on
the  Hill   \etal  (1998)      system.   This  difference    in    the
reddening--corrected  distance moduli  results  from a  0.02 mag  mean
offset in the V--band, and a 0.04 mag  mean offset in the I--band. The
more  recent calibrations are based   on a more extensive  calibration
data  set than that  available in  the  Hill  \etal or the  Saha \etal
analyses,  and they result in  galaxy distance moduli that are closer.
The main  reason for this difference   is that the earlier  Hill \etal
``long'' versus ``short'' zero points determined for globular clusters
(bright stars on faint  sky) turned out  to  be inappropriate for  the
Cepheid fields (faint stars on bright sky) because the combinations of
flux  dependence and background  dependence were  different in the two
situations.    Stetson   (private communication)   indicates   that  a
0.02--0.03  mag uncertainty remains due  to this  effect.  The Stetson
CTE correction is in agreement with  Dolphin (2000) and Whitmore \etal
(1999):  the  Stetson  zero   point  results  in  reddening--corrected
distance  moduli  that   agree within 1.5\%   (0.03  mag)  of the  new
calibration   by Dolphin  (2000).   Although  Stetson  did not find  a
significant time dependence as seen in the more recent studies, in all
studies,  the  temporal variation  of the CTE  ramps  are found  to be
negligible for the high background long exposures for the Key Project.

In this paper,  we have adopted  the WFPC2 calibration due  to Stetson
(1998),  and applied a   --0.07   $\pm$ 0.04  mag  correction to   the
reddening--corrected  distance  moduli.  The  uncertainty reflects the
remaining differences in  the published WFPC2 calibrations,  and their
impact on the distance moduli, when corrected for reddening (Equations
3,4).  As we shall  see later in \S\ref{systematics}, the  uncertainty
due  to   the  WFPC2 photometric zero   point   remains  a significant
systematic  error affecting  the measurement of  \hub.  Unfortunately,
until  linear,  well-calibrated detectors can  be  applied  to the Key
Project   reference  stars, this      uncertainty is unlikely    to be
eliminated.

\section{  The Cepheid Distance Scale }
\label{cephdist}

The Cepheid period--luminosity  relation remains the most important of
the  primary distance indicators for  nearby  galaxies.  The strengths
and  weaknesses of Cepheids have  been reviewed extensively (\eg Feast
\& Walker 1987;  Madore \& Freedman 1991;  Jacoby \etal 1992; Freedman
\& Madore  1996;  Tanvir 1999).   However, since the  Cepheid distance
scale  lies at the heart  of the \hub  Key  Project, we summarize both
its advantages and disadvantages briefly here again.

The strengths of Cepheids  are, of course,  many: they are amongst the
brightest  stellar indicators, and   they are relatively young  stars,
found in abundance in spiral galaxies.  Thus, many independent objects
can be observed   in a single   galaxy.   Their large  amplitudes  and
characteristic   (saw--tooth)  light  curve   shapes facilitate  their
discovery and identification; and they have  long lifetimes and hence,
can  be   reobserved at other  times,   and  other wavelengths (unlike
supernovae, for example).  The Cepheid period--luminosity relation has
a small  scatter (\eg in the  I--band, the dispersion  amounts to only
$\sim \pm$0.1 mag: Udalski \etal  1999).  Moreover, Cepheids have been
studied and theoretically modelled  extensively; the reason  for their
variability is well-understood to be a consequence of pulsation of the
atmosphere,  resulting    from a   thermodynamic,  valve-like  driving
mechanism  as (primarily)  helium  is cycled  from  a singly to doubly
ionized state, and the opacity increases with compression.

There  are  also difficulties    associated  with  measuring   Cepheid
distances.  First,  since Cepheids are  young stars, they are found in
regions  where there   is dust scattering,  absorption and  reddening.
Corrections must be  made for extinction, requiring assumptions  about
the universal behavior of Cepheids at different wavelengths, and about
the    universality of the  Galactic   extinction  law.  Extinction is
systematic, and its effects must either be removed by multicolor data,
or  minimized by observing at long  wavelengths, or both.  Second, the
dependence of the PL  relation  on chemical composition  (metallicity)
has  been very  difficult to  quantify.   Third, an accurate geometric
calibration of the PL relation, at  any given metallicity, has not yet
been   established.  Fourth, as  the distance  of the galaxy increases
(and the  resolution decreases),   finding and   measuring  individual
Cepheids  becomes increasingly   difficult  due to  crowding  effects.
Finally, the reach of  Cepheids  is currently  (with HST)  confined to
spiral galaxies   with  distances less   than  about  30 Mpc.   Hence,
Cepheids alone cannot be observed at sufficient distances to determine
H$_0$ directly,  and an  accurate determination   of \hub requires  an
extension to other methods.

\subsection{ Adopted Method for Measuring Cepheid Distances }
\label{reddening}

The  application of the  PL relation  for the Key  Project follows the
procedure  developed in  Freedman   (1988), and extended in  Freedman,
Wilson \& Madore (1991).  The Large Magellanic Cloud (LMC) PL relation
has been used as fiducial, and  a distance modulus  of $\mu_0$ = 18.50
mag (a  distance of 50 kpc), and  a mean reddening  of  E(V--I) = 0.13
(E(B--V) = 0.10) mag (Madore \& Freedman 1991) have been adopted.  The
LMC V-- and  I--  band PL relations   are fit by least-squares to  the
target spiral data to determine apparent distance moduli in each band.
A  reddening--corrected distance  modulus and  differential absorption
with    respect   to   the  LMC   are  obtained    using  a  ratio  of
total-to-selective absorption $ R =  A_V / (A_V - A_I)  = 2.45 $  (\eg
Cardelli,  Mathis \& Clayton 1989).  This   procedure is equivalent to
defining a reddening--free   index W =  V --  R (V--I) (Madore   1982;
Freedman 1988; Freedman, Wilson \& Madore 1991).


\subsection{ Effect of Metallicity on the Cepheid Period--Luminosity Relation}
\label{cephmet}

A longstanding uncertainty in the Cepheid distance  scale has been the
possibility that the zero  point of the   PL relation is  sensitive to
chemical composition (Freedman \& Madore 1990 and references therein).
It  is  only within  the last decade  or  so, that major observational
efforts   to  address the metallicity  issue   for  Cepheids have been
undertaken.  Accurately establishing the  size of a metallicity effect
for Cepheids  alone has proven to  be very challenging,  and the issue
has not  yet been definitively resolved (see  Freedman  \etal 2001 and
references therein).  However,  although neither the  magnitude of the
effect nor its wavelength dependence have yet been firmly established,
the observational  and theoretical evidence  for an effect is steadily
growing.  Published   empirical values for    the index $\gamma$  (see
Equation 5  in \S\ref{adopted}  below) range from  0 to  --1.3 mag/dex
(with most  values between 0   and --0.4), but these published  values
have been  derived  using  a   variety of different   combinations  of
bandpasses.     Since   the     effects    of       metallicity    are
wavelength-dependent, it  is critical  that the appropriate correction
for a given dataset be applied.

Some recent theoretical models (\eg  Chiosi,  Wood \& Capitanio  1993;
Sandage, Bell  \& Tripicco 1999;  Alibert \etal 1999; Bono \etal 1999,
2000) suggest that at the VI  bandpasses of the  \hub Key Project, the
effect of metallicity is   small,  $\gamma_{VI} \sim$  -0.1   mag/dex.
Unfortunately, the sign of the effect is still uncertain. For example,
Caputo,  Marconi \& Musella (2000) find  a slope of 0.27 mag/dex, with
the opposite sign.  Thus, for the present, calibrating the metallicity
effect based on models alone is not feasible.

A  differential, empirical test for  the effects of metallicity on the
Cepheid distance  scale  was first  carried out  by Freedman \& Madore
(1990) for the  nearby galaxy M31.    As part of  the  Key Project, we
carried out a  second differential test   comparing two fields  in the
face--on  galaxy, M101 (Kennicutt \etal 1998).   These two studies are
consistent with there being a shallow  metallicity dependence, but the
statistical significance  of each test    is individually low.   As  a
follow-on to the optical study,  H-band NICMOS observations have  been
obtained for the two fields previously observed in the optical in M101
(Macri \etal 2001).  A comparison of the  VIH photometry for the inner
and outer field is consistent with a metallicity sensitivity of the PL
relations, but artificial star tests  in the inner field indicate that
crowding is  significant, and  precludes an accurate  determination of
the magnitude of the effect.  Other recent studies (\eg Sasselov \etal
1997; Kochanek 1997) conclude that a metallicity effect is extant, and
all of the empirical studies  agree on the sign,  if not the magnitude
of  the effect.  Considering all  of the  evidence presently available
and  the   (still considerable)   uncertainties, we    therefore adopt
$\gamma_{VI}$ = --0.2 $\pm$  0.2 mag/dex, approximately the mid--range
of   current empirical  values, and    correct  our Cepheid  distances
accordingly.

\subsection{ Adopted Period--Luminosity Relations }
\label{adopted}

For  earlier  papers in this  series,  we adopted the  slopes and zero
points for the LMC  V-- and I--  PL relations from Madore  \& Freedman
(1991), based on 32 Cepheids.  These  PL relations are consistent with
those published  by Feast \& Walker  (1987).  However, the OGLE survey
has recently  produced a significantly  larger sample of $\sim$650 LMC
Cepheids (Udalski  \etal    1999).  This sample  has   extensive phase
coverage at BVI magnitudes and covers the period range  of 0.4 $<$ log
P $<$ 1.5.  As part of the Key Project, we also undertook observations
of a sample  of  105 LMC  Cepheids  (Sebo  \etal 2001), and  these  PL
relations are in very good statistical agreement with those of Udalski
\etal adjusting to a common distance to the LMC.  For about 60 objects
common to both   samples, with P$>$8  days  and having  both  V and  I
magnitudes, the offsets are  -0.004 $\pm$ 0.008 mag  in I,  and +0.013
$\pm$ 0.010 mag in  V (Sebo \etal).   The Sebo \etal sample extends to
longer periods   ($\sim$40 days),  and has   10 Cepheids  with periods
greater than 30 days, the limit of  the Udalski \etal sample. These 10
Cepheids are all  well fit by,  and  all lie within  1-$\sigma$ of the
period--luminosity slopes defined by  the  Udalski \etal sample.   The
Udalski \etal data are clearly the most extensive to date, and we thus
adopt their  apparent PL relations  as fiducial  for the reanalysis in
this paper.

The  Udalski \etal (1999) PL  calibration adopts a distance modulus of
18.2 mag,   based  on  a  distance  determined   using the red   clump
technique, whereas, as discussed above, in this paper, we adopt a true
distance modulus to the LMC  of 18.50 mag.  With  this modulus and the
reddening--corrected Udalski  \etal Cepheid data  to define the slopes
and errors, our adopted M$_V$ and M$_I$ PL relations become:

\begin{equation}
\rm
M_V  = -2.760 ~[\pm 0.03] ~(log P - 1) - 4.218 ~[\pm 0.02]  ~~~(\sigma_V = \pm 0.16)
\end{equation}
\begin{equation}
\rm
M_I  = -2.962 ~[\pm 0.02] ~(log P - 1)  - 4.904 ~[\pm 0.01]  ~~~(\sigma_I = \pm 0.11)
\end{equation}

\noindent
In the absence  of a metallicity dependence,  and  correcting only for
reddening, the  true distance moduli  ($\mu_0$) can be calculated from
the apparent V and I distance moduli ($\mu_V$ and $\mu_I$) as follows:

\begin{equation}
\rm
\hspace{-2cm} \mu_0 = \mu_W = \mu_V - R(\mu_V - \mu_I) =  2.45\mu_I - 1.45\mu_V
\end{equation}
\begin{equation}
\rm
~~~~~~~~~~~~~~~~~~~~~~~~= W + 3.255 ~[\pm 0.01] ~(log P - 1) + 5.899 ~[\pm 0.01]  ~~~(\sigma_W = \pm 0.08)
\end{equation}

As  discussed in more  detail    in 
\S\ref{revised}, it is the change  in slope of  the I-band PL relation
that has the most impact on the resulting distances.

Allowing   for    a   correction  term     $\delta    \mu_z$  for    a
metallicity-dependence  of the  Cepheid PL relation  in   terms of the
observed HII   region abundance of oxygen   relative  to hydrogen (see
\S\ref{metallicity}), the true distance modulus becomes:

\begin{equation}
\rm
\mu_0 = \mu_V - R(\mu_V - \mu_I) + \delta \mu_z  
\end{equation}

\noindent
where $\delta    \mu_z$  = $\gamma_{VI}$ ([O/H]    - [O/H]$_{LMC}$) is
applied to the reddening-corrected (VI)  modulus, and $\gamma_{VI}$ is
measured in mag/dex (where a dex  refers to a  factor of 10 difference
in metallicity).

\subsection{ New Revised Cepheid Distances }
\label{revised}

Over the 6   years that we have  been   publishing data  from the  Key
Project, our analysis methods, as well as the photometric calibration,
have evolved and improved.  Hence, the sample of published Key Project
distances  has not been  analyzed  completely homogeneously.  In  this
paper, we have redetermined the true moduli to each galaxy used in the
Key Project.  These distances are calculated  with the new calibration
described above, and with    attention  to  minimizing bias at     the
short--period end   of the  PL  relation, as  described below   and by
Freedman \etal 1994b; Kelson \etal 1994; Ferrarese \etal 2000b.

In this  analysis we have  (1) consistently adopted only the published
Cepheid photometry  which  were  reduced  using the  ALLFRAME  stellar
photometry reduction package,  whose  phase points  were  converted to
mean    magnitudes   using   intensity-weighted    averages  (or their
template-fitted equivalents).   \footnote{For  Key Project   galaxies,
both phase--weighted and intensity--weighted magnitudes were generally
calculated  for each of  the galaxies,  and found  to  be in very good
agreement.   This is  to be  expected,  since  the optimal  scheduling
results in  well--sampled phase coverage.}   (2) To compensate for the
small ($\sim$ 0.01 mag)  mean bias in  the PL fits (see the discussion
in \S\ref{bias} and Appendix  A), we have also  applied period cuts to
the  PL relations, thereby   eliminating the  shortest-period Cepheids
where magnitude incompleteness effects become important.  In two cases
(NGC 3368 and NGC 300), a single  long-period Cepheid was also dropped
because of stochastic effects at  the bright (sparsely populated)  end
of the  PL relation,  which can  similarly bias  solutions.   The mean
correction for  this magnitude-limited   bias  is  small  ($+$1\%   in
distance),  but it  is  systematic, and  correcting  for it results in
larger distances than are determined without this faint-end debiasing.
(3) We  have adopted a --0.07 mag  correction to the Hill \etal (1998)
WFPC2  calibration  to be consistent with  Stetson  (1998) and Dolphin
(2000).  Finally,  (4) we have   adopted the published  slopes of  the
Udalski \etal (1999) PL relations.

The adoption  of the new Udalski \etal  (1999)  PL slopes alone  has a
dramatic, and unanticipated effect on the previously published Cepheid
distances  based on the Madore  \&  Freedman (1991) calibration.  Most
importantly, the change is distance dependent.   The V and I PL slopes
for the Madore \& Freedman calibration, based on  32 stars, are --2.76
$\pm$ 0.11 and --3.06 $\pm$ 0.07, respectively.  The new Udalski \etal
(1999) values for these same   quantities are  --2.76 $\pm$ 0.03   and
--2.96 $\pm$ 0.02, (Equations  1 and 2).   Although the V--band slopes
agree  identically, and the   I--band slopes differ  by only  0.1, the
impact on   the  derived  reddenings,   and  therefore  distances,  is
significant.   The  new  calibration predicts higher   reddenings, and
therefore smaller distances.  In addition,  because of the  difference
in  (V--I) slope, the   new  relation predicts  systematically  larger
reddenings  for  Cepheids of  increasing  period.  As   a result,  the
differences  in distance between the  previous and the new calibration
will  be  largest for galaxies  at greater  distances,  where the mean
period   of  the  samples is  larger   (since  a  greater  fraction of
shorter--period Cepheids will  fall  below the detection  threshold in
the most distant targets).

Expressing the  divergence of the two  calibrations as a correction to
the true modulus (in the sense of Udalski \etal (1999) minus Madore \&
Freedman (1991)):

\begin{equation}
\rm
\Delta\mu_0 = -0.24(log P - 1.0) ~~mag
\end{equation}

\noindent
The two calibrations agree at around 10 days in period.  At 20 days
the correction amounts to less than a 4\% decrease in distance.  At 30 
days, this difference is 6\%, and it rises to 9.5\% (or --0.19 mag in
distance modulus) at 60 days.

In Table \ref{tbl:cephdist}, this new calibration is applied to all KP
galaxies and  other    Cepheid  distances  from   HST    observations.
Corrections for  metallicity are applied in Table \ref{tbl:finaldist}.
In addition, we present revised  VI moduli for  M33, M31, IC 1613, NGC
300, and I-band for NGC 2403.  These galaxies were previously observed
from the ground, and with the exception of IC 1613 (which was observed
with NICMOS), have also been used as calibrators for secondary methods
for the Key Project.  We have  not included other dwarf galaxies (like
NGC 6822 or  WLM) which are not calibrators  for the secondary methods
adopted in this paper.   The  fits were done  using the  same standard
procedure described in \S\ref{reddening}, and adopting Equation 3.  To
make  it  clear  where   the differences  lie   compared to   previous
calibrations, we list in columns  1--4 the galaxies, distance  moduli,
errors, and  number of Cepheids fit, based  on the Madore  \& Freedman
(1991) LMC PL relations, and ALLFRAME magnitudes,  for an LMC distance
modulus of 18.50 mag.  In columns 5 and 6, we list distance moduli and
errors for fits to the same Cepheid samples adopting the Udalski \etal
(1999) PL slopes.   In columns 7, 8, and   9, distance moduli, errors,
and number  of  Cepheids fit  are  given, after   imposing period cuts
correcting for PL bias as described above.  Finally, the references to
the  sources for the  Cepheid  photometry is  given in  column 10.  In
Table \ref{tbl:finaldist}, we  list the galaxy name,  apparent V and I
distance moduli  and  PL-fitting  (random) errors,   E(V--I), distance
moduli on  the   new calibration  for the  case   where no metallicity
correction has   been applied ($\delta    \mu_z$  = 0),  and where   a
correction of $\delta \mu_z$ = -0.2  mag/dex is adopted.  In addition,
we list  the distance in  Mpc  and the  metallicities for the  Cepheid
fields.   For    ease   of   comparison,     columns   7  of     Table
\ref{tbl:cephdist}, and 8  of Table \ref{tbl:finaldist}, are the  same
distance moduli values, uncorrected for metallicity.


The errors  on the Cepheid distances  are  calculated as follows.  The
random uncertainties, $\sigma_{random}^2$ are  given by: $$ \sigma_W^2
/  (N-1) $$ where   N is the number   of Cepheids observed  in a given
galaxy. The error in W, $\sigma_W^2$ includes the random errors in the
photometry minus the correlated  scatter along a reddening  trajectory
(from   equation  3).     The  systematic   errors    are  given   by:
$$\sigma_{systematic}^2   =    \sigma_{zp}^2     +     \sigma_{z}^2  +
\sigma_{WFPC2}^2 +\sigma_{apcorr}^2 $$ with corresponding terms due to
the  uncertainty in the LMC  zero  point, metallicity, the photometric
zero point and  aperture corrections. A  further  discussion of errors
can be found in Madore \etal 1999; Ferrarese \etal 2000b.


There are 3 interesting effects of the differential distance-dependent
effect in  adopting the new Udalski  \etal (1999) calibration.  First,
the absolute magnitudes of  the  Type Ia supernovae, which  previously
produced  lower values of  the Hubble  constant  in comparison to  the
other  Key  Project secondary   distance   indicators  now  come  into
systematically   better correspondence  (\S\ref{sn}).  Second, another
apparent divergence in the Cepheid distance  scale is also ameliorated
by this new calibration; that of the  difference between the maser and
the Cepheid    distance   to  NGC  4258.  As    discussed   further in
\S\ref{n4258}, adopting the  Key Project fitting methodology, ALLFRAME
photometry, template--fitted  magnitudes, and the new calibration, the
Cepheid distance  to NGC  4258 comes into  better  agreement  with the
maser distance to this galaxy (Herrnstein \etal 1999). And third, the
reddening solutions for  two galaxies, NGC 300  and IC 4182 previously
yielded negative values.  The adoption of the new Udalski \etal (1999)
slopes results in positive reddening solutions for both these (and now
all) galaxies with measured Cepheid distances.

\section{ The Local Flow Field }
\label{flowfd}

Before proceeding with a determination of the Hubble constant, we turn
our attention to the question of the  local flow field, recalling that
\hub requires a solid knowledge of both distances and velocities.  The
large-scale distribution of matter in the nearby universe perturbs the
local  Hubble flow,  causing peculiar  motions.   If uncorrected  for,
these perturbations  can be a  significant  fraction of  the  measured
radial velocity, particularly  for the  nearest  galaxies.  The  local
flow  field has been  modeled extensively by  a number of authors (\eg
Tonry  \etal 2000). In  general,  there is good qualitative  agreement
amongst different studies.  On average,  these peculiar motions amount
to $\sim$200--300 km/sec (Tonry \etal; Giovanelli \etal 1999), but the
flow field is complicated locally  by the presence of massive,  nearby
structures, most  notably,  the Virgo Cluster.   At 3,000  km/sec, the
peculiar  motion for  an individual object  can  amount to a  7--10\%
perturbation,  whereas for Type   Ia  supernovae (which reach  out  to
30,000 km/sec), these effects drop to less than 1\%, on average.

For  the nearest galaxies, the  effects of the local peculiar velocity
field, and the resultant  uncertainty in \hub  can be quite large. For
example, a recent  study by Willick  \&  Batra (2000) finds  values of
\hub = 85 $\pm$ 5 and 92 $\pm$ 5  \h0units based on applying different
local velocity  models     to 27 Cepheid galaxies     within  $\sim$20
Mpc. However, the velocity model of Han \& Mould  (1990) applied to 12
Cepheid distances fits best with \hub $\sim$  70 \h0units (Mould \etal
1996). Some of this difference reflects a difference in calibration of
the  surface--brightness--fluctuation method. However, the  remaining
large  discrepancies serve to emphasize  that the Key Project strategy
of extending secondary   distance measurements beyond  100 Mpc,  where
recession velocities have  become large,  is  preferable to any  local
determination.

For the Key Project, we  have corrected the observed galaxy velocities
for  the  local   flow field as    described in  Mould  \etal  (2000a,
2001).\footnote{Note that the signs in  Equation A2 published in Mould
\etal 2000a are wrong in  the text; however, they  were correct in the
code used to do the calculations.}  A linear  infall model composed of
3 mass concentrations (the Local   Supercluster, the Great  Attractor,
and  the   Shapley  concentration)  is   constructed  with  parameters
estimated  from    existing catalogs of   Tully-Fisher  distances  and
velocities.  In \S\ref{bulk},   we return to  the  question of whether
there  is evidence  for  a  bulk (or   non-converging) flow on  larger
scales.

\section{ Cepheid Hubble Diagram }
\label{cephh0}

\medskip

A Hubble diagram  for 23 galaxies with  Cepheid distances is shown  in
Figure   \ref{fig:veldistceph}.  The   galaxy velocities   have   been
corrected for the flow field model described above.  The error bars in
this plot reflect  the  difference between the predictions   from this
flow field and those of Tonry \etal (2000). A fit to the data yields a
slope of   75 $\pm$ 10 \h0units, excluding   systematic errors.  As we
shall see in  \S\ref{combineh0} below, the  scatter is larger in  this
Hubble diagram than for the secondary  methods that operate at greater
distances; however, the mean value  of \hub for  nearby galaxies is in
very  good  agreement    with    the   distant sample.    In     Table
\ref{tbl:cephvel}, we  give the  uncorrected, heliocentric  velocities
for the Cepheid galaxies, and the velocities as successive corrections
are added: corrections for  the  Local Group,  the Virgo cluster,  the
Great  Attractor,  and the    Shapley concentration.   The  velocities
plotted  include  all of these   corrections.  For comparison, we also
list the velocities calculated from the Tonry \etal (2000) flow model,
using our  Cepheid  distances, and  assuming \hub  = 78 \h0units,  and
$\Omega_m$  = 0.2,  as in their   paper.   There are some  differences
between the simple flow model that we have adopted and the Tonry \etal
model, most significantly, the Fornax  cluster galaxies.  Our  adopted
triple attractor model yields a quieter flow at Fornax, and reproduces
the  cosmic microwave background frame.   The  agreement for the Virgo
cluster,  however,  is excellent.   Again this comparison demonstrates
the   importance  of  measuring    \hub  at   large  distances   where
uncertainties in the velocities become unimportant.



\section{ Relative Distance Methods and H$_0$ }
\label{h0}

For the determination of H$_0$, a given method for measuring distances
should satisfy   several basic criteria  (\eg  Freedman  1997): (a) It
should  exhibit high  internal precision;  (b)  have a solid empirical
calibration; (c) ideally it should  be  applicable to large  distances
(and therefore not  subject    to  significant systematics    due   to
large-scale  flows);  and  also  (d)  ideally  it should  be  based on
straightforward physics.  As discussed further  below, based on  these
criteria, each of the relative distance indicators  has its own merits
and  drawbacks.  For example, Type  Ia supernovae (SNIa) have a number
of advantages relative to other methods: currently they can be applied
at the greatest distances ($\sim$  400 Mpc) and the internal precision
of  this    method is very  high.   But   finding   them is difficult:
supernovae are   rare objects, and  separating  the supernova from the
background light of the galaxy is challenging in  the inner regions of
galaxies. Moreover, for nearby galaxies, surveying for supernovae is a
time-consuming  process  that must  be  done  on  a galaxy--by--galaxy
basis.   The  internal precision of the surface-brightness-fluctuation
(SBF) method is also very high, but this method currently has the most
limited  distance range,  of   only $\sim$70 Mpc.   Of somewhat  lower
internal precision is the  Tully-Fisher (TF) relation,  but it can  be
applied   out   to  intermediate  distances    ($\sim$150  Mpc).   The
fundamental  plane (FP) for   elliptical  galaxies can be  applied, in
principle,  out to  to z$\sim$1, but  in   practice, stellar evolution
effects limit this method  to z$\ltsim$0.1 ($\sim$400 Mpc).  Moreover,
since elliptical galaxies do not contain  Cepheids, the FP calibration
currently   relies on  less direct group/cluster   distances.  Each of
these distance indicators is  now discussed briefly.  The results from
these methods are then combined in \S\ref{combineh0}.

\subsection{ Type Ia Supernovae }
\label{sn}

One of the most promising cosmological distance indicators is the peak
brightness of Type~Ia   supernovae.  Of longstanding interest   (e.g.,
Kowal~1968;  Sandage  \&  Tammann    1982), this  secondary  indicator
currently    probes further  into  the  unperturbed   Hubble flow, and
possesses the   smallest intrinsic scatter  of  any  of the indicators
discussed thus  far.  A simple lack   of Cepheid calibrators prevented
the     accurate  calibration of type     Ia  supernovae prior to HST.
Substantial improvements to the supernova distance scale have resulted
both from recent dedicated, ground-based supernova search and followup
programs yielding  CCD  light curves  (e.g.,  Hamuy \etal  1995, 1996;
Riess \etal  1998, 1999), as well  as a campaign  to  find Cepheids in
nearby galaxies which  have been host to  Type Ia supernovae  (Sandage
\etal 1996; Saha \etal 1999).

An  ALLFRAME analysis of  the  Cepheid distances  to Type Ia supernova
hosts, and   a  comparison with    the published  DoPHOT  results  was
undertaken by Gibson \etal  (2000) as part of the  Key Project.  Using
the same pipeline reduction methods that we applied to  all of the Key
Project galaxies, we independently derived  Cepheid distances to seven
galaxies  that were hosts  to  Type Ia  supernovae.  We found  that on
average, our new   distance  moduli were 0.12$\pm$0.07\,mag   (6\%  in
distance) smaller than  those  previously published (see  Gibson \etal
Table 4).  Adopting the recalibrated  distances, and applying these to
the reddening--corrected Hubble relations of  Suntzeff et~al.  (1999),
Gibson    \etal determined  a  value  of  \hub  =68  $\pm$ 2$_{\rm r}$
$\pm$5$_{\rm s}$\,km\,s$^{-1}$\,Mpc$^{-1}$.  In general, the published
DoPHOT Cepheid photometry and our ALLFRAME analysis agrees quite well,
at or significantly better than the 1--$\sigma$ level, with the I-band
data tending to show poorer agreement.  Thus, photometric reduction is
not  the  major source  of the   difference.    A variety  of reasons,
detailed by Gibson {\it et al.}, lead  to the differences in the final
distance moduli.

In principle, one could average  the distances  determined by the  two
groups.  However, in some cases,  there are very clear--cut reasons to
prefer  the  Gibson  \etal  results.  For   example,   in the case  of
NGC~4536, the WFC2 chip  results are discrepant (by  0.66 mag)  in the
Saha et~al.  (1996a) DoPHOT analysis, whereas  the Saha \etal analysis
of the other three chips agrees with our ALLFRAME analysis of all four
WFPC chips.   Parodi \etal (2000) have  attributed  this difference to
uncertainties in aperture corrections,   and  continue to  prefer   to
average  all 4 chips together.   However,  given their quoted aperture
correction uncertainties  (0.10  to 0.15 mag), and  the  fact that our
analysis  reveals  no  such difference   in aperture correction,  this
appears to be an unlikely explanation.  For the case of NGC~4639, Saha
\etal (1997)  introduced a different weighting  scheme for that galaxy
only; however, in our analysis we  find no significant difference in a
weighted  or unweighted fit.  Our preferred  approach is  to treat the
fitting of all of the galaxies and their reddening determinations in a
consistent    manner,  rather  than   adopting different   schemes for
individual galaxies.

The  supernova Hubble relation  calibrated by Gibson  \etal (2000) was
that of Suntzeff \etal (1999), based upon a subsample of 35 supernovae
from Hamuy \etal (1996) and Riess \etal (1998).  A larger total sample
of nearby  supernovae  is now available  as  a result of  the  ongoing
search program  of Riess \etal  (1999).  In this paper,  we add  21 of
these  additional 22 supernovae to  the original Hamuy \etal sample of
29; only  SN1996ab is not  considered further, as   its redshift is in
excess of the regime over which the Hamuy \etal (1993) $k$-corrections
are  applicable.   For     completeness,  in  the   first    panel  of
Figure~\ref{fig:sn1a}, we  show the  raw,  uncorrected,  B,  V, and  I
Hubble diagrams for this full set of 50 supernovae.


Following    Jha    \etal   (1999),   in   the     middle   panel   of
Figure~\ref{fig:sn1a}, we show the B, V, and I Hubble diagrams for the
subset of 36 supernovae  having 3.5 $<\log$(cz)$_{\rm  CMB}$ $<$4.5, and
peak  magnitude colors $|$B$_{\rm   max}-$V$_{\rm  max}|\le$0.20.   In
addition,  a correction for the  internal reddening of the host galaxy
E(B$-$V)$_{\rm Host}$, from  Phillips \etal (1999), has  been applied.
In the third panel of Figure~\ref{fig:sn1a}, our  adopted subset of 36
supernovae have had  their peak magnitudes  corrected for their  light
curve shape, via  application of a simple linear  fit to  the relation
between   decline rate $\Delta   m_{15}$(B) and  peak magnitude.  This
correction echoes  that adopted  in the  original  Hamuy \etal  (1996)
analysis, as  opposed to the quadratic  fits adopted by Phillips \etal
(1999) and Gibson \etal (2000); however, we find no difference in the
result whether a linear or quadratic fit is adopted.

Adopting our default Hubble relations (Figure~\ref{fig:sn1a}), coupled
with   the zero points    provided  by our  revised Cepheid  distances
(applying a  metallicity  correction of --0.2  $\pm$  0.2 mag/dex)  to
NGC~4639,    4536,  3627,   3368,   5253,  and    IC~4182  from  Table
\ref{tbl:finaldist},   yields a value of  \hub  = 71 $\pm$ 2$_{\rm r}$
$\pm$6$_{\rm s}$\,km\,s$^{-1}$\,Mpc$^{-1}$.    This  value    can   be
compared to that from Gibson  \etal (2000) of \hub  = 68 $\pm$ 2$_{\rm
r}$ $\pm$5$_{\rm   s}$\,km\,s$^{-1}$\,Mpc$^{-1}$.   The  difference in
\hub compared to Gibson \etal comes from the new calibration of the PL
relation, a  metallicity correction, and our  adoption of  an expanded
supernovae sample. An  identical error  analysis  to that employed  by
Gibson  \etal   was assumed  here.    The velocities,  distances, \hub
values, and uncertainties for  the 36 type  Ia supernovae used in this
analysis are listed in Table \ref{tbl:snveldist}.



\noindent

\subsection{ The Tully--Fisher Relation }

For spiral   galaxies,  the total (corrected   to face-on inclination)
luminosity is strongly  correlated with the maximum  rotation velocity
of the galaxy (corrected to edge-on  inclination), which is useful for
measuring  extragalactic distances  (Tully  \& Fisher  1977;  Aaronson
\etal 1986;    Pierce \& Tully  1988;  Giovanelli   \etal  1997).  The
Tully-Fisher relation at present is the most commonly applied distance
indicator: thousands of distances are now  available for galaxies both
in the general field, and in groups and clusters.  The scatter in this
relation is approximately  $\pm$0.3 mag (Giovanelli \etal 1997;  Sakai
\etal   2000; Tully \&  Pierce 2000),  or $\pm$15\% in  distance for a
single galaxy.  In a broad   sense, the Tully-Fisher relation can   be
understood in  terms of the  virial  relation applied  to rotationally
supported disk   galaxies,     under the assumption    of   a constant
mass-to-light  ratio (Aaronson, Mould    \& Huchra 1979).   However, a
detailed self-consistent,    physical  picture that   reproduces   the
Tully-Fisher relation (\eg Steinmetz \& Navarro 1999), and the role of
dark matter in   producing  almost universal  spiral galaxy   rotation
curves (Persic, Salucci,   \&  Stel 1999;  McGaugh   \etal 2000) still
remain a challenge.

Macri \etal  (2000) obtained new  BVRI photometry, and using published
data, remeasured   line  widths  for the   Cepheid  galaxies that  are
Tully-Fisher calibrators.  Sakai \etal (2000) applied this calibration
to a sample of 21 clusters out to 9,000  km/sec observed by Giovanelli
\etal (1997), and to an  H-band sample  of  10 clusters from  Aaronson
\etal (1982, 1986).  With  an adopted distance  to the LMC of 50  kpc,
Sakai \etal  determined a value of \hub  = 71 $\pm$ 4$_r$  $\pm$ 7$_s$
\h0units.  Based on  the  same set of  Key  Project Cepheid calibrator
distances, the same LMC zero point, and a compilation of BRIK data for
Tully-Fisher  cluster galaxies from  the  literature, Tully  \& Pierce
(2000) determined a value  of \hub = 77 $\pm$  8 \h0units (at a quoted
95\%  confidence level). In  the I-band,  where  there is good overlap
with Tully and Pierce, Sakai \etal found \hub=73$\pm$2$_r$ $\pm$ 9$_s$
km/sec/Mpc.   Based on analyses  using  an earlier available subset of
Cepheid calibrators, Giovanelli \etal  (1997) concluded that \hub = 69
$\pm$  5 \h0units, consistent with  Madore  \etal (1998), who obtained
\hub = 72 $\pm$ 5$_r$ $\pm$ 7$_s$ \h0units.  However, for a consistent
set  of calibrators, the difference in  these values probably reflects
some of the   systematic  uncertainties inherent in   implementing the
Tully-Fisher technique.  Tully \&   Pierce discuss at length  possible
reasons for the source  of  the differences amongst  various published
values of \hub based  on the Tully-Fisher  relation, but they conclude
that the reason for much of this discrepancy remains unresolved.

Adopting the same  Tully--Fisher (BVIH)   galaxy sample discussed   in
Sakai \etal (2000), applying the  new PL calibration, and adopting the
metallicity--corrected   distances  for  the Tully--Fisher calibrators
given in  Table \ref{tbl:finaldist}, results in  a value of H$_0$ = 71
$\pm$3$_r$  $\pm$   7$_s$  \h0units, with no     net  change from that
published by Sakai \etal The adopted  distances and velocities for the
Tully-Fisher   clusters   used in this    analysis  are given in Table
\ref{tbl:tfveldist}.  Also tabulated are  the velocities in the cosmic
microwave background frame, and \hub values and uncertainties.


\subsection{  Fundamental Plane for Elliptical Galaxies }
\label{fp}

For elliptical galaxies,   a  correlation exists between  the  stellar
velocity dispersion and  the intrinsic  luminosity (Faber \&   Jackson
1976),  analogous  to   the relation   between   rotation velocity and
luminosity  for  spirals.  Elliptical galaxies  are  found to occupy a
`fundamental    plane'     (r$_e         \propto      \sigma^{\alpha}$
$<$I$>$$^{\beta}_e$) wherein a defined galaxy effective radius (r$_e$)
is  tightly  correlated with   the surface  brightness  (I$_e$) within
r$_e$,   and  central velocity   dispersion of  the  galaxy ($\sigma$)
(Dressler \etal 1987; Djorgovski \& Davis 1987); and $\alpha \sim$ 1.2
and $\beta \sim$  --0.85 (Djorgovski \& Davis  1987).  The scatter  in
this relation is approximately  10--20\% in distance for an individual
cluster.

Jorgensen,  Franx \&  Kjaergaard (1996) have  measured the fundamental
plane  for 224 early-type galaxies in   11 clusters spanning cz $\sim$
1,000    to 11,000 km/sec.   Kelson  \etal  (2000)  provided a Cepheid
calibration for the distant clusters based on Key Project distances to
spiral galaxies in the Leo I group, and the Virgo and Fornax clusters,
yielding  \hub =  78 $\pm$  5$_r$ $\pm$ 9$_s$  \h0units.   The revised
Cepheid distances presented  in this paper result  in new distances to
the  Virgo cluster,  the Fornax cluster,  and the  Leo I  group (Table
\ref{tbl:clusters}).  The  galaxies in these  objects are  amongst the
most distant  in the Key   Project  sample, and  they also  have  high
metallicities.   Hence, the  new  calibration impacts  the fundamental
plane more than for  the other secondary  methods analyzed here.   The
new calibration  yields \hub =  82  $\pm$ 6$_r$  $\pm$ 9$_s$ \h0units,
adopting a metallicity  correction  of --0.2  $\pm$ 0.2  mag/dex.  The
numbers of galaxies,  adopted distances,  velocities, \hub values  and
uncertainties for  the clusters in  this analysis  are  given in Table
\ref{tbl:fpveldist}.



\subsection{ Surface Brightness Fluctuations }

Another  method  with high  internal precision, developed  by Tonry \&
Schneider  (1988); Tonry \etal 1997; and  Tonry \etal (2000) makes use
of the fact  that the resolution of  stars within galaxies is distance
dependent.  This  method is  applicable to  elliptical galaxies or  to
spirals with prominent bulges.  By normalizing to the mean total flux,
and correcting for an observed color dependence, relative distances to
galaxies can be  measured.  The  intrinsic  scatter of this method  is
small: a factor of three  improvement compared to the Tully-Fisher and
D$_n-\sigma$ relations makes the   method an order of  magnitude  less
susceptible to Malmquist biases.  Application  of the method  requires
careful  removal of  sources  of noise   such  as  bad pixels  on  the
detector, objects  such  as  star  clusters, dust   lanes,  background
galaxies, and  foreground stars.  With  HST, this method is  now being
extended  to  larger  distances  (Lauer  \etal  1998);  unfortunately,
however, only 6  galaxies  beyond the  Fornax cluster have   published
surface-brightness fluctuation distances, with only 4 of them accurate
enough to be of  interest for cosmology.   Furthermore, all lie within
the very narrow range $cz =$ 3800 to 5800 km/s, where local flow-field
contributions  to  the observed   velocities are  still non-negligible
($\sim$15\% $v_{CMB}$).

As part  of the Key Project, Ferrarese  et al.  (2000a) applied an HST
Cepheid calibration   to the 4   Lauer \etal (1998) SBF  galaxies, and
derived $H_0$ = 69 $\pm$ 4$_r$  $\pm$ 6$_s$ km/s/Mpc.  The results are
unchanged if  all  6 clusters  are included.   The largest sources  of
random  uncertainty are the    large--scale  flow corrections  to  the
velocities,    combined with the    very   sparse sample of  available
galaxies.    Most of the systematic   uncertainty  is dominated by the
uncertainty in the Cepheid calibration of the method itself (Ferrarese
et  al.  2000a, Tonry et al.   2000).  These three factors account for
the 10\%  difference between the SBF-based  values of $H_0$ derived by
the KP and  that by Tonry et  al. (2000).  Flow--corrected velocities,
distances, and  \hub values for the  6 clusters  with SBF measurements
are    given  in   Table   \ref{tbl:sbfveldist}.    Applying  our  new
calibration, we  obtain $H_0$  = 70 $\pm$  5$_r$  $\pm$ 6$_s$ km/s/Mpc
applying a metallicity correction  of  --0.2 mag/dex, as described  in
\S\ref{cephdist}.

\subsection{ Type II Supernovae }
\label{snii}

Type II supernovae result  from massive stars.   They are fainter, and
show  a wider variation in   luminosity than  the type Ia  supernovae.
Although not ``standard   candles'',   type II supernovae  can   yield
distances through application    of the Baade-Wesselink technique   to
their   expanding atmospheres.  By   following the  time  evolution of
spectra for  the  expanding atmosphere    (yielding the radius   as  a
function of  time and velocity), in  combination with  the photometric
angular size (yielding the ratio of the radius  to the distance of the
supernova),  the distance to  the supernova  can  be obtained.  Recent
applications  of this technique have  been undertaken by Schmidt \etal
(1994) and Eastman \etal (1996) using   detailed model atmospheres  to
correct  for the  scattering in   the  atmosphere.  In principle,  the
method can  be applied independent  of the  local  calibration of  the
extragalactic distance scale.  The diversity  of different methods  is
critical  in  constraining  the   overall  systematic   errors  in the
distances measured as  part of the Key  Project,  since the underlying
physics of expanding  supernova atmospheres is completely  independent
of the Cepheid  distance scale and its  calibration.  Based on 16 Type
II supernovae, covering a range of redshifts  from cz = 1100 to 14,600
km/sec, Schmidt  \etal (1994)  determine a value  of  \hub = 73  $\pm$
6$_r$ $\pm$ 7$_s$ \h0units.

In Table \ref{tbl:snii}, we list  the 3 galaxies currently having both
Cepheid  and  Type  II supernovae  (SNII)    distances.  The Type   II
supernovae distances are from  Schmidt,  Kirshner, \& Eastman  (1994).
The distances  from   the two  methods  agree well  within the  quoted
errors,  and a weighted  fit for the  three calibrators  yields a mean
difference in the distance moduli of 0.09 $\pm$ 0.14 mag, in the sense
of the Cepheids giving slightly  shorter distances.  A fourth  galaxy,
NGC 3627, also  has  both a Cepheid and  a  Type II distance, but  the
latter has a quoted  uncertainty of $\pm$1.00 mag.  We did not include
the observed SNII for M81, M100, or NGC 1559 because Schmidt, Kirshner
\& Eastman comment that these supernovae are peculiar SNII.  There are
4 galaxies in the Schmidt \etal (1994) sample having velocities in the
range  $\sim$2000  $<$ v$_{CMB}$ $<$ 14,000.    If we apply  a Cepheid
calibration based on the distances  to the LMC,  M101, and NGC 7331 to
these distant SNII, for which we adopt velocities corrected to the CMB
frame, we find  \hub =  72 $\pm$  9$_r$  $\pm$ 7$_s$ \h0units.    This
result does not  change  if the  Cepheid  distances are corrected  for
metallicity since two  of the calibrators (the  LMC and M101) are  not
affected by  the  metallicity term,   and  the difference in  distance
modulus for NGC  7331  is only 0.03  mag.  Hence,  the value   of \hub
remains unchanged after    applying a metallicity  correction   to the
Cepheid distances for SNII.

We  note  that our  results agree  very well  with Schmidt, Eastman \&
Kirshner (1994), despite the  5\% difference in  the distances seen in
Table \ref{tbl:snii}.  However,  we have limited  our \hub analysis to
galaxies beyond cz = 1500 km  s$^{-1}$, whereas 10  of the 14 galaxies
in  the  Schmidt \etal sample   are within  this  limit.  The  nearest
supernovae (where flow field effects are largest) yield a higher value
of    \hub.


\section{ Combining the Results and a Value for H$_0$}
\label{combineh0}

In Table \ref{tbl:kpresults}, we list the values of H$_0$ obtained for
each   of  the secondary   methods  which   are based on   our Cepheid
distances,   updated   using  the   new   calibration    described  in
\S\ref{revised}.   For each method, the   formal random and systematic
uncertainties    are given.   We   defer until  \S\ref{systematics}, a
detailed discussion of the systematic uncertainties that affect all of
these methods equally; however, the dominant overall systematic errors
include the   uncertainty in  the WFPC2 photometric   calibration, the
uncertainty in the adopted distance to  the LMC, metallicity, and bulk
motions of galaxies on large scales (cz $\gtsim$ 10,000 km/sec).


We next  address the question of  how  to combine  the values  of \hub
obtained  using the different   secondary methods, given 5 independent
measurements, H$_i$, with errors $\sigma_i$.  All of these methods are
based on a common Cepheid  zero point, although with different subsets
of Cepheid calibrators.  We now treat  the combination of these values
using the  quoted  internal errors.  The  secondary methods themselves
are largely independent of each  other (for example, the kinematics of
spiral disks represented by the Tully--Fisher relation are independent
of the  physics of the  explosions of carbon--oxygen white dwarfs that
give rise  to type  Ia   supernovae, and in  turn independent   of the
physics relating   to the luminosity fluctuations   of red giant stars
used by SBF).  We  use 3 methods to combine  the results:  a classical
(frequentist) analysis,  a Bayesian analysis,  and a  weighting scheme
based on numerical simulations.  Because of the relatively small range
of the  individual determinations  (\hub =  70 to  82 km/sec/Mpc, with
most of the values clustered toward the low end of  this range), all 3
methods  for  combining the \hub  values are  in  very good agreement.
This  result alone gives us  confidence that the  combined  value is a
robust one, and  that   the choice  of   statistical method does   not
determine the   result,  nor does  it strongly   depend upon choice of
assumptions and priors.

In the Bayesian data  analysis, a conditional probability distribution
is calculated, based  on a model or prior.  With a Bayesian formalism,
it is  necessary to be concerned about   the potential subjectivity of
adopted priors and whether they influence the  final result.  However,
one of the advantages of  Bayesian techniques is that the  assumptions
about the distribution of probabilities are  stated up front, whereas,
in fact,     all  statistical methods  have    underlying,  but  often
less--explicit assumptions, even   the  commonly  applied  frequentist
approaches (including  a  simple weighted  average,   for example).  A
strong advantage of  the Bayesian  method is  that it  does not assume
Gaussian distributions.  Although more common, frequentist methods are
perhaps not always the appropriate statistics  to apply.  However, the
distinction is  often  one  of nomenclature  rather  than subjectivity
(Gelman \etal 1995; Press 1997).

In Figure \ref{fig:frequentist}, we plot probability distributions for
the individual  \hub  determinations (see Table  \ref{tbl:kpresults}),
each represented by  a Gaussian of  unit area, with a dispersion given
by  their individual $\sigma$ values.   The cumulative distribution is
given by the solid  thick line.  The  frequentist solution, defined by
the median is  \hub =  72 $\pm$  3$_r$  ($\pm$ 7$_s$) \h0units.    The
random   uncertainty is  defined   at the   $\pm$34\%  points of   the
cumulative distribution.   The   systematic uncertainty is   discussed
below.  For our  Bayesian analysis, we  assume that the priors on \hub
and on  the probability  of any  single measurement being  correct are
uniform.  In this case,  we find \hub =  72 $\pm$ 2$_r$  ($\pm$ 7$_s$)
\h0units.   The formal uncertainty on  this result  is very small, and
simply  reflects  the fact that 4   of  the values  are clustered very
closely,  while  the  uncertainties   in   the FP  method   are large.
Adjusting  for the differences in  calibration, these results are also
in    excellent  agreement with  the    weighting  based  on numerical
simulations  of  the errors by Mould   \etal (2000a)  which yielded 71
$\pm$ 6  \h0units, similar  to   an earlier frequentist   and Bayesian
analysis  of Key  Project  data (Madore \etal  1999)  giving \hub = 72
$\pm$ 5  $\pm$  7 \h0units,  based on  a  smaller subset  of available
Cepheid calibrators.


As evident from  Figure 3, the value  of \hub based on the fundamental
plane is  an outlier.  However, both the  random and systematic errors
for this method are larger than for the  other methods, and hence, the
contribution to the combined value  of \hub is relatively low, whether
the results are weighted by the random or systematic errors. We recall
also  from   Table \ref{tbl:calibrators}    and  \S\ref{h0},  that the
calibration of the fundamental plane  currently rests on the distances
to only  3  clusters.  If  we  weight  the fundamental  plane  results
factoring in  the small   numbers   of calibrators and the    observed
variance of this method, then the fundamental  plane has a weight that
ranges from 5 to 8 times smaller than any of  the other 4 methods, and
results   in a combined, metallicity--corrected  value  for \hub of 71
$\pm$ 4$_r$ \h0units.

Figure  \ref{fig:veldist}   displays  the  results graphically    in a
composite Hubble diagram  of  velocity  versus  distance for  Type  Ia
supernovae (solid squares), the Tully-Fisher relation (solid circles),
surface-brightness      fluctuations (solid diamonds), the fundamental
plane (solid triangles), and  Type II supernovae  (open squares).   In
the  bottom panel,  the  values of H$_0$ are  shown  as a  function of
distance.  The Cepheid distances have  been corrected for metallicity,
as  given in Table   \ref{tbl:finaldist}.  The Hubble line plotted  in
this figure has a slope  of 72 \h0units, and  the adopted distance  to
the LMC is taken to be 50 kpc.


\section{ Overall Systematic Uncertainties }
\label{systematics}

There  are   a number of   systematic   uncertainties that affect  the
determination  of H$_0$ for  all  of the  relative distance indicators
discussed in the  previous sections.   These  errors differ  from  the
statistical and systematic   errors  associated   with each of     the
individual  secondary methods, and  they cannot  be  reduced by simply
combining the results from different  methods.  Significant sources of
overall systematic error include the  uncertainty in the zero point of
the Cepheid PL  relation, the effect  of reddening and metallicity  on
the  observed  PL relations, the effects  of   incompleteness bias and
crowding on the  Cepheid  distances, and velocity  perturbations about
the Hubble flow on  scales comparable to,  or larger than, the volumes
being  sampled.  Since the  overall accuracy  in  the determination of
H$_0$ is  constrained by these factors, we  discuss each  one of these
effects in turn below.  For readers  who may wish  to skip the details
of   this  part  of   the discussion,  we     refer them  directly  to
\S\ref{overall} for a summary.

\subsection{ Zero Point of the PL Relation }

It  has  become     standard  for  extragalactic  Cepheid     distance
determinations    to  use the   slopes  of  the  LMC period-luminosity
relations   as fiducial,    with   the zero   point    of the  Cepheid
period-luminosity  relation tied  to  the LMC  at  an adopted distance
modulus  of  18.50 mag (\eg  Freedman  1988).  However, over  the past
decade, even with more accurate and sensitive detectors, with many new
methods for measuring distances, and with many individuals involved in
this effort, the full range of the most  of distance moduli to the LMC
remains at approximately 18.1 to 18.7 mag (\eg Westerlund 1997, Walker
1999, Freedman 2000a, Gibson 2000), corresponding to a  range of 42 to
55 kpc.

For the purposes of the present discussion, we can compare our adopted
LMC  zero  point with other   published  values.   We show  in  Figure
\ref{fig:lmcdist},   published  LMC   distance  moduli   expressed  as
probability density distributions, primarily for the period 1998-1999,
as  compiled by Gibson (2000).  Only  the single  most recent revision
from a  given author  and method is  plotted.   Each determination  is
represented by a Gaussian of unit  area, with dispersions given by the
published errors.   To facilitate viewing the individual distributions
(light dotted lines), these have been scaled up by a factor of 3.  The
thicker solid line shows the cumulative distribution.

It is   clear from the  wide  range of moduli   compared to the quoted
internal errors in   Figure \ref{fig:lmcdist}  that systematic  errors
affecting individual  methods are still  dominating the determinations
of  LMC   distances.   Some  of  the   values at  either  end  of  the
distribution  have error bars that do  not  overlap (at several sigma)
with other methods.   At the current time,  there is no  single method
with  demonstrably lower  systematic errors,   and  we find no  strong
reason to prefer one   end of the  distribution over  the  other.  For
example, while  systematics in the Cepheid period--luminosity relation
have   been subjected  to scrutiny    for many   decades, no  accurate
photometric zero point has  yet been established based on  astrometric
distances and the  zero   point  is still  in  debate  (\eg  Feast  \&
Catchpole 1997;  Madore   \& Freedman 1997; Groenewegen   \& Oudmaijer
2000).  The  absolute astrometric  calibration is   statistically more
reliable for the red clump method, but compared to many other methods,
this method is still relatively new, and the systematics have not been
studied in as much detail (Udalski 2000; Stanek \etal 2000).

In addition to   the  frequentist probability distributions,   we have
computed Bayesian probability distributions, assuming a uniform prior.
The Bayesian and median or average frequentist methods yield excellent
agreement  at  18.45 and 18.47   mag, respectively.    Another way  of
estimating the overall uncertainty  is simply to estimate  the overall
average and the standard error  of the mean,  based on a mean distance
for different methods, and  giving  each technique  unit  weight.   An
advantage of this procedure is that it simply averages over all of the
inherent systematic uncertainties that affect any given method.  There
are  7 independent methods  for  measuring distances that are commonly
applied to  the LMC; these include Cepheids,  the red clump, eclipsing
binaries, SN1987A light echoes, tip of the red giant branch (TRGB), RR
Lyraes, and Miras.  The mean values of the LMC distance moduli and the
standard  error of the  mean for  each  technique  are given in  Table
13, for  the  Gibson   (2000)   and   Westerlund (1997)
compilations.   For  the Gibson  compilation,  these averaged distance
moduli range from 18.27 to  18.64 mag, with an  overall mean of  18.45
mag, and an $rms$ dispersion of $\pm$0.15 mag.   The standard error of
the mean therefore amounts to  $\pm$0.06 mag.  The  mean based on  the
Westerlund data is in excellent agreement at 18.46 $\pm$ 0.05 mag.

From the above  discussion, it can be  seen that there still remains a
range in distance moduli to the LMC based  on a wide range of methods.
However,  our adopted Cepheid modulus  of 18.50 $\pm$  0.10 mag agrees
with the mean and median of the  distribution for other methods at the
2.5\% level. \footnote{In two recent Key Project  papers, we adopted a
distance modulus uncertainty to the  LMC of $\pm$0.13 mag (Mould \etal
2000a,  and   Freedman 2000b).   This  value   defined  the 1-$\sigma$
dispersion based on a   histogram of the  distance moduli  compiled by
Gibson (2000).   However, the  standard  error   of the  mean  is  the
relevant statistic in  this case.}  Given the remaining uncertainties,
and the  good agreement with  other methods, we  do not believe that a
change in  zero point is warranted at  the  current time.  However, we
note that the uncertainty  in the distance to  the  LMC is one  of the
largest remaining  uncertainties in the overall  error  budget for the
determination of \hub.  We  note that if the  distance modulus  to the
LMC is 18.3 mag, there will be a  resulting 10\% increase in the value
of \hub to 79 km/sec/Mpc.




It would be extremely useful to have a calibration that is independent
of  the distance to the LMC.   Very recently, a  new distance has been
independently measured to the maser galaxy,  NGC 4258, a nearby spiral
galaxy also useful for  calibrating the extragalactic  distance scale,
which  can provide  an external  check on the  adopted LMC zero--point
calibration.  We briefly  summarize the distance determination to  NGC
4258 and its implications below.

\subsubsection{NGC 4258: Comparison of a Maser and Cepheid Distance}
\label{n4258}

Given the  current  uncertainties and systematics affecting  the local
distance scale, it would be highly desirable to have geometric methods
for measuring  distances,   independent  of the    classical  distance
indicators.   A  very promising  new  geometric technique has recently
been developed and applied  to the galaxy, NGC 4258,  a galaxy with an
inner  disk  containing H$_2$O masers  (Herrnstein  \etal 1999).  Five
epochs  of measurements are  now available for  these masers, and both
radial and transverse motions of the  maser system have been measured.
Assuming a circular, Keplerian   model for the disk, Herrnstein  \etal
derive a distance to  the galaxy of 7.2 $\pm$  0.3 Mpc, with the error
increasing  to $\pm$ 0.5  Mpc allowing for systematic uncertainties in
the model.

To provide a   comparison with the maser  distance,  Maoz \etal (1999)
used $HST$ to discover a sample of 15 Cepheids in NGC 4258. Adopting a
distance modulus for the LMC of 18.50 mag,  these authors determined a
Cepheid distance  to NGC 4258 of  8.1 $\pm$ 0.4  Mpc, or  12\% further
than the maser distance.  These authors  noted that the difference was
not highly significant, amounting  to only 1.3$\sigma$.  However, with
the new  LMC PL    relations given  in  equations 1   and 2,   and the
correction to the   WFPC2  zero point  discussed  in  \S\ref{cal}, the
revised  Cepheid distance  is in somewhat  better   agreement with the
maser distance at  7.8 $\pm$ 0.3$_r$  $\pm$ 0.5$_s$  Mpc (Newman \etal
2000).
Allowing for a metallicity correction  of --0.2 mag/dex results
in a Cepheid distance  of 8.0 Mpc.  Based on  the new calibration, the
Cepheid  distance agrees to  within  1.2-$\sigma$ of the maser
distance. Unfortunately, however, the situation  remains that there is
currently only one maser  galaxy with which  to make this  comparison.
For  the future, increasing the  sample  of maser  galaxies for  which
distance measurements can be made (for example, with ARISE, a proposed
radio interferometer in space) would be extremely valuable.

\subsubsection{Resolving the Cepheid Zero--Point Discrepancy}

Given   the  range    of  published   LMC    distance moduli   (Figure
\ref{fig:lmcdist}),  and  the subtle systematic   errors  that must be
affecting  some (or all) of  the distance methods, it appears unlikely
that  this zero-point uncertainty  will  be resolved  definitively any
time soon.   Upcoming  interferometry  (NASA's  SIM and   ESA's  GAIA)
missions will  deliver a few  microarcsec astrometry, reaching fainter
limits   than Hipparcos ($\sim$20 mag);    NASA's  FAME will reach  50
microarcsec accuracy.  These missions  are  capable of delivering  1\%
distances  to  many Galactic   Cepheids. They  will   be critical   for
establishing a more  accurate extragalactic distance scale zero point,
and  should provide accurate  parallaxes for statistically significant
samples of many distance indicators currently in use (\eg Cepheids, RR
Lyrae stars,   red giant stars,  red clump  stars).   In addition, SIM
(currently scheduled for   launch   in 2008) may   provide  rotational
parallaxes for  some of the nearest  spiral galaxies, thereby allowing
the calibration to bypass the LMC altogether.

\subsection{ Reddening }

As described in \S\ref{reddening}, the standard approach to correcting
Cepheid magnitudes for reddening by  dust is to  use a combination  of
bandpasses (V  and I in  the case of  the \hub Key Project), and solve
for the reddening using   a Galactic extinction law (Freedman   1988).
For a  value of the LMC  reddening  appropriate to  the Cepheid sample
considered of  E(V--I) = 0.13 mag, the  reddenings  in the HST Cepheid
target fields range from E(V--I)  = 0.04 to  0.36 mag, with an average
of  0.19 mag  (Table   \ref{tbl:finaldist}).  As a check   on possible
systematic  errors   in the   reddening determinations,  recent H-band
(1.6$\mu$m)  photometry has  been  obtained for  a total  sample of 70
Cepheids in 12 galaxies, including  IC 1613, M31,  M81, M101, NGC 925,
NGC 1365,  NGC 2090, NGC 3198,  NGC 3621, NGC  4496A, NGC 4536, and IC
4182 (Macri \etal 2001;  Freedman \etal 2001).  The Galactic reddening
law of Cardelli \etal (1989) predicts E(V--H) = 1.98$\pm$0.16 E(V--I).
For the galaxies  with both NICMOS  H--band and optical VI data  (from
the ground  or HST), the slope of  the correlation between the optical
and  near-infrared  reddenings   yields E(V--H)  =    [2.00 $\pm$ 0.22
E(V--I)] +  0.02 $\pm$ 0.04.  This  relation is based  on the same VI
Udalski \etal (1999)  data used in the current  paper.   Hence, the IR
data  confirm  the reddenings derived  from   the optical data  alone,
ruling   out   a   significant systematic    error    in the reddening
determinations.

\subsection{ Metallicity }
\label{metallicity}

As discussed in \S\ref{cephmet}, recent empirical results suggest that
there is a small dependence of the Cepheid period--luminosity relation
on metallicity.   In this paper, we  have adopted a correction of -0.2
mag (10\% in distance) for  a factor of   10 in abundance (O/H).   The
observed fields in Cepheid-calibrating galaxies have  a range in (O/H)
abundance  of about  a factor  of 30  (Ferrarese  \etal 2000b).  These
abundances are those of HII regions in the Cepheids fields, calibrated
on  the scale of  Zaritsky \etal (1994).  The  mean  abundance of this
sample (12 + log(O/H)) is 8.84 $\pm$0.31 dex.  This is higher than the
LMC abundance  of  8.50 dex  (Kennicutt \etal 1998).   The mean offset
between the metallicity--corrected, and the uncorrected Cepheid moduli
in Table \ref{tbl:finaldist} amounts to 0.07 mag or 3.5\% in distance.
We adopt this difference  as the uncertainty  due to metallicity.  The
effect is systematic, and  with  the exception  of type II  supernovae
(\S\ref{snii}), if no correction for metallicity is applied, the value
of \hub is increased by $\sim$4\% ($\sim$3 km/sec/Mpc). Conversely, if
the slope of the metallicity relation is --0.4 mag/dex, then the value
of   \hub   is decreased   by 3    km/sec/Mpc.     We show  in  Figure
\ref{fig:cephmet},  histograms   of abundance  distributions  for  the
Cepheid calibrators for the secondary methods.


\subsection{   Completeness / Bias Effects  }
\label{bias} 

An  issue of recurring concern   regarding the application of distance
indicators   is the extent   to which incompleteness   in the observed
samples could  lead to a bias in  the derived distances.   This effect
has been discussed extensively in the  literature, particularly in the
context of the  Tully-Fisher  relation  (\eg Schechter  1980,  Willick
1994; Giovanelli \etal 1997; Tully \& Pierce 2000).  For Cepheids, the
concern derives from  the fact that magnitude  cut-offs in the Cepheid
samples (imposed  by  the decreasing  signal-to-noise ratios at  faint
magnitudes) will tend to select against the faintest variables thereby
leading to systematically small (biased)   moduli.  The fact that  the
bias operates most strongly at the  shortest periods will also tend to
produce  a  flattening of the  observed  PL relation (\eg  see Sandage
1988); elimination of the shortest--period Cepheids from a sample will
generally result in increased mean moduli, less affected by this bias.

This effect is illustrated in Figure  A1 in Appendix  A, along with an
analytic derivation of the size of  the bias.  As more distant objects
are observed, a brighter intrinsic magnitude cutoff will occur for the
same apparent magnitude.  The observed  erroneous flattening of the PL
slope extends to longer and longer  periods including more and more of
the available  Cepheid sample.  Similar biases  occur for any standard
candle possessing an  intrinsic  dispersion in luminosity:  the larger
the   intrinsic dispersion of the    relation being truncated, and the
shallower the range of  apparent  magnitude being sampled, the  larger
the bias will be.  For the Key Project application of the Tully-Fisher
relation   (Sakai \etal  2000),    we  adopted  corrections for   this
incompleteness bias, based  on simulations similar to those undertaken
by  Giovanelli  \etal (1998).  For  Type   Ia supernovae, the observed
dispersion in the  Hubble diagram amounts  to only $\sim$0.15 mag (\eg
Riess \etal 1998; Hamuy  \etal 1996); hence, incompleteness biases are
very small for this technique.

In the  case  of Cepheids,  incompleteness  biases are expected  to be
small  for the methodology   that we have   adopted here.  First,  the
intrinsic  scatter   is    small,    and second,  as    discussed   in
\S\ref{revised},  we have applied  a period cutoff  above the limiting
magnitude cutoff at the short-period end  to reduce the incompleteness
bias (Freedman  \etal 1994b; Kelson  \etal 1994; Ferrarese \etal 1996,
2000b).  The  scatter in  the observed  V-band PL relation  amounts to
$\pm$0.16  mag (equation  1); the  scatter in the   I-band PL relation
amounts  to $\pm$0.11 mag (equation 2);  however, much of this scatter
is  physically correlated between  bandpasses, so that  the scatter in
the  W PL  relation is  small.    After correcting for reddening,  the
correlated scatter   in the combined relation   for the  true distance
modulus   (or   equivalently, W:  \S\ref{reddening})   is smaller, and
amounts to only $\pm$0.08 mag (equation 3).  Hence, the resulting bias
on the final  distance modulus is negligible  for most of the galaxies
in the sample.  As can be seen from the  differences between columns 5
and 7  in Table \ref{tbl:cephdist},  typically, the  size  of the bias
corrections   amounts to  only  a  few  hundredths  of  a  mag in  the
reddening--corrected (true)  modulus;  in 2  cases (M81 and  NGC 4414)
they are as   large  as  0.08 and   -0.08   mag  (4\%  in   distance),
respectively, but the  mean correction  for the  sample is only  +0.01
mag.

\subsection{ Crowding / Artificial Star Tests }
\label{artificial}

One of the most direct  ways of assessing  the quantitative effects of
chance superpositions on the photometry  is by adding artificial stars
with known  input magnitudes and colors  into the actual $HST$ images,
and then recovering those stars using exactly the same techniques used
to  perform the original  analysis (\ie  ALLFRAME  and DoPHOT).  While
these experiments cannot provide numerical crowding corrections to the
real Cepheids  in  the frames, they are   powerful  in quantifying the
vulnerability  of  the photometric  methods    to crowding under  each
individual set of circumstances.

Artificial  star tests for two Key  Project galaxies have been carried
out by Ferrarese \etal (2000c). Their analysis indicates that the bias
due  to crowding  in   individual WFPC2  frames,  can  be significant,
ranging from 0.05 mag in a relatively  uncrowded field of NGC 2541 (at
12  Mpc), to 0.2 mag  for a crowded field  in one of  the most distant
galaxies, NGC 1365. We note that the artificial  stars in these frames
were  not inserted with random  positions.  Each field  was divided up
into  a 10$\times$10  array  of cells each   65 pixels on a side;  the
probability that an artificial star would be added within a given cell
was  therefore proportional to the number  of  real stars in the cell.
The measured bias goes in the expected sense of resulting in recovered
magnitudes that are too bright,  and it is   a direct function of  the
stellar density in the field.  However,  when using the multi-epoch, V
and  I-band observations,   and then  imposing  the same  criteria  on
variable  star selection  as for the  actual Cepheid  sample (\eg same
error flags  for deviant data points,  same magnitude range applicable
to the period  range for  the known Cepheids,  the  same procedure for
reddening correction,  etc.),  the effect  of  this bias on  the final
determination of distances drops  significantly, amounting to only 1\%
for ALLFRAME, and 2\% for DoPHOT.

The Ferrarese \etal (2000c) results are consistent with an independent
study  by Saha  \etal (2000),  who  have  investigated the effects  of
crowding in the galaxy NGC  4639 (at a  distance of $\sim$25 Mpc,  the
largest distance  measured  by either   the  Key Project  or Type   Ia
supernova teams).  For this galaxy,  the crowding bias in single epoch
observations is found   to  be 4\% (0.07  mag).   Saha   \etal do  not
explicitly  extend their  results  to  multi--epoch   observations  of
Cepheids, for which, as noted above,  the effect would be reduced even
further. A different  approach to placing  limits on  crowding effects
comes from Gibson  \etal (2000) and  Ferrarese \etal (2000c), who have
looked  for    a correlation  with   distance  of   residuals   in the
Tully-Fisher relation.  No significant  effect is found.  Gibson \etal
also see   no systematic  effects  as  would be  expected  for type Ia
supernova peak  magnitudes,  nor  a   difference in the    PC-- versus
WFC--based Cepheid distance moduli.

Very different  conclusions  have been  reached recently  by Stanek \&
Udalski (1999) and Mochesjska (2000).  These authors have specifically
investigated the influence of  blending on the Cepheid distance scale.
Blending is   the close association   of  a Cepheid   with one or more
intrinsically  luminous stars.  Since  Cepheids  are young stars, they
may be  preferentially associated near other  young  stars.  Stanek \&
Udalski conclude that this effect ranges from a few percent for nearby
galaxies to $\sim$15-20\% for galaxies at 25 Mpc.  However, the Stanek
\& Udalski  results are  based  on an extrapolation  from high-surface
brightness regions in the bar of the LMC, and they  do not make use of
photometric reduction programs (like  DoPHOT and ALLFRAME),  which are
designed  for photometry in crowded fields.   These authors simply sum
additional contributions to  the   total flux to   simulate  crowding.
Moreover, they do  not allow for  underlying background contamination,
which become increasingly  important for galaxies at larger distances.
Hence, at  present, it  is  not  possible  to  compare  these  results
directly to the analysis of \hub Key Project data.

In  a comparison of   ground  based images of    M31 with HST  images,
Mochejska \etal (2000) found that the  median V-band flux contribution
from  luminous companions was about 12\%  of the flux  of the Cepheid.
They  argued that  ground based  resolution in  M31 corresponds to HST
resolution  at  about  10   Mpc,  and that    blending will  lead   to
systematically low distances for  galaxies at such distances.  A  more
recent study by this group for the galaxy M33, which has a much larger
sample of stars, indicates that this effect  amounts to only about 7\%
(Mochejska \etal 2001).

The exact  size  of this effect  will  depend  on  the true underlying
distribution of stars in the frame, and the extent to which the actual
Cepheids being measured are affected.  We note  that the galaxies with
HST Cepheid distances for which blending effects are likely to be most
severe are the inner field of M101, the high surface brightness galaxy
NGC 3627,  and the most   distant galaxies searched, for  example, NGC
4639.

To assess quantitatively the impact  of unresolved blending effects on
the final   Cepheid distances would  require  simulations based on the
distribution  of Cepheids in  a  galaxy field unaffected by  blending.
This distribution  could be scaled with distance  and  inserted at the
same surface  brightness levels  encountered  in each of   the Cepheid
target  frames, and then recovered  using the same  techniques as used
originally to analyse the original data frames. Ideally, several input
distributions could be tested. Such a study is beyond the scope of the
present  paper,  but is being applied,  for  example, to  the Cepheids
observed in  M101 with NICMOS  (Macri \etal  2001).   

Naive tests,  which, for example,  assume  constant surface brightness
between the Cepheid  fields in nearby and  distant galaxies, will not,
in general,  correctly   simulate  the Key  Project,  where  generally
low--surface--brightness  fields    were      deliberately   selected.
Examination of the statistics of the  number densities of stars in the
vicinity of Cepheids in the Key Project frames bear this out.  For the
present time,  we   view the  2\% effect measured   by Ferrarese \etal
(2000c) as a lower limit on the effects of crowding and blending, and,
adopt a conservative uncertainty of $^{+5}_{-0}$\% (1--$\sigma$).

\subsubsection{Contamination from Companion Stars}

We  note that Cepheids   can  be located   in binary systems,  and the
presence of  {\it true, physical companions}  has been established for
Cepheids in both the Galaxy and the LMC.   For Cepheids in the Galaxy,
as well   as for  early (B-type)    stars,  the mass   distribution of
companions has been studied intensively, and is strongly peaked toward
low masses  (\eg Evans   1995).   The presence  of  binaries will  add
increased  scatter   to  the underlying  period--luminosity  relation,
including that  for  the LMC,   where  the  binaries  are  unresolved.
However, unless the frequency of Cepheid binaries varies significantly
from galaxy to  galaxy,  the relative  distances to  galaxies will  be
unaffected.

\subsection{\bf Does the Measured Value of H$_0$  Reflect the True, Global Value? }
\label{bulk}

Locally,  variations   in   the expansion  rate  due   to  large-scale
velocities  make measurement of the  true value  of H$_0$ problematic.
Thus, for an  accurate determination of  H$_0$, a large enough  volume
must be observed to  provide a fair  sample of the universe over which
to average.  How large is large  enough?  Both theory and observations
can provide constraints.

A number of theoretical studies have addressed this question recently.
Given a model for structure formation, and therefore a predicted power
spectrum for density fluctuations, local measurements  of H$_0$ can be
compared with the global value of H$_0$ (Turner, Cen \& Ostriker 1992;
Shi \& Turner 1997; Wang, Spergel \& Turner 1998).  Many variations of
cold  dark matter (CDM) models have  been investigated,  and issues of
both the required  volume and sample  size for  the distance indicator
have been  addressed.  The most  recent models predict that variations
at  the level  of 1--2\%   in  $<(\delta$H/H$_0)^2>^{1/2}$) are to  be
expected for the current  (small) samples of  Type Ia supernovae which
probe out  to 40,000 km/sec, whereas  for methods  that extend only to
10,000 km/sec, for small samples, the cosmic variation is predicted to
be 2--4\%.

There are also observational constraints that can test the possibility
that  we live  in  an  underdense  region locally.   These include the
observational determinations that  the expansion is  linear on 100  to
1000  Mpc scales, and measurements  of temperature fluctuations in the
cosmic microwave background.  The linearity  of the Hubble diagram has
been established by  many means, including work   by Sandage \&  Hardy
(1973) and Lauer \&   Postman  (1992) on brightest  cluster  galaxies,
recent studies   of  supernovae at  velocity-distances  out  to 30,000
km/sec (Zehavi et al 1998), and extension of the Tully Fisher relation
to  15,000 km/sec (Giovanelli \etal   1999;  Dale \etal 1999).   These
results limit  the difference between the  global and  local values of
the Hubble constant to  a few percent.   For example, Giovanelli \etal
provide limits to the amplitude of a possible distortion in the Hubble
flow within 70  h$^{-1}$ Mpc of $\delta$H/H  = 0.010 $\pm$ 0.022.  The
rarity of low density bubbles is also attested by the microwave dipole
anisotropy on degree  scales.  Wang \etal (1998)  find a robust  upper
limit on  the global deviation from  the local 10$^4$ km/sec sphere of
10.5\% in H$_0$ with 95\% confidence.

A stronger constraint will come from galaxy counts in redshift shells.
If the  local density were   deficient within 150 Mpc  by  $\delta$n/n
 = $\delta \rho/ \rho$ $\ltsim$ --0.2, the effect on H$_0$ would be

$$\delta H/H = {1 \over 3} \Omega_m^{0.6} \delta \rho/\rho $$

\noindent
For example, for $\Omega_m$  = 0.2,  this is  consistent with  1.0 $>$
H(global)/H(local) $\gtsim$ 0.97.  These  results limit the difference
between the  global and local values  of the Hubble  constant to a few
percent.    This is consistent with    the results cited above.   (For
comparison, with  $\Omega_m$  =  1,  it is  consistent with   1.0  $>$
H(global)/H(local) $\gtsim$ 0.93).

There  are  two sources of data  on  $\delta$n/n.  The slope of galaxy
counts versus magnitude is a relatively  weak constraint, as excellent
knowledge of the luminosity function of galaxies  is required in order
to infer   a  density.  Redshift  survey  data   is superior; however,
selection effects must be well--understood  before $\delta$n/n can  be
determined.  Improved constraints will   soon be forthcoming  from the
2dF, Sloan, 2MASS, and 6dF surveys.

The   overall   conclusion   derived from   these   studies   is  that
uncertainties due to inhomogeneities in the galaxy distribution likely
affect  determinations of H$_0$  only at the  few percent level.  This
must be  reflected  in the  total uncertainty in  H$_0$;  however, the
current distance  indicators  are now  being  applied  to sufficiently
large depths, and in many   independent directions, that large  errors
due to this source of  uncertainty are statistically unlikely.   These
constraints will  tighten  in the  near future  as larger   numbers of
supernovae are discovered,  when  all--sky  measurements of  the   CMB
anisotropies  are made  at smaller  angular  scales,  and when  deeper
redshift surveys have been completed.

\subsection{Overall Assessment of Systematic Uncertainties}
\label{overall}

We now briefly summarize the  sources of systematic error discussed in
the previous  section.  The standard  error of  the mean for  the zero
point of the LMC PL relation is $\pm$0.06 mag, and is currently set by
an average over several  independent methods. Conservatively, we adopt
a value of $\pm$0.1 mag, corresponding to  $\pm$5\% in the uncertainty
for the  distance to  the LMC.   Systematic  errors in  the  reddening
determinations  are    small,  amounting  to  less  than    1\%.  Both
observational and theoretical studies  of Cepheids suggest  that there
is   a  small  metallicity dependence  of   the  PL relation.  Cepheid
galaxies have  a range of  metallicities that are  in  the mean, a few
tenths of a  dex  greater that of  the  LMC.  Adopting  a  metallicity
correction  results in values of  \hub that are  lower in  the mean by
4\%.  We  take  this  difference  between corrected   and  uncorrected
distances  to be indicative  of the  uncertainty  due to  metallicity.
Cepheid distances can be affected by incompleteness  biases at the few
percent level, but these   are  minimized by adopting  a  conservative
choice for the lower period limit, and by the fact that the dispersion
in the  reddening--corrected    PL relation is   only   $\pm$0.08 mag.
Ultimately this  sample bias  effect  contributes less  than  $\pm$1\%
uncertainty    to the   final  results.  Based      on artificial star
experiments, crowding effects  on the final  distances also contribute
at a 1-2\% level.  Allowing  for unresolved blending effects, we adopt
an overall uncertainty  of $+$5, 0\%.   Finally, based on a  number of
both empirical and   theoretical studies, bulk  motions  on very large
scales are likely to contribute less than $\pm$5\%.

Correcting for the effects of   bias and metallicity decrease \hub  by
1\% and 4\%,  respectively, whereas the effect  of the new  WFPC2 zero
point is to increase  \hub by 3.5\%.   The effect of adopting  the new
Udalski \etal (1999) PL slopes   differs  from galaxy to galaxy   (and
therefore differs in the magnitude of the effect on the zero point for
each secondary  method).  Adopting the  new slopes  results in a  mean
decrease in distance from  the Madore and Freedman calibration  (1991)
of 7\% for the  galaxies listed in  Table \ref{tbl:cephdist}, but each
individual method is impacted slightly  differently depending on  what
subset of calibrators is  applicable to that  method.  The sign of the
uncertainty  due to a  possible  bulk flow component  to  the velocity
field is, of  course, unknown.  In this  paper, we have  not applied a
correction  for crowding,  but  incorporate this  uncertainty into the
final error budget.  These corrections individually   amount to a  few
percent,  but with differing signs so  that  the overall impact on the
mean value of the Hubble constant agree at the 1\% level with those in
Mould \etal (2000a) and Freedman (2000b).

We list    the  major identified  systematic   uncertainties  in Table
\ref{tbl:h0errors2};  these can be combined in  quadrature to yield an
overall systematic  uncertainty  of $\pm$10\%  (or 7  \h0units.)   Our
current \hub value incorporates  four refinements discussed in  detail
above: (a) adopting the slopes of the PL relations as given by Udalski
\etal  (1999), (b) using  the WFPC2 photometric zero-point calibration
of  Stetson (1998), (c) applying   a  metallicity correction of  --0.2
$\pm$ 0.2  mag/dex, and  (d) correcting for  bias  in the PL relation.
Applying the resulting Cepheid  calibration   to 5 secondary   methods
gives \hub $ = 72 \pm$ 3$_r$ $\pm$ 7$_s$ \h0units.

\section{ H$_0$ From Methods Independent of Cepheids }
\label{othermethods}

A detailed discussion of  other  methods is beyond  the scope  of this
paper; however, we briefly compare our results with two other methods:
the Sunyaev--Zel'dovich (SZ) technique, and measurement of time delays
for gravitational   lenses.    Both   of these  methods    are  entirely
independent of the local extragalactic distance scale, and they can be
applied directly at  large distances.  Currently their accuracies  are
not yet  as high  as   has recently been  achieved  for  the classical
distance measurements, but  both methods hold considerable promise for
the future. We show in Figure \ref{fig:szlens} values of \hub published
based on these two methods from 1991 to the present. 


\subsection{ The Sunyaev-Zel'dovich Effect }

For clusters of galaxies, the   combination  of a measurement of   the
microwave background decrement (the SZ effect), the X-ray flux, and an
assumption of spherical symmetry, yield  a measurement of the distance
to the cluster    (\eg Birkinshaw 1999;  Carlstrom  \etal  2000).  The
observed microwave decrement (or more  precisely, the shift of photons
to higher   frequencies)   results  as low-energy   cosmic   microwave
background  photons are scattered off the  hot  X-ray gas in clusters.
The SZ effect  is independent of distance,  whereas the  X-ray flux of
the  cluster is distance--dependent: the combination  thus can yield a
measure of the distance.

There are also, however,   a number of astrophysical complications  in
the    practical application  of  this method   (\eg  Birkinshaw 1999;
Carlstrom 2000).  For example, the gas distribution in clusters is not
entirely uniform: clumping of the gas, if significant, would result in
a decrease in the  value of $\rm  H_0$.  There may also  be projection
effects: if the  clusters  observed are prolate and  seen  end on, the
true $\rm  H_0$ could be  larger than inferred  from spherical models.
(In a flux--limited sample, prolate clusters could  be selected on the
basis of brightness.)    Cooling   flows  may also be     problematic.
Furthermore, this method assumes  hydrostatic equilibrium, and a model
for the  gas  and electron densities.  In   addition,  it is  vital to
eliminate potential contamination from  other sources.  The systematic
errors incurred from all of these effects are difficult to quantify.

Published values of $\rm H_0$ based on  the SZ method have ranged from
$\sim$40  - 80  km/sec/Mpc (\eg  Birkinshaw   1999).  The most  recent
two--dimensional interferometry  SZ  data  for well-observed  clusters
yield H$_0$  = 60 $\pm$  10 km/sec/Mpc.   The systematic uncertainties
are still  large, but  the near--term  prospects  for this  method are
improving rapidly (Carlstrom  2000)  as additional clusters  are being
observed,  and  higher-resolution X-ray  and   SZ  data  are  becoming
available (\eg Reese \etal 2000; Grego \etal 2000).

\subsection{ Time Delays for Gravitional Lenses }

A    second method for    measuring  H$_0$  at  very large  distances,
independent   of the   need for  any  local   calibration, comes  from
gravitational lenses.  Refsdal  (1964, 1966) showed that a measurement
of  the  time delay, and   the angular separation  for gravitationally
lensed images of a  variable object, such as  a quasar, can be used to
provide a  measurement   of $\rm H_0$  (\eg  see  also the  review  by
Blandford \& Narayan  1992).  Difficulties with  this method stem from
the fact that the underlying  (luminous or dark) mass distributions of
the  lensing galaxies are  not  independently known.  Furthermore, the
lensing galaxies  may be sitting in more  complicated group or cluster
potentials.  A degeneracy exists between  the mass distribution of the
lens and the   value of H$_0$   (Schechter \etal 1997;  Romanowsky  \&
Kochanek  1999;  Bernstein  \&  Fischer 1999).   In  the case   of the
well-studied   lens 0957+561, the  degeneracy  due  to the surrounding
cluster can  be broken with the  addition of weak lensing constraints.
However,  a careful analysis by   Bernstein \& Fischer emphasizes  the
remaining uncertainties in the mass models for both the galaxy and the
cluster which  dominate the overall errors  in  this kind of analysis.
H$_0$ values based on this technique appear  to be converging to about
65 km/sec/Mpc (Impey  \etal  1998; Franx  \& Tonry  1999; Bernstein \&
Fischer; Koopmans \& Fassnacht 1999; Williams \& Saha 2000).

\subsection{ Comparison with Other Methods }

It  is encouraging  that to  within the  uncertainties, there is broad
agreement  in \hub values   for completely independent  techniques.  A
Hubble  diagram   (log  d   versus  log  v)   is   plotted   in Figure
\ref{fig:veldistfar}.  This  Hubble  diagram covers  over 3 orders  of
magnitude, and includes distances obtained locally from Cepheids, from
5 secondary methods, and for 4 clusters with recent Sunyaev--Zel'dovich
measurements   out  to  z  $\sim$ 0.1.   At    z  $\gtsim$ 0.1,  other
cosmological  parameters  (the matter  density,   $\Omega_m$, and  the
cosmological constant, $\Omega_\Lambda$) become important.


\section{ Implications for Cosmology }
\label{cosmology}

One    of the  classical  tests   of  cosmology is   the comparison of
timescales.  With a knowledge of \hub,  the average density of matter,
$\rho$,  and   the  value  of  the cosmological   constant, $\Lambda$,
integration of the Friedmann equation

\begin{equation}
H^2 = 8 \pi G \rho / 3  - k / r^2 + \Lambda  / 3
\end{equation}

\noindent 
yields a measure of the expansion  age of the universe. This expansion
age can be compared with other independent estimates of the age of the
Galaxy and its oldest stars, t$_0$, and thus  offers a test of various
possible cosmological models.  For example, the dimensionless product,
\hub t$_0$,  is   2/3 in  the  simplest  case where  $\Omega_m$   = 1,
$\Omega_\Lambda$ = 0 (the Einstein--de  Sitter model), and the product
is 1 for  the case of  an empty universe where  the matter and  energy
density are zero.

An  accurate determination of the  expansion age requires not only the
value    of \hub, but   also  accurate measurements  of $\Omega_m$ and
$\Omega_\Lambda$.  At  the time when  the  Key Project  was begun, the
strong   motivation  from inflationary  theory  for   a flat universe,
coupled with a strong theoretical preference for $\Omega_\Lambda$ = 0,
favored the Einstein--de Sitter  model (\eg Kolb  \& Turner 1990).  In
addition, the ages of  globular cluster stars  were estimated at  that
time  to be $\sim$15 Gyr  (VandenBerg, Bolte \& Stetson 1996; Chaboyer
\etal 1996).  However, for a value of H$_0$ = 72 \h0units, as found in
this  paper,  the Einstein--de   Sitter  model  yields a   very  young
expansion age of only 9 $\pm$ 1 Gyr,  significantly younger than the
globular cluster and other age estimates.

Over the  past  several years,   much progress has  been made   toward
measuring  cosmological parameters, and  the Einstein--de Sitter model
is not currently favored.   For example, estimates of cluster velocity
dispersions, X-ray masses,  baryon fractions, and weak lensing studies
all  have   provided      increasingly   strong   evidence   for     a
low--matter--density ($\Omega_m$) universe  (\eg Bahcall \& Fan 1998).
In addition, strong new evidence for a flat  universe has emerged from
measurements of  the position of  the  first acoustic peak  in  recent
cosmic microwave background anisotropy experiments (de Bernardis \etal
2000; Lange \etal 2000).   Together  with evidence  for a low   matter
density, and with   recent data from  high--redshift supernovae (Riess
\etal  1998;  Perlmutter   \etal   1999), evidence    for  a  non-zero
cosmological  constant    has  been  increasing.   Moreover,  the  age
estimates for globular  clusters  have been revised downward  to 12-13
Gyr, based on a new calibration from the Hipparcos satellite (Chaboyer
1998; Carretta \etal 2000).    A non--zero value of  the  cosmological
constant helps  to avoid a discrepancy between   the expansion age and
other age  estimates.   For \hub  =  72  \h0units,  $\Omega_m$ =  0.3,
$\Omega_\Lambda$   = 0.7, the   expansion   age is 13   $\pm$  1  Gyr,
consistent to  within the uncertainties,  with recent globular cluster
ages.   In   Table \ref{tbl:t0lambda},  we   show expansion   ages for
different values of \hub and a range of flat models.

In Figure \ref{fig:h0t012} H$_0$t$_0$   is  plotted as a function   of
$\Omega$. Two curves are shown: the solid curve is  for the case where
$\Lambda$  = 0,  and the dashed  curve  allows for non-zero  $\Lambda$
under the  assumption  of  a    flat  universe.   The $\pm$ 1--    and
2--$\sigma$  limits are plotted for \hub  = 72  \h0units, t$_0$ = 12.5
Gyr, assuming independent uncertainties of $\pm$10\% in each quantity,
and adding the uncertainties in quadrature.  These data are consistent
with either a  low-density ($\Omega_m \sim$ 0.1)  open universe, or  a
flat   universe with  $\Omega_m \sim$  0.35,  $\Omega_\Lambda$ = 0.65;
however,  with these data alone,  it  is not possible to  discriminate
between an open or flat universe.  As  described above, recent studies
favor $\Omega_{total}$ = 1, a low--matter--density universe ($\Omega_m
\sim$ 0.3), and a non-zero value of  the cosmological constant.  Note,
however,  that   the open circle at  $\Omega_m$   = 1,  $\Lambda$ = 0,
represents the Einstein--de Sitter  case, and is inconsistent with the
current values of \hub and t$_0$ only at a $\sim$2--$\sigma$ level.


\section{ Summary }
\label{summary}

We   have used HST  to measure  Cepheid distances  to 18 nearby spiral
galaxies.   Based on a new, larger  sample  of calibrating Cepheids in
the Large Magellanic   Cloud, an improved photometric  calibration for
the HST Wide Field and Planetary Camera 2, attention to incompleteness
bias in the Cepheid  period--luminosity relation, and a correction for
Cepheid metallicity, we have presented here  a set of self-consistent,
revised Cepheid distances to  31 galaxies.  The total sample  includes
previously--published ground-based  photometry, and   additional   HST
studies.  The {\it  relative}   Cepheid distances are   determined  to
$\sim\pm$5\%.

Calibrating 5 secondary  methods with these revised Cepheid distances,
we find \hub  = 72 $\pm$ 3 (random)  $\pm$ 7 (systematic) \h0units, or
\hub = 72 $\pm$ 8  \h0units, if we simply combine  the total errors in
quadrature.  Type  Ia supernovae currently  extend out to the greatest
distances, $\sim$400 Mpc.  All  of the methods  are in extremely  good
agreement: four of  the methods yield a  value of \hub between  70--72
\h0units,  and the fundamental  plane gives  \hub  = 82 \h0units.  The
largest remaining  sources of error result   from (a) uncertainties in
the   distance  to   the  Large  Magellanic   Cloud,  (b)  photometric
calibration of  the  HST  Wide  Field and   Planetary Camera   2,  (c)
metallicity calibration  of the Cepheid  period--luminosity  relation,
and (d) cosmic scatter in the  density (and therefore, velocity) field
that could lead to  observed variations in  \hub on very large scales.
A value of \hub = 72 \h0units yields an  expansion age of $\sim$13 Gyr
for a  flat  universe (consistent with   the  recent cosmic  microwave
background anisotropy results) if $\Omega_m$ = 0.3, $\Omega_\Lambda$ =
0.7. Combined with the current best estimates of  the ages of globular
clusters ($\sim$12.5  Gyr),  our results favor a  $\Lambda$--dominated
universe.

\bigskip
\noindent
\acknowledgments

There  are an enormous  number of people  who have  contributed to and
supported this project over the years.  It is  a pleasure to thank our
collaborators  Fabio  Bresolin, Mingsheng   Han, Paul Harding,  Robert
Hill, John Hoessel,  Myung Gyoon Lee, Randy  Phelps, Abhijit Saha, Kim
Sebo, Nancy Silbermann, and Anne Turner for their contributions to the
project.  We fondly remember and acknowledge the contributions of Marc
Aaronson who led our initial effort  until his untimely death in 1987.
Daya Rawson and Charles Prosser are also kindly remembered and missed.
In particular we thank  our   HST program coordinator  throughout  our
entire  (complicated) observing  process,  Doug Van Orsow,  and former
STScI director, Robert Williams.  We also  thank Andrew Dolphin, Sandy
Faber,  Riccardo  Giacconi,  Riccardo  Giovanelli,   Jim Gunn,  Martha
Haynes,  David Leckrone,   Mark Phillips, Brian   Schmidt, John Tonry,
Brent Tully,  and  Ed Weiler.   We  further gratefully acknowledge the
many  years of assistance of the  NASA and STScI support and technical
staff.  The work presented in this paper is based on observations with
the NASA/ESA  Hubble Space Telescope,  obtained by the Space Telescope
Science  Institute, which   is   operated by AURA,   Inc.  under  NASA
contract No.  5-26555.   Support  for this  work was  provided by NASA
through grant  GO-2227-87A  from STScI.  This  project  could not have
been completed in a  timely   fashion without the generous   financial
support that  we received.   SMGH and PBS  are   grateful to NATO  for
travel   support via a Collaborative   Research Grant (CGR960178).  LF
acknowledges   support  by   NASA  through   Hubble  Fellowship  grant
HF-01081.01-96A,   and through  grant   number  NRA-98-03-LTSA-03.  SS
acknowledges  support   from  NASA    through the  Long     Term Space
Astrophysics Program,  NAS-7-1260.  BG  acknowledges support  from the
NASA Long-Term Space  Astrophysics  Program (NAG5-7262), and  the FUSE
Science Team  (NAS5-32985).  LMM acknowledges  partial support through
Gemini Fellowship No.  GF-1003-95.  WLF  acknowledges support from the
NSF for the  ground-based   calibration in the  early  phases  of this
project under grants AST-87-13889 and AST-91-16496.  This research has
made  use of  the   NASA/IPAC Extragalactic  Database  (NED) which  is
operated by  the  Jet Propulsion  Laboratory, Caltech,  under contract
with the National Aeronautics and Space Administration.

\clearpage

\section{Appendix A: Magnitude-Limited Bias}
\label{appendix_a}

\medskip

We present here an analytic derivation of the bias introduced into the
PL fits  imposed  by magnitude-limited cuts   on extragalactic Cepheid
samples.    We note    that  in \S\ref{revised},   the  application of
short-period cuts to the observed PL relations (to compensate for this
bias) resulted in slightly increased  moduli.  Depending on the sample
size  and its period  distribution, these empirical corrections ranged
from $<$1\%  up  to  4\%  in distance.  We   now provide  an  analytic
solution to reinforce our understanding of the degree and direction of
this bias.

Consider Figure A1, which is meant to represent a uniform distribution
of Cepheids defining a  period-luminosity relation of finite width CD.
It is clear that the dataset defined by the parallelogram ABMN will be
unbiased with respect to  a fit EF (dashed  ridge line), where EF  has
the predefined slope.  It is fit to  the data distribution within ABMN
using  least squares,  assuming that  all of the   variance  is in the
vertical direction, and that errors on the periods are negligible.

Now if a magnitude limiting cut-off to the Cepheid data is imposed (by
line GH), it is  clear that an asymmetry  in the data distribution will
be introduced with a bias in the fitted zero point ensuing: the region
ACD is uncompensated  for by the  exclusion of its complementary region
ADB. Of course the  full  bias introduced by ACD   will depend on  the
relative numbers of stars in this section as compared  to those in the
unbiased  area. However, if the  section ACD is uniformly populated it
can be shown that a fixed-slope solution (to that portion only) fit by
least squares would introduce a bias  toward brighter magnitudes by an
amount  $\Delta m  =  w/6$, where   $w$ is  the  full  magnitude width
(measured at fixed period) of the instability strip.

The minimization of the least-squares conditioning

$$ {\partial \over \partial x_o} \Biggl[ \int_0^w (1.0 - x/w)(x-x_0)^2dx\Biggr] = 0$$

\par\noindent
gives $x_o = w/3$ such that the difference between the biased solution
$x_o$ and the unbiased solution at $w/2$ is $x_o - w/2 = -w/6$

For the reddening-free W-PL relation, where the Key Project fitting is
done, the intrinsic scatter in  the relation is   $\sigma = 0.08$  mag
(Udalski et al.  1999) giving $ w = 4  \times\sigma \sim 0.3$ mag, and
a  predicted  (maximum) bias of  0.05~mag, or  less than  about 3\% in
distance.   This  bias  term  will  of course  be  diluted  in  direct
proportion with  the relative numbers of stars  outside  of the biased
zone. This is in complete agreement  with the quantitative corrections
found in the  main text, where   the typical correction is +0.02  mag,
with the largest correction leading to an increase of 4\% in distance.

\clearpage

\figcaption[veldistceph.ps]{  Velocity  versus distance  for  galaxies
with Cepheid distances.  Velocities  in this plot have  been corrected
using the flow  model described  in Mould  \etal  (2000).  The Cepheid
distances have been corrected for  metallicity. A formal fit to  these
data yields  a  slope of \hub  = 75  $\pm$ 10$_r$ km/sec/Mpc,  in good
agreement, to within the uncertainties,  of the value of \hub obtained
for methods which extend to much greater distances.
\label{fig:veldistceph}}

\medskip

\figcaption[sn1a.eps]{Three  sets of Hubble relations constructed from
the Cal\'an--Tololo (Hamuy \etal  1996)  and CfA-2 (Riess \etal  1999)
Type Ia supernova  samples.  \it Left  Panel\rm: The full sample of 50
supernovae, with   peak   magnitudes corrected   only  for  foreground
Galactic reddening.   All tabulated heliocentric  velocities have been
corrected to the cosmic microwave background reference frame using the
velocity  calculator  available   in  the NASA  Extragalactic Database
(NED).  \it  Middle Panel\rm:  Our  adopted sample of   36 supernovae,
excluding those  with  peak B$-$V colors  in  excess  of 0.20  mag and
velocities with respect to the  cosmic microwave background below 3100
km\,s$^{-1}$.  Both foreground   Galactic and  host galaxy   reddening
corrections were applied.   \it Right  Panel\rm: The Hubble  relations
adopted for this   paper. Same as for  the  middle panel,  but  now an
additional correction for the  light  curve shape (linear  in  $\Delta
m_{15}$(B)) has been applied.  All   slopes $a$, zero points $b$,  and
dispersions $\sigma$ are noted  in their relevant panels.   Foreground
Galactic reddening   corrections E(B$-$V)$_{\rm  Gal}$ are  based upon
COBE/DIRBE  data (Schlegel,   Finkbeiner  \& Davis  1998). To   retain
consistency with the Key Project series of papers, we employed a ratio
of total-to-selective   absorption R$_{\rm   V}=3.3$ and the  Cardelli
\etal (1989) extinction law throughout.
\label{fig:sn1a}}

\medskip

\figcaption[frequentist.ps] {Values  of H$_0$ and  their  uncertainties
for type  Ia  supernovae, the  Tully-Fisher  relation, the fundamental
plane, surface  brightness  fluctuations, and type  II supernovae, all
calibrated   by  Cepheid variables.  Each   value  is represented by a
Gaussian  curve (joined  solid dots) with  unit  area and a 1-$\sigma$
scatter equal to the random uncertainty.  The systematic uncertainties
for each method are indicated by the horizontal bars  near the peak of
each Gaussian.  The upper curve  is obtained by summing the individual
Gaussians.  The cumulative  (frequentist) distribution  has a midpoint
(median)  value of H$_0$  = 72 (71) $\pm$  4 $\pm$  7 km/sec/Mpc.  The
overall systematic error    is obtained   by adding the     individual
systematic errors in quadrature.   
\label{fig:frequentist}}

\medskip

\figcaption[veldist.ps]{[Top  panel]:  A  Hubble diagram  of  distance
versus  velocity  for   secondary distance indicators   calibrated  by
Cepheids.  Velocities  in this plot are corrected  for the nearby flow
model of Mould  \etal  2000a.  The  symbols  are as follows:   Type Ia
supernovae -- squares, Tully-Fisher clusters (I--band observations) --
solid circles,    Fundamental Plane  clusters  --  triangles,  surface
brightness fluctuation galaxies -- diamonds,  Type II supernovae (open
squares).  A slope of H$_0$ = 72 is shown, flanked by $\pm$10\% lines.
Beyond 5,000 km/sec  (indicated by the  vertical line), both numerical
simulations and  observations  suggest that   the effects of  peculiar
motions are  small.   The Type Ia  supernovae  extend to  about 30,000
km/sec and  the Tully-Fisher and Fundamental  Plane clusters extend to
velocities  of about 9,000 and  15,000 km/sec, respectively.  However,
the current  limit for surface  brightness fluctuations is about 5,000
km/sec.  [ Bottom panel: ] Value of H$_0$ as a function of distance.
\label{fig:veldist}}

\medskip

\figcaption[lmcdist.ps]{The distribution of  LMC distance moduli as
compiled by  Gibson  (2000) plotted  as a   continuous probability
density distribution  built up from  the sum of individual
unit-area gaussians  centered at the  quoted modulus, and broadened by
the  published internal  random   error.  
\label{fig:lmcdist}}

\medskip

\figcaption[cephmet.ps]{ Histograms for  the  distributions  of oxygen
abundances  (12 +  log (O/H) )  from  Ferrarese \etal  (2000b) for the
galaxies with Cepheid distances that calibrate type Ia supernovae, the
Tully--Fisher relation,  the fundamental plane, and surface brightness
fluctuations.  The metallicity of the LMC  in these units is 8.50 dex.
The total  distribution   is also  shown.  In  the  mean, most  of the
Cepheid fields observed have higher abundances than the LMC.
\label{fig:cephmet}}

\medskip

\figcaption[szlens.ps]{Values  of the Hubble constant determined using
the Sunyaev--Zel'dovich effect  (open squares) and gravitational lens
time delays (asterisks) from 1990 to the present. From the compilation
of   Huchra   (http://cfa-www.harvard.edu/$\sim$huchra)  for   the Key
Project.  
\label{fig:szlens}}


\medskip

\figcaption[veldistfar.ps]{Logarithm    of distance    in  Mpc  versus
logarithm  of redshift for Cepheids,  the Tully--Fisher relation, type
Ia supernovae, surface brightness fluctuations, fundamental plane, and
type II supernovae, calibrated as part of the Key Project. Solid black
circles are from   Birkinshaw (1999), for   nearby Sunyaev-Zel'dovich
clusters with  cz $<$  30,000 (z$<$0.1)  km/sec,  where the  choice of
cosmological model does not have  a significant effect on the results.
The SZ   clusters are Abell  478, 2142,  and  2256, and are  listed in
Birkinshaw's Table 7.  The solid line is for  \hub = 72 \h0units, with
the dashed lines representing $\pm$10\%.
\label{fig:veldistfar}}

\figcaption[h0t012.ps]{ H$_0$t$_0$  versus  $\Omega$   for \hub =   72
  \h0units, t$_0$  = 12.5 Gyr,  and uncertainties of $\pm$10\% adopted
  for both quantities.  The  dark line indicates  the  case of  a flat
  Universe with $\rm \Omega_\Lambda + \Omega_m$  = 1.  The abscissa in
  this case  corresponds to $\rm  \Omega_\Lambda$.   The lighter curve
  represents a Universe with $\rm  \Omega_\Lambda$ = 0.  In this case,
  the abcissa   should  be read  as  $\rm \Omega_m$.   The dashed  and
  dot-dashed    lines  indicate   1-$\sigma$   and 2-$\sigma$  limits,
  respectively  for values of \hub  = 72 and t$_0$ =   12.5 Gyr in the
  case  where both  quantities are assumed   to be known  to $\pm$10\%
  (1-$\sigma$).  The large open  circle denotes values of H$_0$t$_0$ =
  2/3   and $\Omega_m$   = 1 ({\it    i.e.,}  those predicted   by the
  Einstein-de Sitter model).  On  the basis of a timescale  comparison
  alone,  it  is  not  possible to  discriminate  between  models with
  $\Omega_m  \sim$ 0.1, $\Omega_\Lambda$ = 0  or  $\Omega_m \sim$ 0.35,
  $\Omega_\Lambda \sim$ 0.65.
\label{fig:h0t012}}

\medskip

\noindent
Figure A1 ---  Illustration of bias due  to a magnitude  cutoff in the
Cepheid period-luminosity relation.  See text for details.

\clearpage

\begin{deluxetable}{lccccc}
\footnotesize
\tablecaption{Numbers of Cepheid Calibrators for Secondary Methods
\label{tbl:calibrators}}
\tablewidth{0pt}
\tablehead{
\colhead{ Secondary Method } &
\colhead{ $\sigma$ } &
\colhead{ N (pre--HST) } &
\colhead{ $\sigma_{mean}$ } &
\colhead{ N (post--HST)} &
\colhead{ $\sigma_{mean}$ }  \\
\colhead{ } &
\colhead{ \% } &
\colhead{ } &
\colhead{ \% } &
\colhead{ } &
\colhead{ \% } 
}
\startdata
\tf relation &  $\pm$20\%  \tablenotemark{a}  & 
	     5 \tablenotemark{b} & 
	     $\pm$10\% &
	     21 &
	     $\pm$5\% \nl
\snia        &  $\pm$8\% \tablenotemark{c} & 
	     0   & 
	     n/a &
	     6 \tablenotemark{d} &
	     $\pm$4\% \nl
Surface brightness fluctuations         &  $\pm$9\% \tablenotemark{e}  & 
	     1  & 
	     $\pm$9\% &
	     6 &
	     $\pm$4\% \nl
Fundamental plane            &  $\pm$14\%  &  
	     0  & 
	     n/a &
	     3 \tablenotemark{f} &
	     $\pm$10\% \nl
\snii        &  $\pm$12\% \tablenotemark{g}   &
	     1  & 
	     $\pm$12\% &
	     4 &
	     $\pm$6\% \nl
\enddata
\tablenotetext{a}{ Giovanelli \etal (1997) }
\tablenotetext{b}{ M31, M33, NGC 2403, M81, NGC 300 (Freedman 1990) }
\tablenotetext{c}{ Hamuy \etal (1996)  }
\tablenotetext{d}{ Using the distances   to  the host galaxies to  SN
1960F, 1972E, 1981B, 1989B, 1990N, 1998bu; but excluding 1895B, 1937C, 1974G }
\tablenotetext{e}{ Tonry \etal (1997)  }
\tablenotetext{f}{ calibration based on Cepheid distances to Leo I group, Virgo and Fornax clusters  }
\tablenotetext{g}{this paper, Schmidt \etal (1994) distant clusters }
\end{deluxetable}

\clearpage

\begin{deluxetable}{ll}
\footnotesize
\tablecaption{List of Cepheid Galaxies / Calibrators}
\label{tbl:listceph}
\tablewidth{0pt}
\tablehead{
\colhead{ Galaxy } &
\colhead{ Secondary Methods Calibrated by a Given Galaxy }
}
\startdata
NGC     224 &  TF, SBF \nl
NGC     300 &  . . .\nl
NGC     598 &  TF \nl
NGC     925 &  TF \nl
NGC   1326A &  FP-Fornax \tablenotemark{1} \nl
NGC    1365 &  TF, FP-Fornax \nl
NGC    1425 &  TF, FP-Fornax  \nl
NGC    2090 &  TF \nl
NGC    2403 &  TF \nl
NGC    2541 &  TF \nl
NGC    3031 &  TF, SBF \nl
NGC    3198 &  TF \nl
NGC    3319 &  TF \nl
NGC    3351 &  TF, FP-Leo \tablenotemark{1} \nl
NGC    3368 &  TF, FP-Leo, SBF, SNIa (1998bu) \nl
NGC    3621 &  TF \nl
NGC    3627 &  TF, SNIa (1989B), SNII \nl
NGC    4258 &  . . . \nl
NGC    4321 &  FP-Virgo \tablenotemark{1} \nl
NGC    4414 &  TF, [SNIa (1974G)]\tablenotemark{2} \nl
NGC   4496A &  FP-Virgo, SNIa (1960F) \nl
NGC    4535 &  TF, FP-Virgo \nl
NGC    4536 &  TF, FP-Virgo, SNIa (1981B) \nl
NGC    4548 &  TF, FP-Virgo, SBF \nl
NGC    4639 &  SNIa, FP-Virgo (1990N) \nl
NGC    4725 &  TF, SBF \nl
NGC    5253 &  SNIa (1972E) \nl
NGC    5457 &  SNII \nl
NGC    7331 &  TF, SBF, SNII \nl
IC    4182  &  [SNIa (1937C)]\tablenotemark{2} \nl
IC 1613 & . . . \nl
\enddata
\tablenotetext{1}{FP-Leo, FP-Virgo, FP-Fornax denote, respectively,  the 
galaxies in  the Leo I  Group, and the  Virgo and Fornax clusters. The
calibration of  the fundamental plane  is  based on these  group/cluster
distances (\S\ref{fp}.)}
\tablenotetext{2}{not used in Gibson \etal (2000) nor this paper's SNIa calibration }
\end{deluxetable}

\clearpage

\begin{deluxetable}{lllrllllrl}
\footnotesize
\tablecaption{Revised Cepheid Distances to Galaxies
\label{tbl:cephdist}}
\tablewidth{0pt}
\tablehead{
\colhead{ Galaxy } &
\colhead{ $\mu^{ ~a}_{\rm old}$}  &
\colhead{ $\sigma$ } &
\colhead{ N$_{\rm ceph}$ } &
\colhead{ $\mu^{ ~b}_{\rm revised}$} &
\colhead{ $\sigma$ } &
\colhead{ $\mu^{ ~c}_{\rm P ~cut}$} &
\colhead{ $\sigma$  } &
\colhead{ N$_{\rm ceph}^{\rm cut}$}  &
\colhead{ Data Source  } 
}
\startdata
NGC     224 \tablenotemark{d}  &   24.41 & 0.08 & 37 & 24.38 & 0.05 &  24.38 & 0.05 & 37 &   Freedman \& Madore (1990)             \nl
NGC     300  \tablenotemark{d} &   26.62 & 0.10 & 16 & 26.53 & 0.07 &  26.53 & 0.07 & 14 &   Freedman \etal (1992)                 \nl
NGC     598  \tablenotemark{d} &   24.58 & 0.10 & 12 & 24.56 & 0.10 &  24.56 & 0.08 & 11 &   Freedman  \etal (1991)                \nl
NGC     925  &   29.94 & 0.04 & 73 & 29.80 & 0.04 &  29.80 & 0.04 & 72 &   Silbermann  \etal (1996)              \nl
NGC   1326A  &   31.16 & 0.10 & 17 & 31.00 & 0.09 &  31.04 & 0.09 & 15 &   Prosser  \etal (1999)                  \nl
NGC    1365  &   31.38 & 0.05 & 52 & 31.20 & 0.05 &  31.18 & 0.05 & 47 &   Silbermann \etal (1999)                \nl
NGC    1425  &   31.73 & 0.05 & 29 & 31.54 & 0.05 &  31.60 & 0.05 & 20 &   Mould  \etal (2000b)                    \nl
NGC    2090  &   30.42 & 0.04 & 34 & 30.27 & 0.04 &  30.29 & 0.04 & 30 &   Phelps  \etal (1998)                   \nl
NGC    2403  \tablenotemark{e} &   27.59 & 0.24 & 10 & 27.48 & 0.24 &  27.48 & 0.24 & 10 &   Freedman \& Madore (1988)              \nl
NGC    2541  &   30.43 & 0.07 & 34 & 30.26 & 0.07 &  30.25 & 0.05 & 29 &   Ferrarese \etal (1998)                 \nl
NGC    3031  &   27.75 & 0.07 & 25 & 27.67 & 0.07 &  27.75 & 0.08 & 17 &   Freedman \etal (1994)                  \nl
NGC    3198  &   30.80 & 0.08 & 42 & 30.64 & 0.08 &  30.68 & 0.08 & 36 &   Kelson \etal (1999)                    \nl
NGC    3319  &   30.79 & 0.09 & 33 & 30.64 & 0.09 &  30.64 & 0.09 & 33 &   Sakai \etal (1999)                     \nl
NGC    3351  &   30.03 & 0.10 & 49 & 29.90 & 0.10 &  29.85 & 0.09 & 48 &   Graham \etal (1997)                    \nl
NGC    3368  &   30.10 & 0.08 & 11 & 29.95 & 0.08 &  29.97 & 0.06 &  9 &   Tanvir \etal (1995) \tablenotemark{f}                    \nl
NGC    3621  &   29.21 & 0.06 & 69 & 29.06 & 0.06 &  29.08 & 0.06 & 59 &   Rawson \etal (1997)                    \nl
NGC    3627  &   29.88 & 0.08 & 35 & 29.71 & 0.08 &  29.86 & 0.08 & 16 &   Gibson \etal (2000) \tablenotemark{e}  \nl
NGC    4258  &   29.49 & 0.07 &  15 & 29.44 & 0.07 &  29.44 & 0.07 &  15 &   Newman \etal (2001)                      \nl
NGC    4321  &   30.93 & 0.07 & 52 & 30.75 & 0.07 &  30.78 & 0.07 & 42 &   Ferrarese \etal (1996)                 \nl
NGC    4414  &   31.37 & 0.09 &  9 & 31.18 & 0.09 &  31.10 & 0.05 &  8 &   Turner \etal (1998)                    \nl
NGC   4496A  &   30.98 & 0.03 & 98 & 30.80 & 0.03 &  30.81 & 0.03 & 94 &   Gibson \etal (2000) \tablenotemark{f}  \nl
NGC    4535  &   31.02 & 0.05 & 50 & 30.84 & 0.05 &  30.85  & 0.05 & 47 &   Macri  \etal (1999)                    \nl
NGC    4536  &   30.95 & 0.04 & 39 & 30.78 & 0.04 &  30.80 & 0.04 & 35 &   Gibson \etal (2000) \tablenotemark{e}  \nl
NGC    4548  &   31.03 & 0.05 & 24 & 30.88 & 0.05 &  30.88 & 0.05 & 24 &   Graham \etal (1999)                    \nl
NGC    4639  &   31.80 & 0.07 & 17 & 31.59 & 0.07 &  31.61 & 0.08 & 14 &   Gibson \etal (2000) \tablenotemark{f}  \nl
NGC    4725  &   30.50 & 0.06 & 20 & 30.33 & 0.06 &  30.38 & 0.06 & 15 &   Gibson \etal (2000) \tablenotemark{e}  \nl
NGC    5253  &   27.60 & 0.10 &  7 & 27.54 & 0.10 &  27.56 & 0.14 &  4 &   Gibson \etal (2000) \tablenotemark{f}  \nl
NGC    5457  &   29.35 & 0.10 & 29 & 29.18 & 0.10 &  29.13 & 0.11 & 25 &   Kelson \etal (1996)                    \nl
NGC    7331  &   30.90 & 0.09 & 13 & 30.81 & 0.09 &  30.81 & 0.09 & 13  &   Hughes \etal (1998)  \nl
IC    4182  &   28.36 & 0.06 & 18 & 28.26 & 0.05 &  28.28 & 0.06 & 16  &   Gibson \etal (2000) \tablenotemark{f} \nl
IC    1613  &   24.29 & 0.14 & 10 & 24.24 & 0.14 &  24.19 & 0.15 &  9  &   Freedman (1988) \tablenotemark{f} \nl
\enddata
\tablenotetext{a} {Adopting Madore \& Freedman (1991) PL slopes; LMC distance modulus 18.50; ALLFRAME intensity-weighted mean magnitudes or Stetson template fits if available; Hill \etal (1998) calibration, except for M31 (NGC 224), M33 (NGC 598), IC 1613, NGC 300, NGC 2403, M81 (NGC 3031), M101 (outer; NGC 5457)}.
\tablenotetext{b} {Adopting Udalski \etal (1999) PL slopes; same
Cepheid sample as for column 2; Stetson (1998) WFPC2 calibration,
except for for M31, M33, IC 1613, NGC 300, NGC 2403, M81.  (To transform
distance moduli from Hill \etal to Stetson, 0.07 mag is subtracted.)}.
\tablenotetext{c} {Same calibration  as for column  5, but applying  a
period  cut at   the  short  period   end   to minimize  bias   in the
period-luminosity relation; where the numbers of Cepheids  in columns 4 
and 9 are equal, no period cut was applied }.
\tablenotetext{d} {For the galaxies M31, M33, NGC 300, observed from the
ground, and for which BVRI photometry are available, distances tabulated
here are based on VI photometry to be consistent with the HST sample galaxies.}
\tablenotetext{e} {I-band data only are available for NGC 2403. A reddening
of E(V--I) = 0.20 $\pm$ 0.10 has been adopted, comparable to that for other
spiral galaxies. See Table \ref{tbl:finaldist}.}
\tablenotetext{f} {Reanalyzed by Gibson \etal (2000).}
\end{deluxetable}

\begin{deluxetable}{llllllllcrlrcl}
\footnotesize
\tablecaption{Final Adopted Distance Moduli, Reddenings,  Distances, Metallicities
\label{tbl:finaldist}}
\tablewidth{0pt}
\tablehead{
\colhead{ Galaxy } &
\colhead{ $\mu_V$ } &
\colhead{ $\sigma_V$\tablenotemark{a} } &
\colhead{ $\mu_I$ } &
\colhead{ $\sigma_I$\tablenotemark{a} } &
\colhead{ E(V--I) } &
\colhead{ $\sigma_{E(V-I)}$ } &
\colhead{ $\mu_0$ } &
\colhead{ $\sigma_0$\tablenotemark{a} } &
\colhead{ D$_0$  } &
\colhead{ $\mu_z$ } &
\colhead{ D$_z$  } &
\colhead{ z\tablenotemark{b} }&
\colhead{ $\sigma_z$\tablenotemark{a}}  \\
\colhead{  } &
\colhead{ (mag) } &
\colhead{  } &
\colhead{ (mag) } &
\colhead{  } &
\colhead{ (mag) } &
\colhead{ (mag) } &
\colhead{ (mag) } &
\colhead{  } &
\colhead{ (Mpc) } &
\colhead{ (mag) } &
\colhead{ (Mpc)  } &
\colhead{ (dex) } &
\colhead{ (dex) } 
}
\startdata
NGC   224 & 25.01 & 0.07 & 24.76 & 0.05 &  0.26 & 0.04 &  24.38 & 0.05 & 0.75 & 24.48 & 0.79 & 8.98 & 0.15 \nl
NGC   300 & 26.60 & 0.05 & 26.57 & 0.04 &  0.04 & 0.03 &  26.53 & 0.07 & 2.02 & 26.50 & 2.00 & 8.35 & 0.15 \nl
NGC   598 & 25.21 & 0.11 & 24.94 & 0.08 &  0.27 & 0.05 &  24.56 & 0.10 & 0.82 & 24.62 & 0.84 & 8.82 & 0.15 \nl
NGC   925 & 30.33 & 0.04 & 30.12 & 0.03 &  0.21 & 0.02 &  29.80 & 0.04 & 9.12 & 29.81 & 9.16 & 8.55 & 0.15 \nl
NGC 1326A & 31.41 & 0.07 & 31.26 & 0.07 &  0.15 & 0.04 &  31.04 & 0.10 & 16.14 & 31.04 & 16.14 & 8.50 & 0.15 \nl
NGC  1365 & 31.69 & 0.05 & 31.49 & 0.04 &  0.20 & 0.02 &  31.18 & 0.05 & 17.22 & 31.27 & 17.95 & 8.96 & 0.20 \nl
NGC  1425 & 32.01 & 0.07 & 31.85 & 0.05 &  0.16 & 0.03 &  31.60 & 0.05 & 20.89 & 31.70 & 21.88 & 9.00 & 0.15 \nl
NGC  2090 & 30.71 & 0.05 & 30.54 & 0.04 &  0.17 & 0.02 &  30.29 & 0.04 & 11.43 & 30.35 & 11.75 & 8.80 & 0.15 \nl
NGC  2403 & . . . &. . . &  27.75 & 0.10 &  0.2\tablenotemark{c} & 0.1  &  27.48 & 0.10 & 3.13 & 27.54 & 3.22 & 8.80 & 0.10 \nl
NGC  2541 & 30.74 & 0.05 & 30.54 & 0.04 &  0.20 & 0.02 &  30.25 & 0.05 & 11.22 & 30.25 & 11.22 & 8.50 & 0.15 \nl
NGC  3031 & 28.22 & 0.09 & 28.03 & 0.07 &  0.19 & 0.05 &  27.75 & 0.08 & 3.55 & 27.80 & 3.63 & 8.75 & 0.15 \nl
NGC  3198 & 31.04 & 0.05 & 30.89 & 0.04 &  0.15 & 0.04 &  30.68 & 0.08 & 13.68 & 30.70 & 13.80 & 8.60 & 0.15 \nl
NGC  3319 & 30.95 & 0.06 & 30.82 & 0.05 &  0.13 & 0.04 &  30.64 & 0.09 & 13.43 & 30.62 & 13.30 & 8.38 & 0.15 \nl
NGC  3351 & 30.43 & 0.06 & 30.19 & 0.05 &  0.24 & 0.04 &  29.85 & 0.09 & 9.33 & 30.00 & 10.00 & 9.24 & 0.20 \nl
NGC  3368 & 30.44 & 0.11 & 30.25 & 0.08 &  0.20 & 0.04 &  29.97 & 0.06 & 9.86 & 30.11 & 10.52 & 9.20 & 0.20 \nl
NGC  3621 & 29.97 & 0.07 & 29.61 & 0.05 &  0.36 & 0.04 &  29.08 & 0.06 & 6.55 & 29.11 & 6.64 & 8.75 & 0.15 \nl
NGC  3627 & 30.44 & 0.09 & 30.20 & 0.07 &  0.24 & 0.03 &  29.86 & 0.08 & 9.38 & 30.01 & 10.05 & 9.25 & 0.15 \nl
NGC  4258 & 29.99 & 0.08 & 29.77 & 0.05 &  0.22 & 0.04 &  29.44 & 0.07 & 7.73 & 29.51 & 7.98 & 8.85 & 0.15 \nl 
NGC  4321 & 31.31 & 0.06 & 31.09 & 0.05 &  0.22 & 0.03 &  30.78 & 0.07 & 14.32 & 30.91 & 15.21 & 9.13 & 0.20 \nl
NGC  4414 & 31.48 & 0.14 & 31.33 & 0.10 &  0.15 & 0.04 &  31.10 & 0.05 & 16.60 & 31.24 & 17.70 & 9.20 & 0.15 \nl
NGC 4496A & 31.14 & 0.03 & 31.00 & 0.03 &  0.14 & 0.01 &  30.81 & 0.03 & 14.52 & 30.86 & 14.86 & 8.77 & 0.15 \nl
NGC  4535 & 31.32 & 0.04 & 31.13 & 0.04 &  0.19 & 0.02 &  30.85 & 0.05 & 14.79 & 30.99 & 15.78 & 9.20 & 0.15 \nl
NGC  4536 & 31.24 & 0.04 & 31.06 & 0.04 &  0.18 & 0.02 &  30.80 & 0.04 & 14.45 & 30.87 & 14.93 & 8.85 & 0.15 \nl
NGC  4548 & 31.30 & 0.07 & 31.12 & 0.04 &  0.18 & 0.04 &  30.88 & 0.05 & 15.00 & 31.05 & 16.22 & 9.34 & 0.15 \nl
NGC  4639 & 31.96 & 0.09 & 31.84 & 0.07 &  0.12 & 0.04 &  31.61 & 0.08 & 20.99 & 31.71 & 21.98 & 9.00 & 0.15 \nl
NGC  4725 & 31.08 & 0.08 & 30.79 & 0.07 &  0.29 & 0.03 &  30.38 & 0.06 & 11.91 & 30.46 & 12.36 & 8.92 & 0.15 \nl
NGC  5253 & 28.01 & 0.17 & 27.83 & 0.12 &  0.19 & 0.08 &  27.56 & 0.14 & 3.25 & 27.49 & 3.15 & 8.15 & 0.15 \nl
NGC  5457\tablenotemark{d} & 29.46 & 0.07 & 29.33 & 0.05 &  0.13 & 0.06 &  29.13 & 0.11 & 6.70 & 29.13 & 6.70 & 8.50 & 0.15 \nl
NGC  7331 & 31.42 & 0.09 & 31.17 & 0.06 &  0.25 & 0.05 &  30.81 & 0.09 & 14.52 & 30.84 & 14.72 & 8.67 & 0.15 \nl
IC  4182 & 28.37 & 0.07 & 28.33 & 0.06 &  0.04 & 0.03 &  28.28 & 0.06 & 4.53 & 28.26 & 4.49 & 8.40 & 0.20 \nl
IC  1613 & 24.44 & 0.09 & 24.34 & 0.10 &  0.10 & 0.05 &  24.19 & 0.15 & 0.69 & 24.06 & 0.65 & 7.86 & 0.50 \nl
\enddata
\tablenotetext{a}{ random uncertainty, not including systematic errors }
\tablenotetext{b}{ 12 + log(O/H) (Ferrarese \etal 2000b) }
\tablenotetext{c}{ adopted reddening; see text}
\tablenotetext{d}{The distance given for M101 is based on data for an
outer field in this galaxy (Kelson 1996), where the metallicity is
very nearly that of LMC.}
\end{deluxetable}

\clearpage
\bigskip

\begin{deluxetable}{lrrrrrrr}
\footnotesize
\tablecaption{Local Velocity Flow 
\label{tbl:cephvel}}
\tablewidth{0pt}
\tablehead{
\colhead{ Galaxy } &
\colhead{ $V_{Helio}$ } &
\colhead{ $V_{LG}$ } &
\colhead{ $V_{CMB}$ } &
\colhead{ $V_{Virgo}$ } &
\colhead{ $V_{GA}$ } &
\colhead{ $V_{Shapley}$ } &
\colhead{ $V_{Tonry}$ } 
}
\startdata
NGC 0300 &  144 &  125 &  -57 &  114 &   92 &  133 & -140\nl
NGC 0925 &  553 &  781 &  398 &  778 &  561 &  664 &  374\nl
NGC 1326A& 1836 & 1749 & 1787 & 1698 & 1742 & 1794 & 1164\nl
NGC 1365 & 1636 & 1544 & 1597 & 1503 & 1544 & 1594 & 1157\nl
NGC 1425 & 1512 & 1440 & 1477 & 1403 & 1417 & 1473 & 1465\nl
NGC 2403 &  131 &  300 &  216 &  343 &  222 &  278 &  193\nl
NGC 2541 &  559 &  646 &  736 &  744 &  674 &  714 &  936\nl
NGC 2090 &  931 &  757 & 1057 &  805 &  869 &  882 &  926\nl
NGC 3031 &  -34 &  127 &   65 &  139 &   43 &   80 &  246\nl
NGC 3198 &  662 &  704 &  890 &  768 &  765 &  772 &  848\nl
NGC 3351 &  778 &  641 & 1117 &  594 &  696 &  642 & 1175\nl
NGC 3368 &  897 &  761 & 1236 &  715 &  823 &  768 & 1238\nl
NGC 3621 &  805 &  615 & 1152 &  557 &  687 &  609 & 1020\nl
NGC 4321 & 1571 & 1469 & 1856 & 1350 & 1501 & 1433 & 1436\nl
NGC 4414 &  716 &  693 &  959 &  586 &  661 &  619 & 1215\nl
NGC 4496A& 1730 & 1575 & 2024 & 1350 & 1518 & 1424 & 1467\nl
NGC 4548 &  486 &  381 &  763 & 1350 & 1460 & 1384 & 1421\nl
NGC 4535 & 1961 & 1826 & 2248 & 1350 & 1530 & 1444 & 1410\nl
NGC 4536 & 1804 & 1642 & 2097 & 1350 & 1521 & 1423 & 1463\nl
NGC 4639 & 1010 &  902 & 1283 & 1350 & 1481 & 1403 & 1448\nl
NGC 4725 & 1206 & 1161 & 1446 & 1040 & 1156 & 1103 & 1225\nl
IC  4182 &  321 &  344 &  513 &  312 &  355 &  318 &  636\nl
NGC 5253 &  404 &  156 &  612 &  160 &  349 &  232 &  800\nl
NGC 7331 &  816 & 1110 &  508 & 1099 &  912 &  999 &  820\nl
\enddata
\end{deluxetable}

\clearpage
\bigskip

\begin{deluxetable}{lrrrr}
\footnotesize
\tablecaption{Type Ia Supernovae Hubble Constant
\label{tbl:snveldist}}
\tablewidth{0pt}
\tablehead{
\colhead{ Supernova } &
\colhead{ $V_{CMB}$ } &
\colhead{ D(Mpc) } &
\colhead{ $H_0^{CMB}$ } &
\colhead{ $\sigma$ } 
}
\startdata
SN1990O  &  9065 & 134.7 & 67.3 & 2.3\nl
SN1990T  & 12012 & 158.9 & 75.6 & 3.1\nl
SN1990af & 15055 & 198.6 & 75.8 & 2.8\nl
SN1991S  & 16687 & 238.9 & 69.8 & 2.8\nl
SN1991U  &  9801 & 117.1 & 83.7 & 3.4\nl
SN1991ag &  4124 &  56.0 & 73.7 & 2.9\nl
SN1992J  & 13707 & 183.9 & 74.5 & 3.1\nl
SN1992P  &  7880 & 121.5 & 64.8 & 2.2\nl
SN1992ae & 22426 & 274.6 & 81.6 & 3.4\nl
SN1992ag &  7765 & 102.1 & 76.1 & 2.7\nl
SN1992al &  4227 &  58.0 & 72.8 & 2.4\nl
SN1992aq & 30253 & 467.0 & 64.7 & 2.4\nl
SN1992au & 18212 & 262.2 & 69.4 & 2.9\nl
SN1992bc &  5935 &  88.6 & 67.0 & 2.1\nl
SN1992bg & 10696 & 151.4 & 70.6 & 2.4\nl
SN1992bh & 13518 & 202.5 & 66.7 & 2.3\nl
SN1992bk & 17371 & 235.9 & 73.6 & 2.6\nl
SN1992bl & 12871 & 176.8 & 72.7 & 2.6\nl
SN1992bo &  5434 &  77.9 & 69.7 & 2.4\nl
SN1992bp & 23646 & 309.5 & 76.3 & 2.6\nl
SN1992br & 26318 & 391.5 & 67.2 & 3.1\nl
SN1992bs & 18997 & 280.1 & 67.8 & 2.8\nl
SN1993B  & 21190 & 303.4 & 69.8 & 2.4\nl
SN1993O  & 15567 & 236.1 & 65.9 & 2.1\nl
SN1993ag & 15002 & 215.4 & 69.6 & 2.4\nl
SN1993ah &  8604 & 119.7 & 71.9 & 2.9\nl
SN1993ac & 14764 & 202.3 & 72.9 & 2.7\nl
SN1993ae &  5424 &  71.8 & 75.6 & 3.1\nl
SN1994M  &  7241 &  96.7 & 74.9 & 2.6\nl
SN1994Q  &  8691 & 127.8 & 68.0 & 2.7\nl
SN1994S  &  4847 &  66.8 & 72.5 & 2.5\nl
SN1994T  & 10715 & 149.9 & 71.5 & 2.6\nl
SN1995ac & 14634 & 185.6 & 78.8 & 2.7\nl
SN1995ak &  6673 &  82.4 & 80.9 & 2.8\nl
SN1996C  &  9024 & 136.0 & 66.3 & 2.5\nl
SN1996bl & 10446 & 132.7 & 78.7 & 2.7\nl
\enddata
\end{deluxetable}

\clearpage
\bigskip

\begin{deluxetable}{lccrccrcr}
\footnotesize
\tablecaption{I--band Tully--Fisher Hubble Constant 
\label{tbl:tfveldist}}
\tablewidth{0pt}
\tablehead{
\colhead{ Cluster/Group } &
\colhead{ $V_{CMB}$ } &
\colhead{ $V_{Flow}$ } &
\colhead{ $\sigma$ } &
\colhead{ D(Mpc) } &
\colhead{ $H_0^{CMB}$ } &
\colhead{ $\sigma$ } &
\colhead{ $H_0^{Flow}$ } &
\colhead{ $\sigma$ } 
}
\startdata
 Abell 1367   & 6709 &  6845 &  88 & 89.2 & 75.2 & 12.5 & 76.7 & 12.8 \nl
 Abell 0262   & 4730 &  5091 &  80 & 66.7 & 70.9 & 11.8 & 76.2 & 12.7 \nl
 Abell 2634   & 8930 &  9142 &  79 &114.9 & 77.7 & 12.4 & 79.6 & 12.7 \nl
 Abell 3574   & 4749 &  4617 &  11 & 62.2 & 76.2 & 12.2 & 74.2 & 11.9 \nl
 Abell 0400   & 7016 &  6983 &  75 & 88.4 & 79.3 & 12.6 & 79.0 & 12.6 \nl
 Antlia       & 3106 &  2821 & 100 & 45.1 & 68.8 & 11.3 & 62.5 & 10.3 \nl
 Cancer       & 4982 &  4942 &  80 & 74.3 & 67.1 & 11.0 & 66.5 & 10.9 \nl
 Cen 30       & 3272 &  4445 & 150 & 43.2 & 75.8 & 12.8 &102.9 & 17.4 \nl
 Cen 45       & 4820 &  4408 & 100 & 68.2 & 70.7 & 11.9 & 64.6 & 10.9 \nl
 Coma         & 7143 &  7392 &  68 & 85.6 & 83.5 & 13.4 & 86.4 & 13.9 \nl
 Eridanus     & 1607 &  1627 &  30 & 20.7 & 77.6 & 12.9 & 78.5 & 13.1 \nl
 ESO 50       & 3149 &  2896 & 100 & 39.5 & 79.8 & 13.0 & 73.3 & 11.9 \nl
 Fornax       & 1380 &  1372 &  45 & 15.0 & 92.2 & 15.3 & 91.7 & 15.2 \nl
 Hydra        & 4061 &  3881 &  50 & 58.3 & 69.6 & 11.1 & 66.5 & 10.6 \nl
 MDL 59       & 2304 &  2664 &  75 & 31.3 & 73.6 & 11.8 & 85.1 & 13.7 \nl
 NGC 3557     & 3294 &  2957 &  60 & 38.7 & 85.0 & 14.4 & 76.3 & 12.9 \nl
 NGC 0383     & 4924 &  5326 &  32 & 66.6 & 73.9 & 11.9 & 80.0 & 12.9 \nl
 NGC 0507     & 4869 &  5257 &  99 & 57.3 & 84.9 & 13.5 & 91.8 & 14.6 \nl
 Pavo 2       & 4398 &  4646 &  70 & 50.9 & 86.3 & 14.2 & 91.2 & 15.0 \nl
 Pegasus      & 3545 &  3874 &  80 & 53.3 & 66.4 & 10.7 & 72.6 & 11.7 \nl
 Ursa Major   & 1088 &  1088 &  40 & 19.8 & 54.8 &  8.6 & 54.8 &  8.6 \nl
\enddata
\end{deluxetable}

\clearpage
\bigskip

\begin{deluxetable}{lclclclc}
\footnotesize
\tablecaption{ Adopted Revised Cepheid Distances to Leo I, Virgo and Fornax
\label{tbl:clusters}}
\tablewidth{0pt}
\tablehead{
\colhead{ Cluster/Group } &
\colhead{ $\mu_0$ (mag) } &
\colhead{ $\pm\sigma$ } &
\colhead{ D (Mpc)} &
\colhead{ $\pm\sigma$ } &
\colhead{ $\mu_z$ (mag)} &
\colhead{ $\pm\sigma$ } &
\colhead{ D$_z$ (Mpc)} 
}
\startdata
 Leo I Group \tablenotemark{a} &      29.90  &  0.10  &   9.5  &  0.4  &  30.01  &  0.09 & 10.0 \nl
 Virgo Cluster \tablenotemark{b}  &      30.81  &  0.04  &  14.6  &  0.3  &  30.92  &  0.05 & 15.3 \nl
 Fornax Cluster \tablenotemark{c}&      31.32  &  0.17  &  18.3  &  1.4  &  31.39  &  0.20 & 19.0 \nl
\enddata
\tablenotetext{a}{ based on distances to NGC 3351 and NGC 3368 }
\tablenotetext{b}{ based on distances to NGC 4321, NGC 4496A, NGC 4535, NGC 4536, NGC 4548 }
\tablenotetext{c}{ based on distances to NGC 1326A, NGC 1365, NGC 1425 }
\end{deluxetable}

\clearpage
\bigskip

\begin{deluxetable}{lclclclclclclc}
\footnotesize
\tablecaption{Fundamental Plane Hubble Constant
\label{tbl:fpveldist}}
\tablewidth{0pt}
\tablehead{
\colhead{ Cluster/Group } &
\colhead{ $N$ } &
\colhead{ $V_{CMB}$ } &
\colhead{ $V_{Flow}$ } &
\colhead{ D(Mpc) } &
\colhead{ $\sigma$ } &
\colhead{ $H_0^{CMB}$ } &
\colhead{ $\sigma$ } &
\colhead{ $H_0^{Flow}$ } &
\colhead{ $\sigma$ } 
}
\startdata
Dorado       &  9 &  1131 &   1064 & 13.8 &   1.4 &  81.9 &  8.7 &  77.0 &  8.2 \nl
GRM 15       &  7 &  4530 &   4848 & 47.4 &   4.7 &  95.6 & 10.0 & 102.2 & 10.7 \nl
Hydra I      & 20 &  4061 &   3881 & 49.1 &   4.7 &  82.8 &  8.4 &  79.1 &  8.0 \nl
Abell S753   & 16 &  4351 &   3973 & 49.7 &   4.2 &  87.5 &  7.9 &  79.9 &  7.2 \nl
Abell 3574   &  7 &  4749 &   4617 & 51.6 &   5.3 &  92.0 & 10.0 &  89.5 &  9.7 \nl
Abell 194    & 25 &  5100 &   5208 & 55.9 &   4.3 &  91.3 &  7.5 &  93.2 &  7.6 \nl
Abell S639   & 12 &  6533 &   6577 & 59.6 &   5.1 & 109.7 &  9.9 & 110.4 & 10.0 \nl
Coma         & 81 &  7143 &   7392 & 85.8 &   5.9 &  83.2 &  6.0 &  86.1 &  6.2 \nl
Abell 539    & 25 &  8792 &   8648 & 102.0 &   7.4 &  86.2 &  6.5 &  84.7 &  6.4 \nl
DC 2345-28   & 30 &  8500 &   8708 & 102.1 &   7.4 &  83.2 &  6.4 &  85.2 &  6.5 \nl
Abell 3381   & 14 & 11536 &  11436 &129.8 &  11.5 &  88.9&  8.3 &  88.1 &  8.2 \nl
\enddata
\end{deluxetable}

\clearpage
\bigskip

\begin{deluxetable}{lclclcl}
\footnotesize
\tablecaption{Surface--Brightness--Fluctuation Hubble Constant
\label{tbl:sbfveldist}}
\tablewidth{0pt}
\tablehead{
\colhead{ Galaxy } &
\colhead{ $V_{Flow}$ } &
\colhead{ $\sigma$ } &
\colhead{ D(Mpc) } &
\colhead{ $\sigma$ } &
\colhead{ $H_0^{Flow}$ } &
\colhead{ $\sigma$ } 
}
\startdata
NGC 4881 &  7441 & 300 & 102.3 & 24.8 & 72.7 & 18.7\nl
NGC 4373 &  3118 & 508 & 36.3 &  3.8 & 85.9 & 17.2\nl
NGC 0708 &  4831 & 300 & 68.2 &  6.7 & 70.8 &  8.6\nl
NGC 5193 &  3468 & 551 & 51.5 &  4.2 & 67.3 & 12.4\nl
IC  0429 &  3341 & 552 & 55.5 &  4.2 & 60.2 & 11.2\nl
NGC 7014 &  5061 & 300 & 67.3 &  4.8 & 75.2 &  7.2\nl
\enddata
\end{deluxetable}

\clearpage
\bigskip

\begin{deluxetable}{llcccc}
\footnotesize
\tablecaption{ Comparison of Nearby Cepheid and Type II SN Distances
\label{tbl:snii}}
\tablewidth{0pt}
\tablehead{
\colhead{ Supernova } &
\colhead{ Host } &
\colhead{ $\mu$ (Cepheid) } &
\colhead{ $\sigma$ } &
\colhead{ $\mu$ (SN~II)} &
\colhead{ $\sigma$ } 
}
\startdata
 SN1970G & M101     & 29.13 & 0.11 & 29.40 & 0.35   \nl
 SN1987A & LMC      & 18.50 & 0.10 & 18.50 & 0.13   \nl
 SN1989L & NGC 7331 & 30.84 & 0.09 & 31.20 & 0.51   \nl
\enddata
\end{deluxetable}

\clearpage


\clearpage

\begin{deluxetable}{lccl}
\footnotesize
\tablecaption{Uncertainties in H$_0$ for Secondary Methods
\label{tbl:kpresults}}
\tablewidth{0pt}
\tablehead{
\colhead{Method  } &
\colhead{ H$_0$  } &
\colhead{Error (\%)  } &
\colhead{References  } 
}
\startdata
36 Type Ia supernovae      & 
71	&  $\pm$2$_r$ $\pm$6$_s$  & Hamuy \etal (1996), Riess \etal(1998),   \nl
4,000 $<$ cz $<$ 30,000 km/sec  & & & Jha \etal (1999),  Gibson \etal (2000) \nl
 &  & &  \nl
21 Tully--Fisher clusters      & 
71	&  $\pm$3$_r$ $\pm$7$_s$  & Giovanelli \etal (1997), Aaronson \etal (1982, 1986),  \nl
1,000 $<$ cz $<$ 9,000 km/sec  & & & Sakai \etal (2000) \nl
 &  & &  \nl
11 FP clusters &    
82	& $\pm$6$_r$ $\pm$9$_s$   & Jorgensen \etal (1996), Kelson \etal (2000)   	  \nl
1,000 $<$ cz $<$ 11,000 km/sec &  &  &  \nl
 & & &  \nl
SBF for 6 clusters   &    
70	  & $\pm$5$_r$ $\pm$6$_s$  &  Lauer \etal (1998),  Ferrarese \etal (2000a) \nl
3,800 $<$ cz $<$ 5,800 km/sec & & &  \nl
 & & &  \nl
4 Type II supernovae  &     
72	& $\pm$ 9$_r$ $\pm$ 7$_s$  &  Schmidt \etal (1994)	  \nl
1,900 $<$ cz $<$ 14,200 km/sec  & & & \nl
\enddata
\tablenotetext{}{Combined values of H$_0$: }
\tablenotetext{}{~~~~~~~~~~~~~~~~~H$_0$ = 72 $\pm$ 2 [random] km/sec/Mpc ~~~[Bayesian] }
\tablenotetext{}{~~~~~~~~~~~~~~~~~H$_0$ = 72 $\pm$ 3 [random] km/sec/Mpc ~~~[frequentist] }
\tablenotetext{}{~~~~~~~~~~~~~~~~~H$_0$ = 72 $\pm$ 3 [random] km/sec/Mpc ~~~[Monte Carlo] }
\end{deluxetable}

\clearpage

\begin{deluxetable}{lccrccr}
\footnotesize
\tablecaption{LMC Distance Moduli for Different Methods}
\label{Table 2}
\tablewidth{0pt}
\tablehead{
\colhead{ Method } &
\colhead{ $<\mu_0>$ (mag) \tablenotemark{a} } &
\colhead{ $\sigma$ (mag)} &
\colhead{ N } &
\colhead{ $<\mu_0>$(mag) \tablenotemark{b} } &
\colhead{ $\sigma$ (mag) } &
\colhead{ N }
}
\startdata
Cepheids  &  18.57    & 
	     $\pm$0.14  & 
	     5 &
	     18.52  &
	     $\pm$0.13  & 
	     15 \nl
Eclipsing variables  &  18.33    & 
	     $\pm$0.05  & 
	     3 &
	     . . .   &
	     . . . & 
	     . . . \nl
SN1987A  &  18.47    & 
	     $\pm$0.08  & 
	     4 &
	     18.50  &
	     $\pm$0.12  & 
	     5 \nl
TRGB  &  18.64    & 
	     $\pm$0.05  & 
	     2 &
	     18.42  &
	     $\pm$0.15  & 
	     1 \nl
Red Clump  &  18.27    & 
	     $\pm$0.11  & 
	     10 &
	     . . .   &
	     . . .  & 
	     . . . \nl
RR Lyrae variables  &  18.30    & 
	     $\pm$0.13  & 
	     7 &
	     18.40  &
	     $\pm$0.19  & 
	     14 \nl
Mira variables  &  18.54    & 
	     $\pm$0.04  & 
	     3 &
	     18.46  &
	     $\pm$0.11  & 
	     4 \nl
\enddata
\tablenotetext{a}{ based on Gibson  (2000) compilation }
\tablenotetext{b}{ based on Westerlund (1997) compilation }
\end{deluxetable}

\clearpage

\begin{deluxetable}{llr}
\footnotesize
\tablecaption{Overall Systematic Errors Affecting All Methods
\label{tbl:h0errors2}}
\tablewidth{0pt}
\tablehead{
\colhead{Source of Uncertainty  } &
\colhead{Description  } &
\colhead{Error (\%)  } 
}
\startdata
LMC zero point   & error on mean from Cepheids, TRGB,   &   	  \nl
 &  SN1987A, red clump, eclipsing binaries & $\pm$ 5\% \nl
WFPC2 zero point & tie-in to Galactic star clusters & $\pm$3.5\% \nl
Reddening & limits from NICMOS photometry & $\pm$1\% \nl
Metallicity & optical, NICMOS, theoretical constraints & $\pm$4\% \nl
Bias in Cepheid PL & short-end period cutoff & $\pm$1\% \nl
Crowding & artificial star experiments & $+$5,$-$0\% \nl
Bulk flows on scales  & limits from SNIa, CMB  & $\pm$5\% \nl
~~~$>$10,000 km/sec &  & \nl
\enddata
\tablenotetext{}{Adopted final value of H$_0$: }
\tablenotetext{}{~~~~~~~~~~~~H$_0$ = 72 $\pm$ 3 [random] $\pm$ 7 [systematic] km/sec/Mpc  }
\end{deluxetable}

\clearpage

\begin{deluxetable}{ccccc}
\footnotesize
\tablecaption{Expansion Ages (in Gyr) for Flat Universes\tablenotemark{a}
\label{tbl:t0lambda}}
\tablewidth{12cm}
\tablehead{
\colhead{ \hub / $\Omega_\Lambda$ } &
\colhead{ 0.0 } &
\colhead{ 0.6 } &
\colhead{ 0.7 } &
\colhead{ 0.8 } 
}
\startdata
55 & 11.9 & \ \ 15.1 & 17.1 & 18.5  \nl
65 & 10.0 & \ \ 12.7 & 14.5 & 16.2  \nl
75 & ~8.7 & \ \ 11.1 & 12.6 & 14.0  \nl
85 & ~7.7 & \ \ ~9.8 & 11.1 & 12.2  \nl
\enddata
\tablenotetext{a} { $\Omega_{Total}$ \ = $\Omega_m$ + $\Omega_\Lambda$ = \ 1.000 }
\end{deluxetable}

\clearpage

\epsscale{1.0}
\plotone{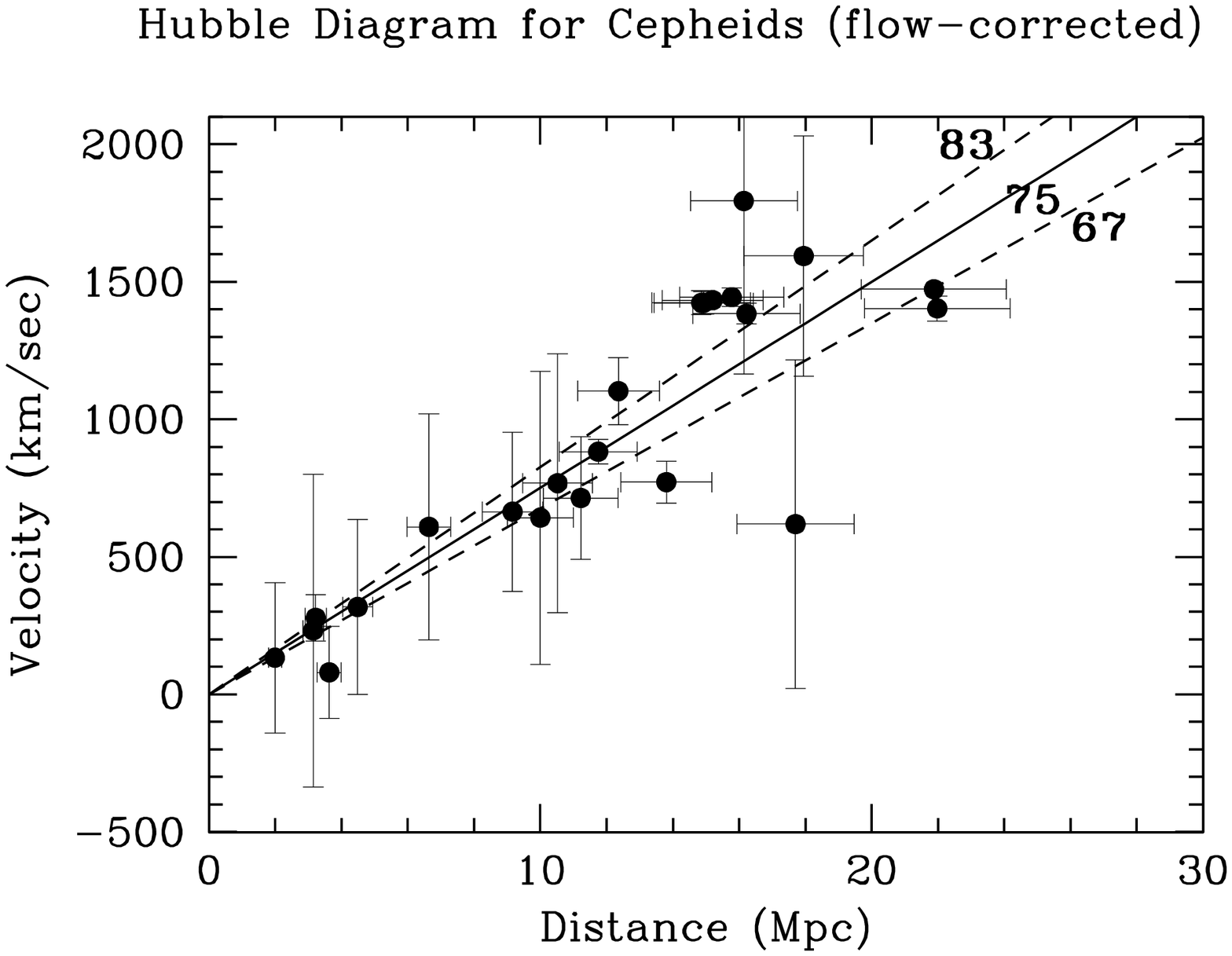}

\clearpage

\epsscale{1.0}
\plotone{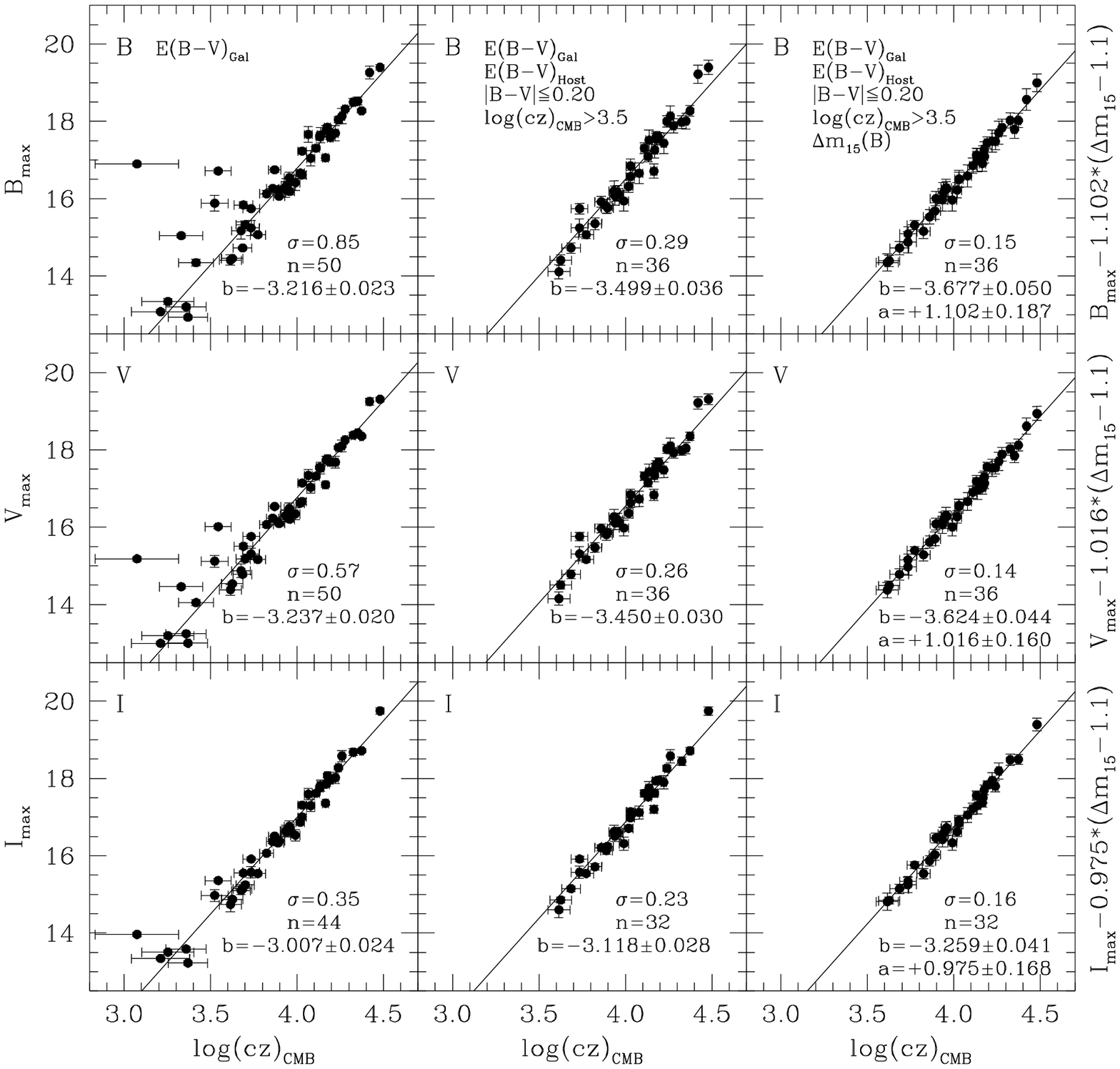}

\clearpage

\epsscale{1.0}
\plotone{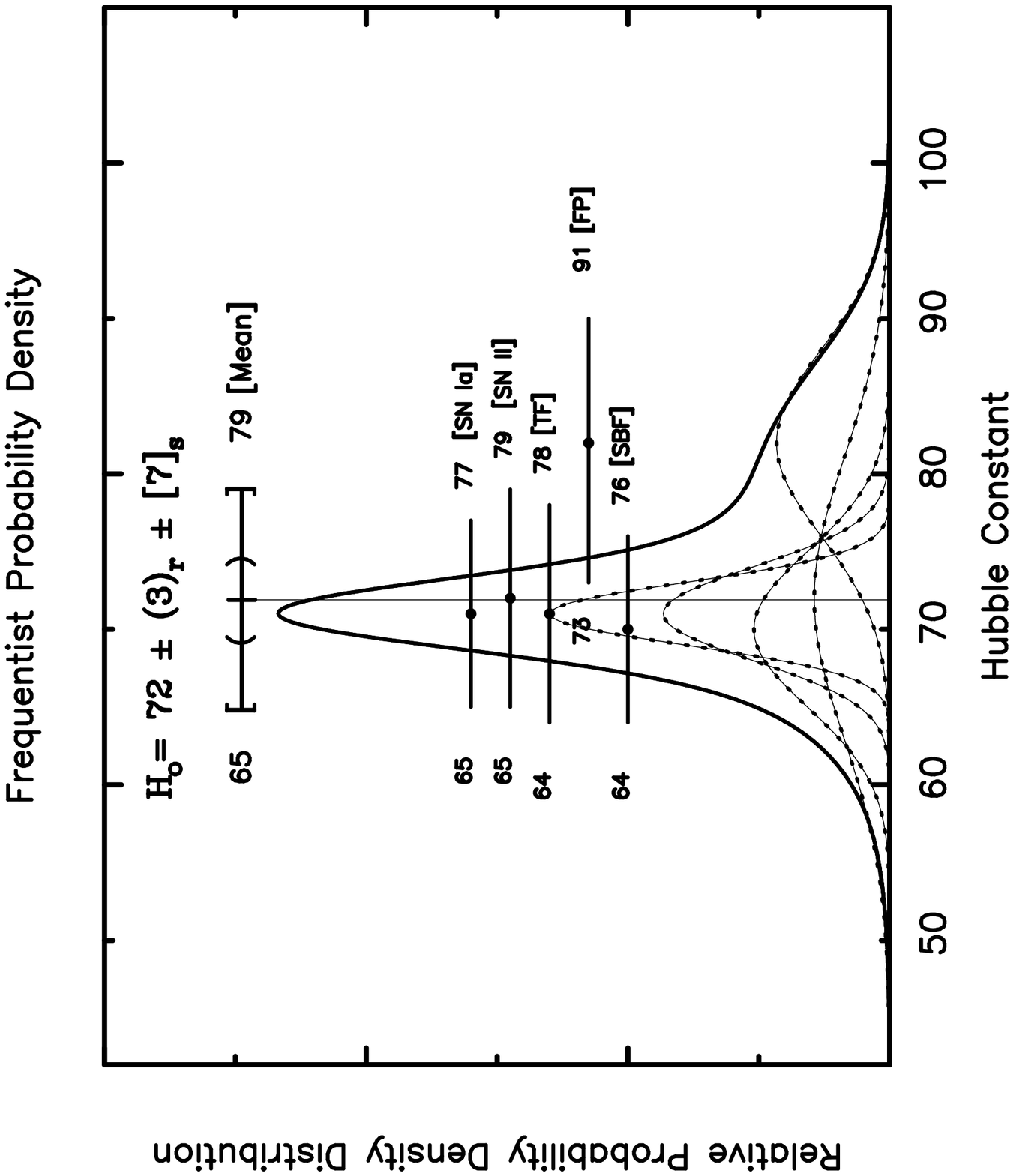}

\clearpage

\epsscale{1.0}
\plotone{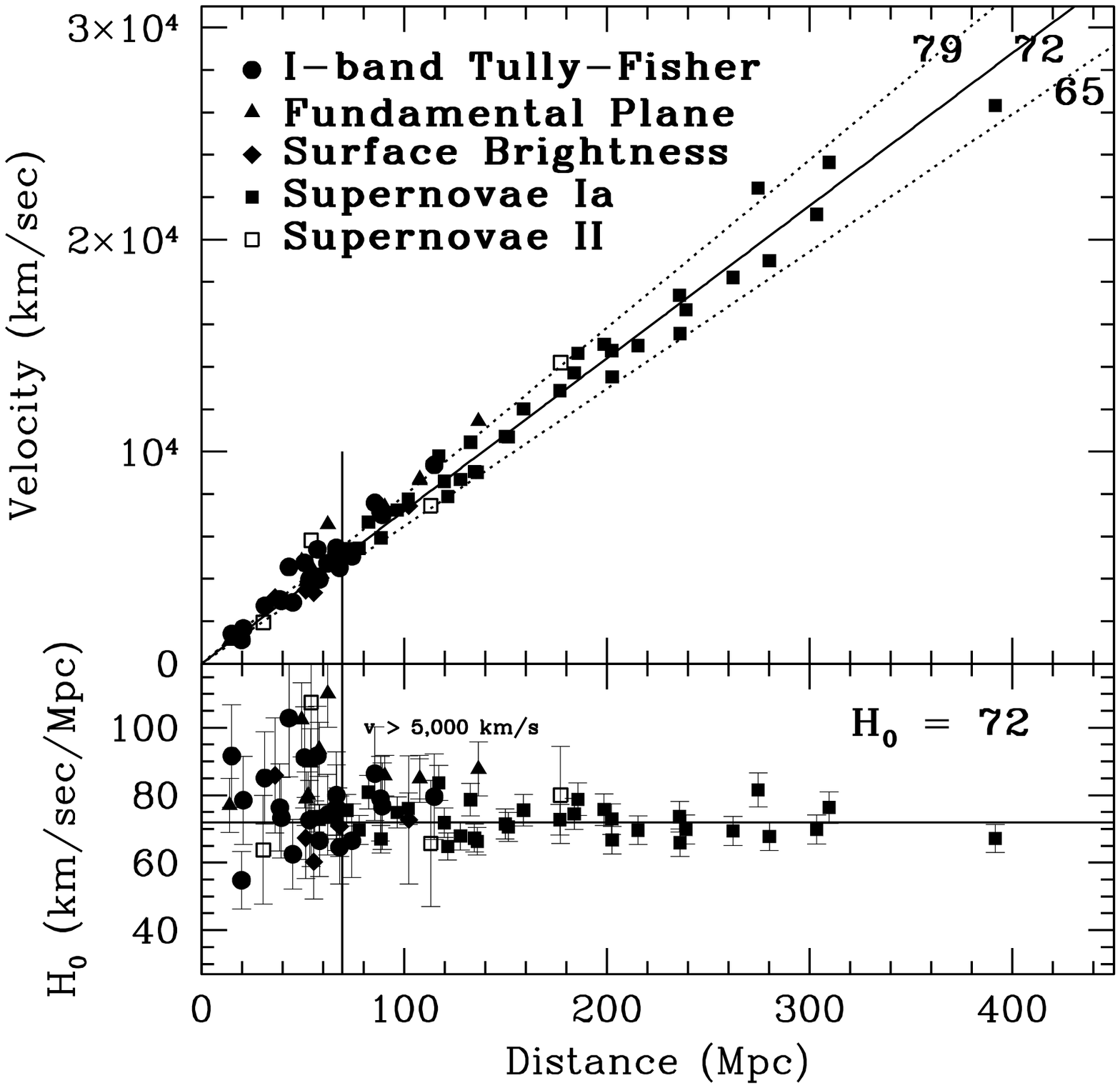}

\epsscale{1.0}
\plotone{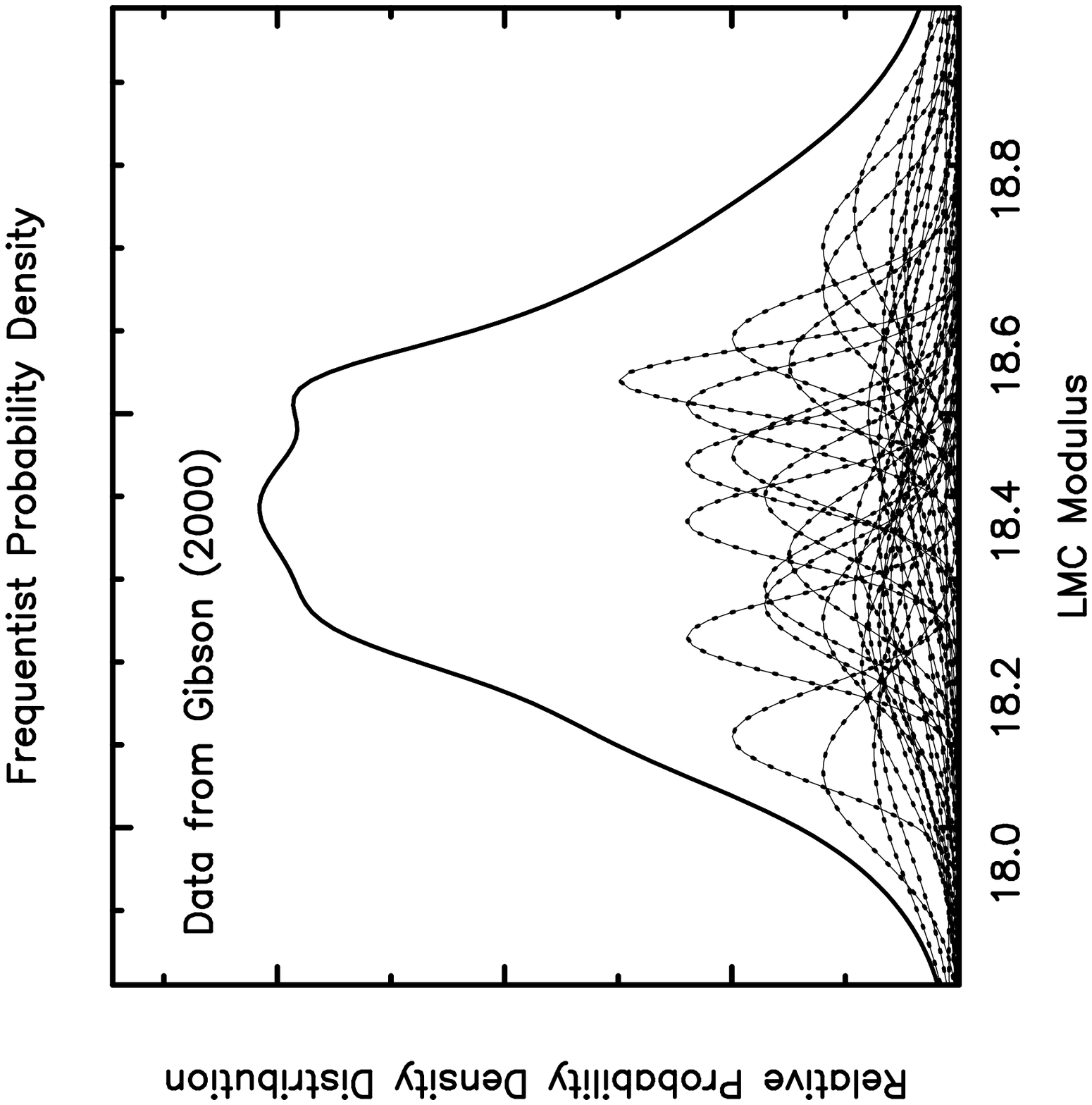}

\clearpage

\epsscale{1.0}
\plotone{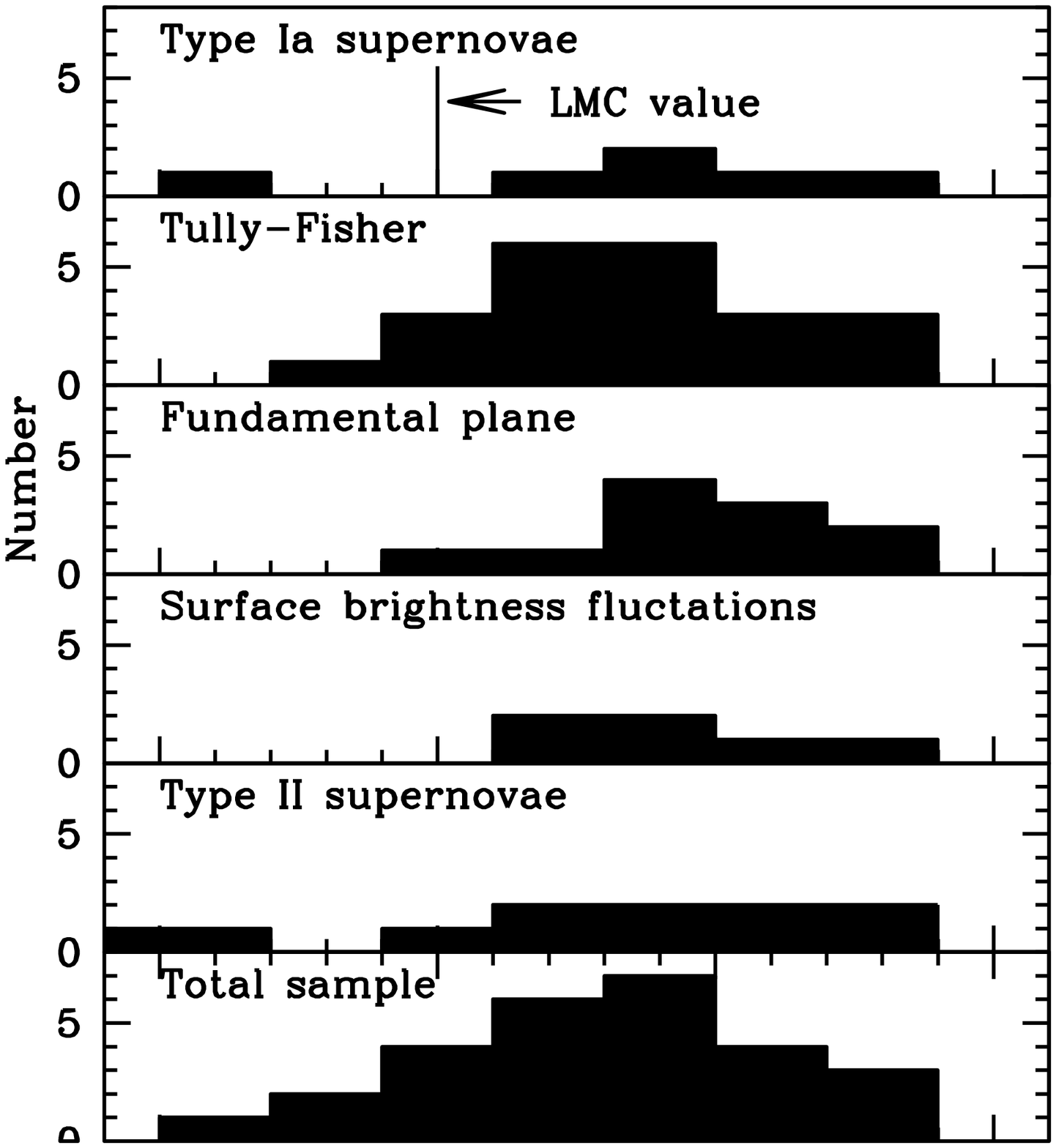}


\epsscale{1.0}
\plotone{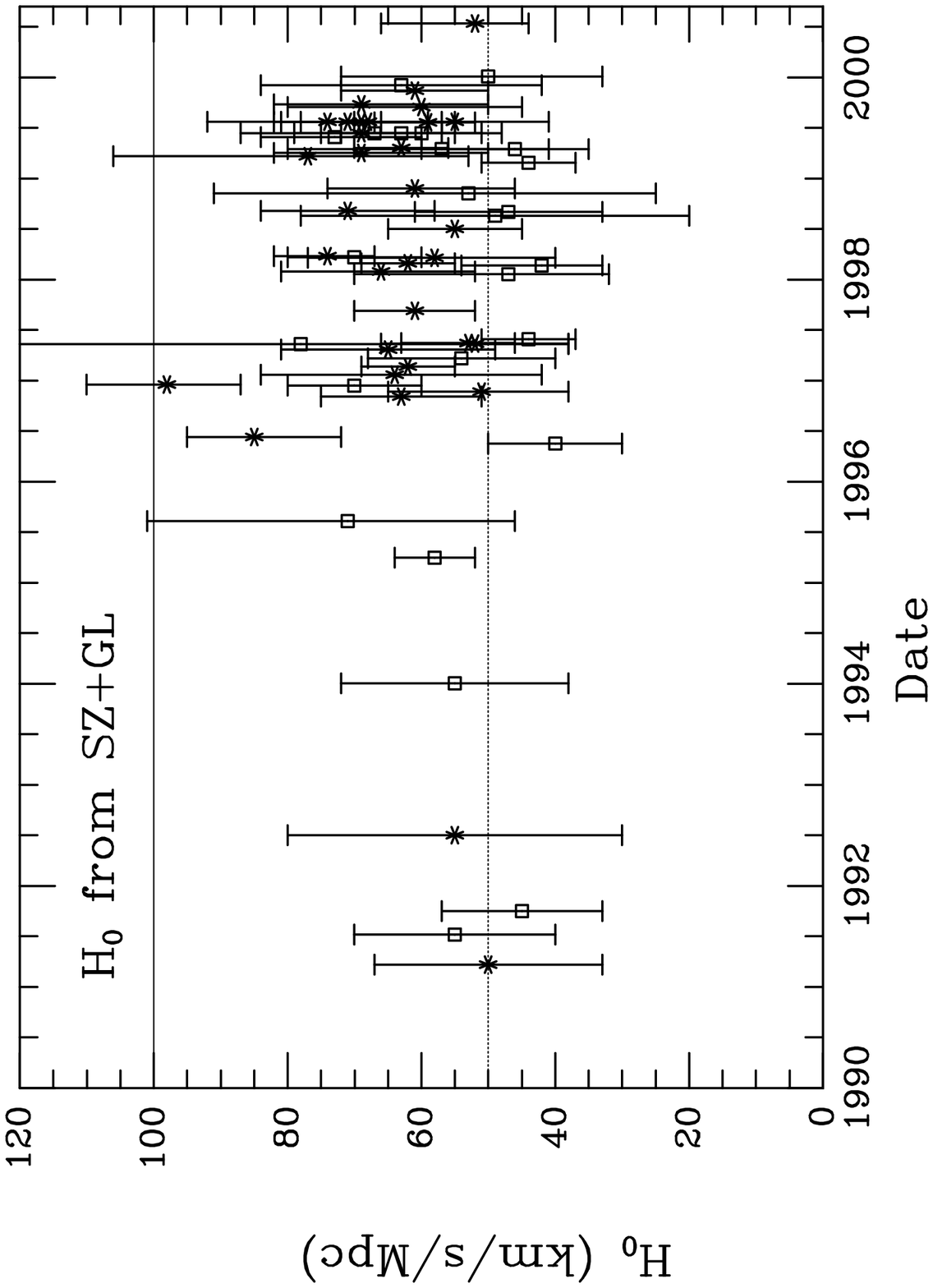}

\clearpage

\epsscale{1.0}
\plotone{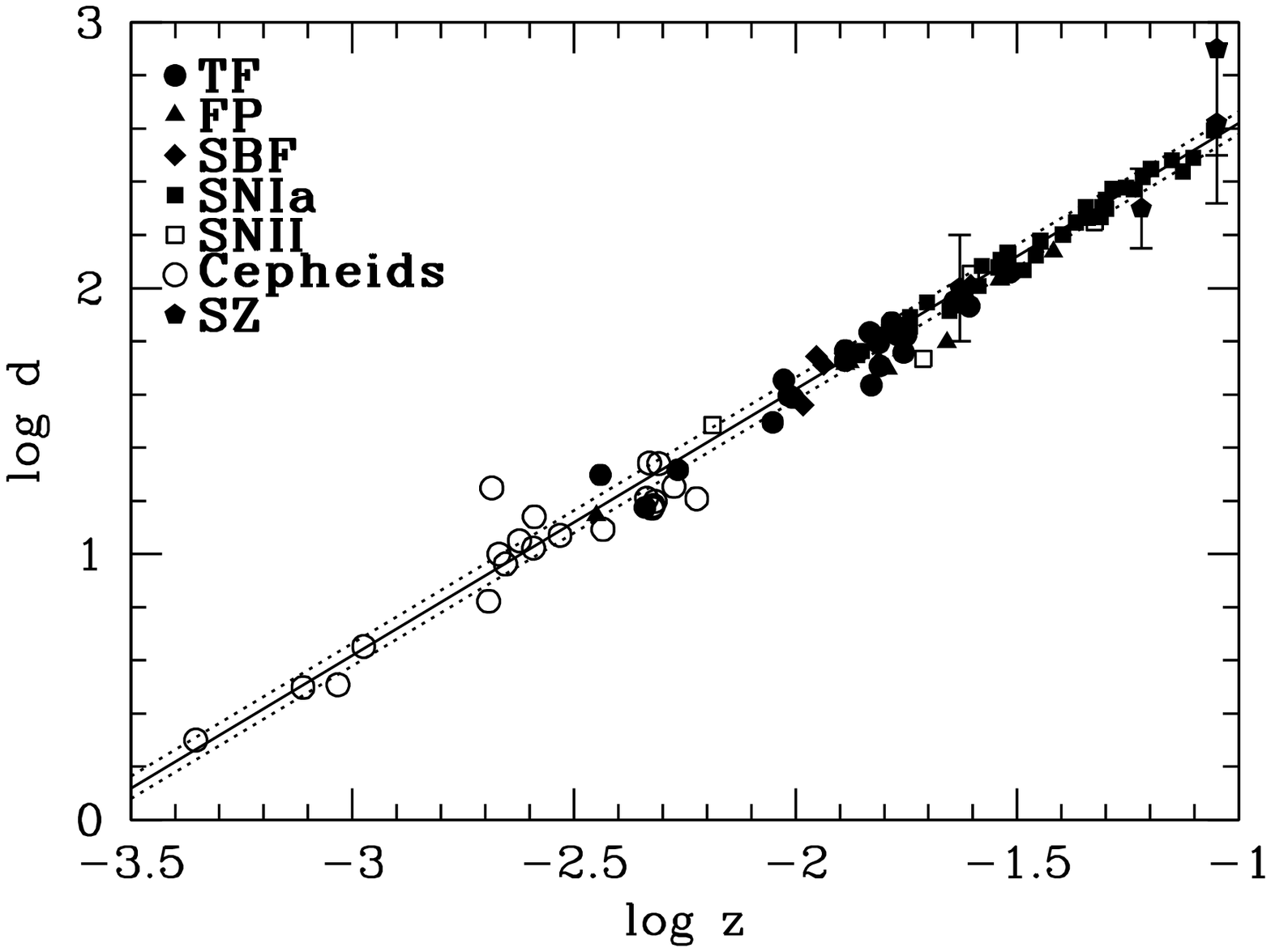}

\clearpage

\epsscale{1.0}
\plotone{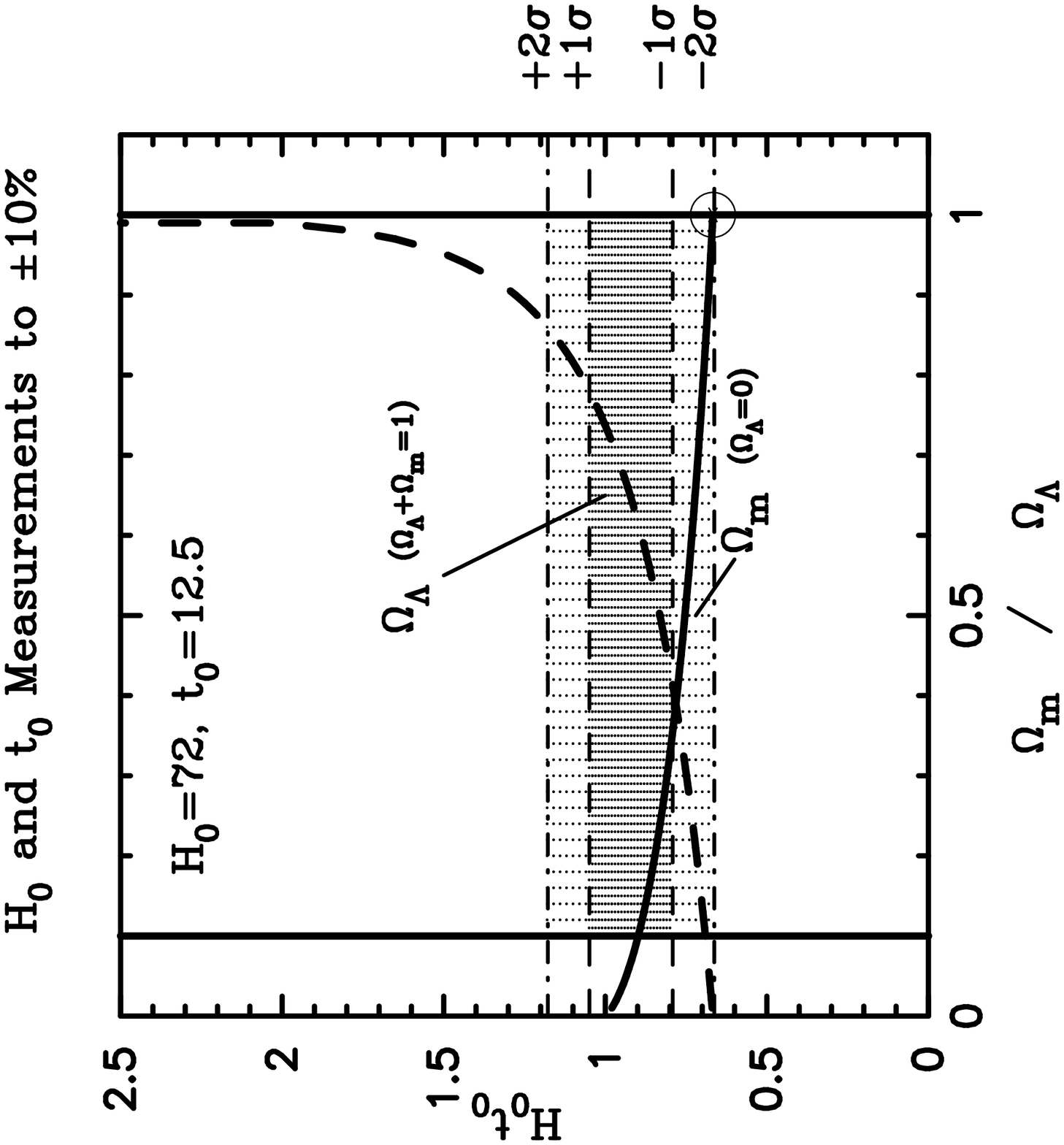}

\clearpage

\epsscale{1.0}
\plotone{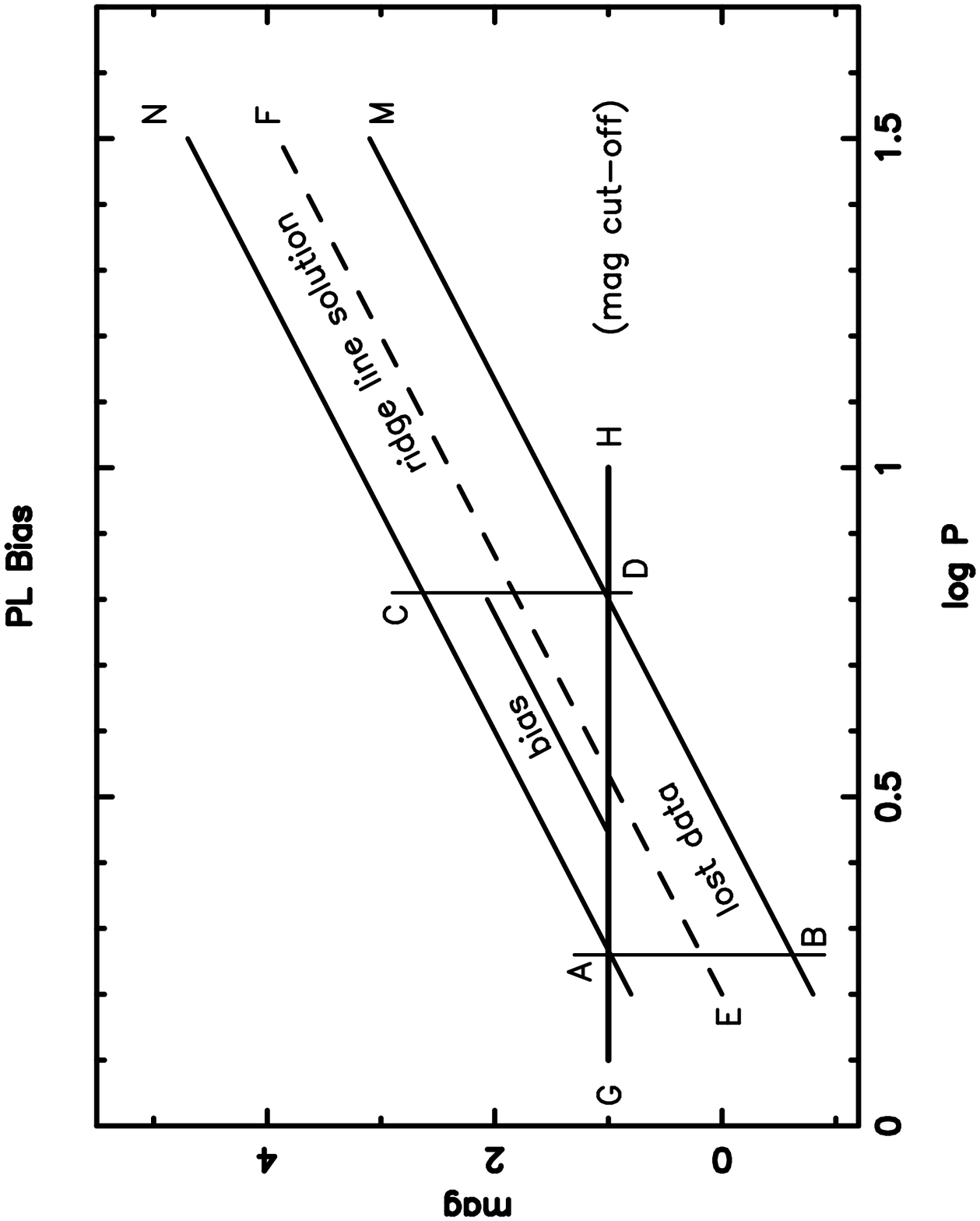}

\end{document}